\documentclass[journal]{IEEEtran}
\IEEEoverridecommandlockouts
\usepackage{cite}
\usepackage[colorlinks=true,linkcolor=black,anchorcolor=black,citecolor=black,filecolor=black,menucolor=black,runcolor=black,urlcolor=magenta]{hyperref}

\usepackage{amsmath,amssymb,amsfonts,mathrsfs}
\usepackage{algorithmic}
\usepackage{graphicx}
\usepackage{textcomp}
\usepackage[defaultcolor=red]{changes}
\usepackage{xcolor}
\usepackage{booktabs}
\usepackage{subeqnarray}
\usepackage{multicol}
\usepackage{multirow}
\usepackage{makecell}
\usepackage{mathtools}
\usepackage{algorithm}
\usepackage{algorithmic}
\usepackage{tikz}
\usepackage{cancel}
\def\BibTeX{{\rm B\kern-.05em{\sc i\kern-.025em b}\kern-.08em
		T\kern-.1667em\lower.7ex\hbox{E}\kern-.125emX}}

\ifCLASSOPTIONcompsoc
  \usepackage[caption=false,font=normalsize,labelfont=sf,textfont=sf]{subfig}
\else
  \usepackage[caption=false,font=footnotesize]{subfig}
\fi

\newcommand{\RE}[1]{\operatorname{Re}\left({#1}\right)}

\newcommand{\argmin}[1]{\underset{#1}{\operatorname{argmin}}}

\newcommand{\minlim}[2]{\underset{\mathrlap{\kern-1em#1}}{\operatorname{min}}\left(#2\right)}
\newcommand{\maxlim}[2]{\underset{\mathrlap{\kern-1em#1}}{\operatorname{max}}\left(#2\right)}

\newcommand{\bs}[1]{\boldsymbol{#1}}
\newcommand{\vc}[1]{\mathrm{vec}\left({#1}\right)}

\begin{document}
\bstctlcite{Settings}

\title{The Choice of Line Lengths in Multiline Thru-Reflect-Line Calibration}
\author{%
	\IEEEauthorblockN{%
		Ziad~Hatab, Michael~E.~Gadringer, and~Wolfgang~B\"osch
	}%
	\thanks{%
       Ziad~Hatab was with the Institute of Microwave and Photonic Engineering, Graz University of Technology, 8010 Graz, Austria, during the development of this work (e-mail: z.hatab@alumni.tugraz.at).%
       \par Michael~E.~Gadringer, and~Wolfgang~B\"osch are with the Institute of Microwave and Photonic Engineering, Graz University of Technology, 8010 Graz, Austria, and with the Christian Doppler Laboratory for Technology-Guided Electronic Component Design and Characterization (TONI), 8010 Graz, Austria (e-mail: \{michael.gadringer, wbosch\}@tugraz.at).%
       \par This work was supported by the Christian Doppler Research Association, the Austrian Federal Ministry for Digital and Economic Affairs, and the National Foundation for Research, Technology and Development.%
       \par Software implementation and measurement data are available at:\\ 
	    \url{https://github.com/ziadhatab/line-length-multiline-trl-calibration}%
	}%
}%
\markboth{This work has been accepted for publication in the IEEE Transactions on Instrumentation and Measurement}{}
\maketitle

\begin{abstract}
    This paper presents an analysis and rigorous procedure for determining the optimal lengths of line standards in multiline thru-reflect-line (TRL) calibration of vector network analyzers (VNAs). The solution is obtained through nonlinear constrained optimization of the eigenvalue problem in multiline TRL calibration. Additionally, we propose a simplified approach for near-optimal length selection based on predefined sparse rulers. Alongside the length calculation, we discuss the required number of lines to meet bandwidth requirements. The proposed methods are validated through measurements of multiple multiline TRL calibration kits on printed circuit boards of different materials and stackups, covering frequencies up to $150\,\mathrm{GHz}$. A measurement-based Monte Carlo uncertainty analysis, using error boxes derived from impedance standard substrate measurements, demonstrates that the proposed line lengths distribute calibration uncertainty more evenly across lines compared to a commercial calibration kit. Practical examples are provided for various applications, including lossy and dispersive lines, as well as banded solutions for waveguides.
\end{abstract}

\begin{IEEEkeywords}
vector network analyzer, calibration, metrology, microwave measurement
\end{IEEEkeywords}

\section{Introduction}
\label{sec:1}

\IEEEPARstart{M}{easurement} traceability for scattering parameters (S-parameters) at microwave frequencies is established through calibration procedures. An important vector network analyzer (VNA) calibration technique is the thru-reflect-line (TRL) method \cite{Engen1979}. This calibration approach leverages the characteristic impedance of the line and thru standard to determine the calibration's reference impedance. In metrology applications, air lines are used for this purpose \cite{stumper2005uncertainty}. Due to their simple mechanical setup, air lines allow linking the lines' characteristic impedance to their material properties (permittivity, permeability, conductivity) and geometrical dimensions, as the isolating material (air) is well known and the geometrical dimensions can be measured traceably to the SI system of units \cite{Davis2019SISystemOfUnitsRevised}. The same approach applies to lines manufactured with different materials and manufacturing technologies, which explains the widespread use of TRL calibration. For broadband measurements, the multiline TRL \cite{Marks1991,Hatab2022} was suggested, which employs multiple transmission lines with identical cross-sectional properties but different lengths.

The classical TRL formulation involves a single eigenvalue problem based on the length difference between the thru and line standards. Selecting this length difference is straightforward, as the eigenvalue problem exhibits optimal stability when the eigenvalue phases are $\pm90^\circ$. The eigenvalues, being phase conjugates, coincide at phases $\pm0^\circ$ or $\pm180^\circ$. The $\pm90^\circ$ phase occurs at a single frequency, typically chosen as the center frequency. The usable bandwidth is then defined by arbitrary phase margins as thresholds, commonly between $\pm20^\circ$ and $\pm30^\circ$ \cite{Microwaves101}. Due to the cyclic nature of the phase, TRL can cover multiple bands, with the fundamental band at quarter-wavelengths. This property enables the use of longer line lengths for waveguides, resulting in ``3/4-wave'' or ``5/4-wave'' TRL kits \cite{Ridler2009,Ridler2019,Ridler2021}.

While length selection in lossless TRL calibration is straightforward, losses introduce complications. With losses, eigenvalues follow a spiral rotation in the complex plane and never exactly coincide at phases $\pm0^\circ$ or $\pm180^\circ$. Higher losses cause the line to more closely resemble a load standard, thereby broadening its bandwidth. This broadening assumes that the thru standard's back-to-back length is not so long as to introduce significant losses, which would impede transmission. The line standard, however, can be very lossy, and in extreme cases can be replaced by a load, resulting in thru-reflect-match (TRM) calibration \cite{Eul1988}.

As modern radio-frequency (RF) and millimeter-wave (mmWave) applications demand wider bandwidth measurements, multiline TRL calibration has become essential. Despite the existence of multiline TRL algorithms for over three decades \cite{Marks1991,DeGroot2002,Williams2003,Hatab2022}, optimal line length selection remains understudied. Current literature primarily focuses on selecting line lengths relative to the thru standard based on phase margin transitions. The Microwaves101 online resource provides a useful spreadsheet \cite{Microwaves101} for determining appropriate lengths based on the relative effective permittivity. Similar approaches are discussed in \cite{Hoer1983, Hoer1987a, Marks1991, Ridler2009, ReynosoHernandez2002, Shin2024}. It is important to note that these methods apply to the lossless multiline TRL case. Including the losses in this evaluation can relax the phase-margin constraints.

A fundamental challenge in optimizing line length selection arises from the multiline TRL implementation method itself. Traditional approaches decompose the problem into several TRL pairs with a common reference line \cite{Marks1991,DeGroot2002}. Since these methods independently select reference lines at each frequency point, they complicate the problem formulation, leading to the current conventional approach of fixing the thru as the reference when computing the lengths.

This article rigorously formulates the line length selection problem based on the multiline procedure from \cite{Hatab2022}. This method reduces the multiline problem to solving a single weighted eigenvalue problem, enabling a formulation similar to the classical TRL calibration.

The remainder of this article is organized as follows: Section~\ref{sec:2} analyzes the frequency limitations in classical TRL and multiline TRL methods, introducing the concept of effective phase for multiline TRL calibration. Section~\ref{sec:3} develops an optimization approach for line length selection based on minimizing eigenvector sensitivity and introduces a simplified line selection method using sparse rulers, as well as a discussion on determining the recommended number of lines for a given frequency bandwidth. Section~\ref{sec:4} validates the proposed methods through printed circuit board (PCB) measurements and a measurement-based Monte Carlo uncertainty analysis, comparing the proposed line lengths against those of a commercial impedance standard substrate. Section~\ref{sec:5} provides examples of computing line lengths for different scenarios. Lastly, Section~\ref{sec:6} offers concluding remarks.


\section{Frequency Limitation in TRL and Multiline TRL Calibrations}
\label{sec:2}

The purpose of this section is to review the frequency limitations in both classical TRL and multiline TRL methods, and to explain the relationship between line length and frequency based on the eigenvalues. Additionally, we introduce the concept of an effective phase for multiline TRL calibration, which aims to provide a phase concept similar to that used in the classical TRL method. The development of this measure will allow us to compare the performance of line length sets even with different numbers of lines.

\begin{figure}[th!]
    \centering
    \includegraphics[width=0.95\linewidth]{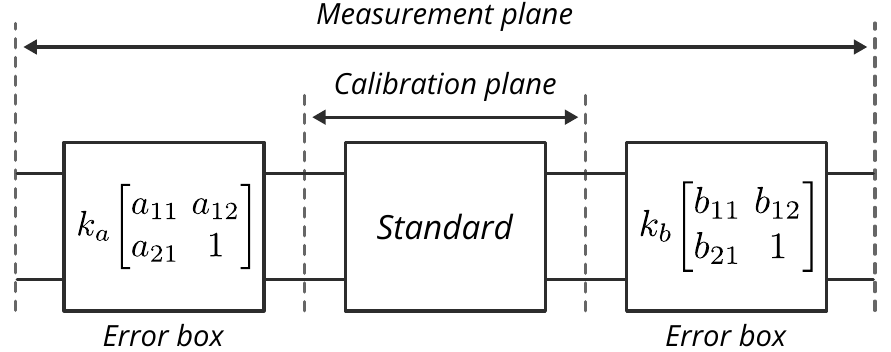}
    \caption{Error box model of a two-port VNA for the measurement of calibration standards.}
    \label{fig:2.1}
\end{figure}

In both classical TRL and multiline TRL, we use the error box model that describes a two-port VNA, as shown in Fig.~\ref{fig:2.1}. This model applies the matrix product for T-parameters to establish the desired relationship:
\begin{equation}
    \bs{M}_i = \underbrace{k_ak_b}_{k}\underbrace{\begin{bmatrix} a_{11} & a_{12} \\ a_{21} & 1\end{bmatrix}}_{\bs{A}}\underbrace{\begin{bmatrix}e^{-\gamma l_i} & 0 \\ 0 & e^{\gamma l_i}\end{bmatrix}}_{\bs{L}_i}\underbrace{\begin{bmatrix} b_{11} & b_{12} \\ b_{21} & 1\end{bmatrix}}_{\bs{B}}
    \label{eq:2.1}
\end{equation}
where $\bs{A}$ and $\bs{B}$ are the left and right error boxes and $k$ is the transmission error term. The matrix $\bs{L}_i$ corresponds to the line model of the $i$th measured line.

\subsection{Eigenvalues in Classical TRL}
\label{sec:2a}
For the discussion that follows, we maintain generality by using a non-zero length line instead of a thru standard, known as line-reflect-line (LRL) calibration~\cite{Hoer1987}. When considering two lines of different lengths, we can formulate an eigenvalue problem by taking the T-parameters measurements of two lines and multiplying the inverse of one by the other:
\begin{equation}
    \bs{M}_i\bs{M}_j^{-1} = \bs{A}\begin{bmatrix}e^{-\gamma l_{ij}} & 0 \\ 0 & e^{\gamma l_{ij}}\end{bmatrix}\bs{A}^{-1}, \qquad l_{ij} = l_i - l_j.
    \label{eq:2.2}
\end{equation}

The eigenvalues depend on the complex propagation constant $\gamma$, which is frequency-dependent and can be expressed in terms of the relative effective permittivity \cite{Marks1992} as follows:
\begin{equation}
    \gamma = \alpha + j\beta = \frac{2\pi f}{c_0}\sqrt{-\epsilon_\mathrm{r,eff}}
    \label{eq:2.3}
\end{equation}
where $\epsilon_\mathrm{r,eff} = \epsilon_\mathrm{r,eff}^\prime - j\epsilon_\mathrm{r,eff}^{\prime\prime}$, $f$ is the frequency, and $c_0$ is the speed of light in vacuum. In \eqref{eq:2.3}, while the frequency dependence of the propagation constant is linear with respect to $\epsilon_\mathrm{r,eff}$, in general $\epsilon_\mathrm{r,eff}$ itself can be frequency-dependent. Thus the propagation constant can be nonlinear with frequency. The special case where $\epsilon_\mathrm{r,eff}$ remains constant occurs only in transmission lines that support a single transverse electromagnetic mode (TEM) \cite{Steer2019}.

\begin{figure}[th!]
    \centering
    \subfloat[]{\includegraphics[width=0.98\linewidth]{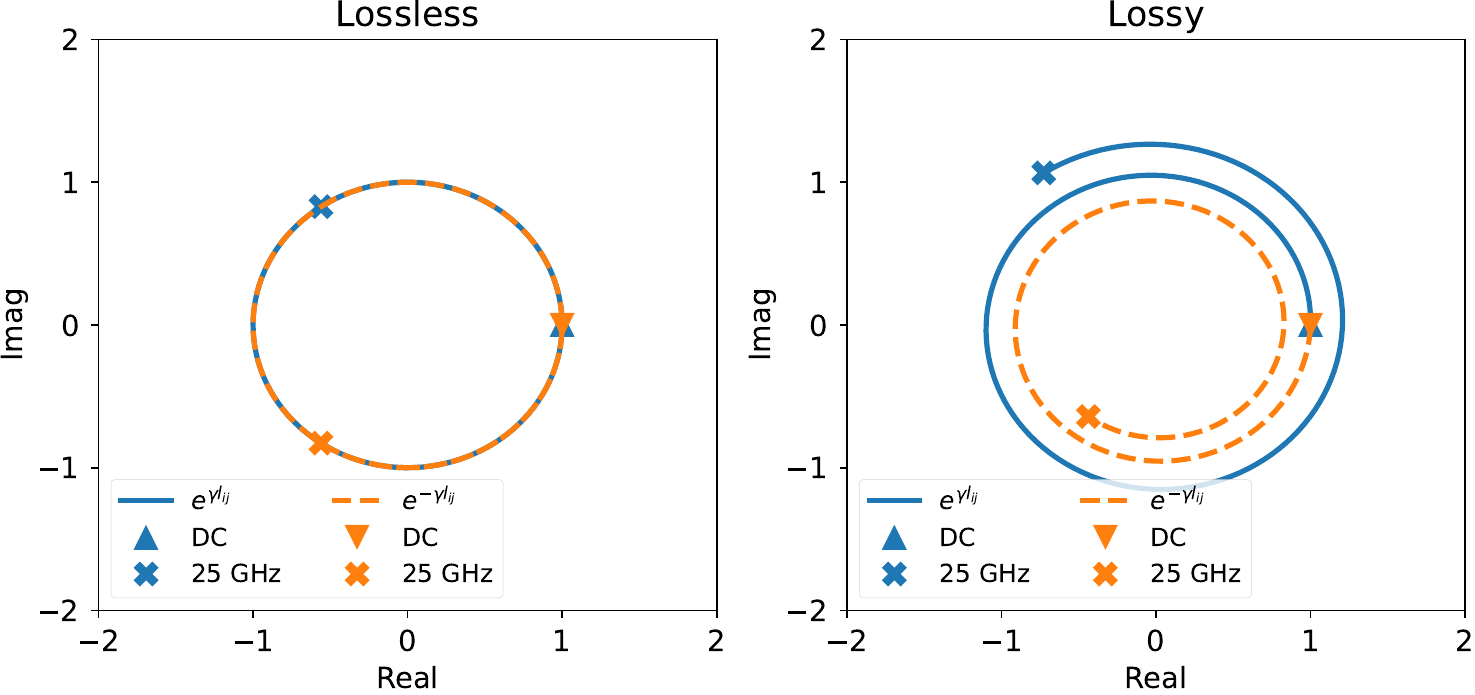}\label{fig:2.2a}}\\[5pt]
    \subfloat[]{\includegraphics[width=1\linewidth]{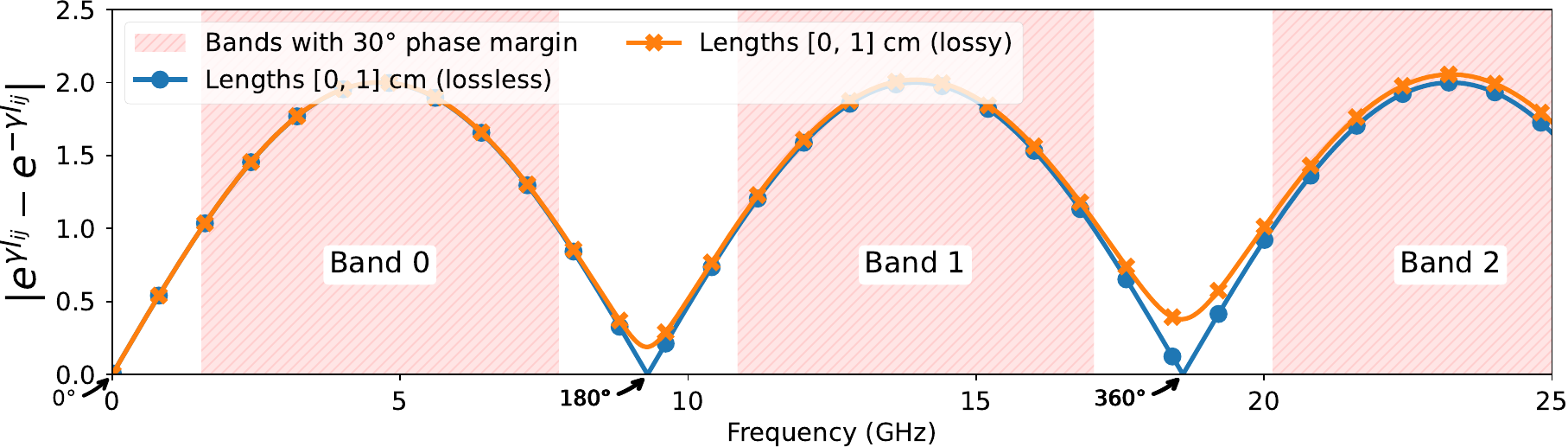}\label{fig:2.2b}}
    \caption{Example illustrating the evolution of the eigenvalues in classical TRL calibration as a function of frequency. (a) Complex plane representation, (b) the eigengap response. Parameters used in this example: length difference of 1~cm, frequency range of 0-25~GHz, and $\epsilon_\mathrm{r,eff} = 2.6$ for the lossless case, while $\epsilon_\mathrm{r,eff} = 2.6(1-0.06j)$ for the lossy case.}
    \label{fig:2.2}
\end{figure}

The real part of $\gamma$ represents attenuation per unit length, while the imaginary part relates to the wave's velocity. As $\gamma$ varies with frequency, the eigenvalues rotate in the complex plane, as illustrated in Fig.~\ref{fig:2.2a} for both lossless and lossy cases.

In classical TRL, frequency limitations are most pronounced in lossless transmission lines. The eigenvalues become identical when the phase $\gamma l_{ij}$ equals $0^\circ$ or $180^\circ$ (or their integer multiples). At these critical points, the eigendecomposition fails, making it impossible to extract the error box parameters. This limitation is most severe in the ideal lossless case, where the propagation constant simplifies to $\gamma_\mathrm{lossless} = j\beta = \frac{2\pi f}{c_0}j\sqrt{\epsilon_\mathrm{r,eff}^\prime}$. Fig.~\ref{fig:2.2b} demonstrates this phenomenon by showing how the eigengap (the error vector magnitude between eigenvalues) reaches zero at these points.

To ensure reliable calibration, the line length difference should maintain eigenvalues within a phase margin, avoiding the $0^\circ$ and $180^\circ$ points. The calibration boundaries, incorporating a phase margin, are:
\begin{equation}
    \pi n + \pi\frac{\varphi}{180} \leq l_{ij}\frac{2\pi f}{c_0}\sqrt{\epsilon_\mathrm{r,eff}^\prime} \leq \pi n + \left(1-\frac{\varphi}{180}\right)\pi
    \label{eq:2.4}
\end{equation}
where $\varphi$ is the phase margin in degrees and $n=0,1,2,\ldots$ indicates the band number, with $n=0$ being the fundamental band (e.g., see Fig.~\ref{fig:2.2b}). This can be rewritten as frequency bounds as follows:
\begin{equation}
    \frac{n + \varphi/180}{2l_{ij}\sqrt{\epsilon_\mathrm{r,eff}^\prime}}c_0 \leq f \leq \frac{n + 1-\varphi/180}{2l_{ij}\sqrt{\epsilon_\mathrm{r,eff}^\prime}}c_0,
    \label{eq:2.5}
\end{equation}
or as minimum and maximum frequencies as follows:
\begin{equation}
    f_\mathrm{min} = \frac{n + \varphi/180}{2l_{ij}\sqrt{\epsilon_\mathrm{r,eff}^\prime}}c_0, \quad f_\mathrm{max} = \frac{n + 1-\varphi/180}{2l_{ij}\sqrt{\epsilon_\mathrm{r,eff}^\prime}}c_0.
    \label{eq:2.6}
\end{equation}

The required line length difference can be determined by solving either of the equations in \eqref{eq:2.6}, leading to:
\begin{subequations}
    \begin{align}
         l_{ij} &= \frac{c_0}{2f_\mathrm{min}\sqrt{\epsilon_\mathrm{r,eff}^\prime}}\left(n + \frac{\varphi}{180}\right),\label{eq:2.7a}\\
         l_{ij} &= \frac{c_0}{2f_\mathrm{max}\sqrt{\epsilon_\mathrm{r,eff}^\prime}}\left(n + 1 - \frac{\varphi}{180}\right),\label{eq:2.7b}
    \end{align}
    \label{eq:2.7}
\end{subequations}

For \eqref{eq:2.7} to be consistent for given minimum and maximum frequencies, the phase wrapping constant $n$ can be determined by equating both equations, which yields:
\begin{equation}
    n = \left\lfloor \frac{q - (q+1)\varphi/180}{1-q} \right\rfloor,
    \label{eq:2.8}
\end{equation}
where $q = f_\mathrm{min}/f_\mathrm{max}$. 

Since $n$ must be an integer, the actual achieved phase margin for given frequency requirements becomes:
\begin{equation}
    \varphi = 180 \frac{nq-n+q}{q+1}.
    \label{eq:2.9}
\end{equation}

For wide bandwidth requirements, the calculation from \eqref{eq:2.8} will result in $n<1$, which forces $n=0$. Consequently, the achieved phase margin may be less than specified. For phase margins $\geq 20^\circ$, it is recommended that $f_\mathrm{max} \leq 8f_\mathrm{min}$. Generally, for phase margins between $0^\circ$ and $90^\circ$ at the fundamental band (i.e., $n=0$), the relationship between minimum and maximum frequencies must satisfy:
\begin{equation}
    f_\mathrm{max} \leq \frac{180-\varphi}{\varphi}f_\mathrm{min}.
    \label{eq:2.10}
\end{equation}

Once $n$ is determined for a given phase margin and frequency limits, the line length difference $l_{ij}$ can be calculated using either \eqref{eq:2.7a} or \eqref{eq:2.7b}. Note that throughout \eqref{eq:2.4}-\eqref{eq:2.10}, $\epsilon_\mathrm{r,eff}^\prime$ is treated as constant. While this is generally not true, we can use either an $\epsilon_\mathrm{r,eff}^\prime$ averaged over the covered frequency range or the DC value for quasi-TEM transmission lines. This constant-permittivity assumption is used only for the analytical length calculation in this section and for the sparse ruler method in Section~\ref{sec:3b}. In the general optimization approach of Section~\ref{sec:3a}, frequency-dependent and lossy models of $\epsilon_\mathrm{r,eff}$ can be supplied directly to the optimizer.

\subsection{Eigenvalues in Multiline TRL}
\label{sec:2b}
This section discusses the multiline TRL formulation from \cite{Hatab2022}. This approach differs from traditional multiline TRL calibration implementations by using a Kronecker product formulation, where the error box model of each line is vectorized using Kronecker product properties, as written below:
\begin{equation}
    \vc{\bs{M}_i} = k(\bs{B}^T\otimes\bs{A})\vc{\bs{L}_i},
    \label{eq:2.11}
\end{equation}
where $\otimes$ is the Kronecker product and $\vc{\cdot}$ is the vectorization operation \cite{Brewer1978}. The terms $\bs{A}$, $\bs{B}$, $k$, and $\bs{L}_i$ are the same as given in \eqref{eq:2.1}.

For $N\geq2$ lines, the equations describing the error boxes for all line measurements are summarized in two equations as given below:
\begin{subequations}
    \begin{align}
        \bs{M} &= k\bs{X}\bs{L},\label{eq:2.12a}\\[5pt] 
        \bs{D}^{-1}\bs{M}^{T}\bs{P}\bs{Q} &= \frac{1}{k}\bs{L}^{T}\bs{P}\bs{Q}\bs{X}^{-1},\label{eq:2.12b}
    \end{align}
    \label{eq:2.12}
\end{subequations}
where 
\begin{subequations}
    \begin{align}
        \bs{M} &= \begin{bmatrix} \vc{\bs{M}_1} & \vc{\bs{M}_2} & \cdots & \vc{\bs{M}_N} \end{bmatrix},\label{eq:2.13a}\\[5pt]
        \bs{L} &= \begin{bmatrix} \vc{\bs{L}_1} & \vc{\bs{L}_2} & \cdots & \vc{\bs{L}_N} \end{bmatrix},\label{eq:2.13b}\\[5pt]
        \bs{X} &= \bs{B}^T\otimes\bs{A},\label{eq:2.13c}\\[5pt]
        \bs{D} &= \mathrm{diag}\left( \begin{bmatrix} \det(\bs{M}_1) & \cdots & \det(\bs{M}_N) \end{bmatrix}\right),\label{eq:2.13d}\\[5pt]
        \bs{P} &= \begin{bmatrix}
            1 & 0 & 0 & 0\\
            0 & 0 & 1 & 0\\
            0 & 1 & 0 & 0\\
            0 & 0 & 0 & 1
        \end{bmatrix}, \quad \bs{Q} = \begin{bmatrix}
            0 & 0  & 0  & 1\\
            0 & -1 & 0  & 0\\
            0 & 0  & -1 & 0\\
            1 & 0  & 0  & 0
        \end{bmatrix}.\label{eq:2.13e}
    \end{align}
    \label{eq:2.13}
\end{subequations}

The eigenvalue problem is formulated by first multiplying an $N\times N$ weighting matrix $\bs{W}$ on the right-hand side of \eqref{eq:2.12a}, and then multiplying this result on the left-hand side of \eqref{eq:2.12b}, resulting in:
\begin{equation}
    \underbrace{\bs{M}\bs{W}\bs{D}^{-1}\bs{M}^T\bs{P}\bs{Q}}_{\bs{F}:\ 4\times4} = \bs{X}\underbrace{\bs{L}\bs{W}\bs{L}^T\bs{P}\bs{Q}}_{\bs{H}:\ 4\times4}\bs{X}^{-1}.
    \label{eq:2.14}
\end{equation}

The equation in \eqref{eq:2.14} represents a similarity transformation between matrices $\bs{F}$ and $\bs{H}$, with $\bs{X}$ as the transformation matrix. By choosing an appropriate $\bs{W}$ as described in \cite{Hatab2022}, matrix $\bs{H}$ becomes diagonal, transforming the similarity problem into an eigenvalue problem:
\begin{equation}
    \bs{F} = \bs{X}\begin{bmatrix}
        -\lambda & 0 & 0 & 0\\
        0 & 0 & 0 & 0\\
        0 & 0 & 0 & 0\\
        0 & 0 & 0 & \lambda
    \end{bmatrix}\bs{X}^{-1},
    \label{eq:2.15}
\end{equation}
where
\begin{equation}
    \lambda = \sum\limits_{\begin{gathered}\\[-18pt]
            \scriptstyle i=1\\[-8pt]\scriptstyle i<j\leq N\end{gathered}}^{N-1}\left|e^{\gamma l_{ij}}-e^{-\gamma l_{ij}}\right|^2 = \frac{1}{2}\left\|\bs{W}\right\|_F^2.
    \label{eq:2.16}
\end{equation}

The details on computing $\bs{W}$ are found in \cite{Hatab2022} and \cite{Hatab2023}. The solution for the calibration coefficients is obtained from the eigenvectors associated with eigenvalues $\pm\lambda$. A solution exists only when $\lambda$ is non-zero. Similar to classical TRL, if we consider only two lines, $\lambda$ becomes zero at $\gamma l_{ij} = 0^\circ$ and $\gamma l_{ij} = 180^\circ$ (or their integer multiples).

Unlike classical TRL, deriving analytical expressions for optimal line lengths is not possible for more than two lines due to the multiple terms in the summation of \eqref{eq:2.16}. In Section~\ref{sec:3}, we present our proposed methods for computing line lengths via a constrained optimization approach and sparse rulers.

\begin{figure}[th!]
    \centering
    \includegraphics[width=1\linewidth]{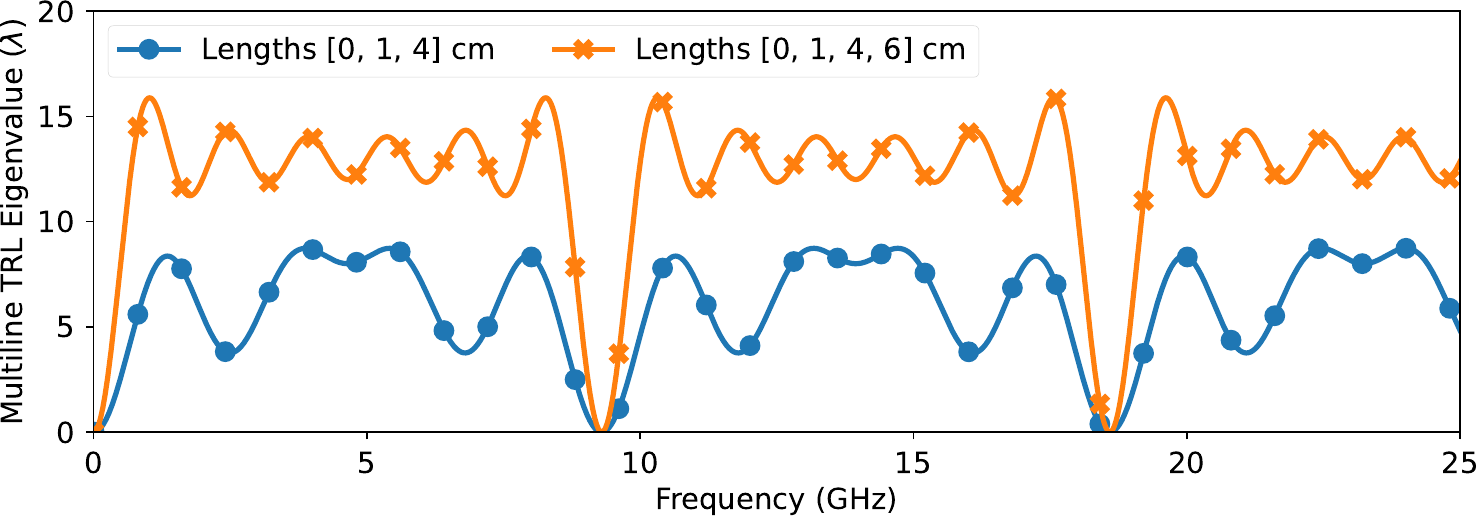}
    \caption{Example illustrating the eigenvalue behavior in multiline TRL calibration for three and four-line configurations. The figure demonstrates how eigenvalues increase with the number of lines used. Relative effective permittivity assumed lossless $\epsilon_\mathrm{r,eff} = 2.6-0j$.}
    \label{fig:2.3}
\end{figure}

Another important consideration is that the eigenvalue in multiline TRL calibration increases significantly with the number of transmission lines used, which is a natural consequence of multiple line pairs, unlike classical TRL, which uses only a single line pair. Fig.~\ref{fig:2.3} illustrates this eigenvalue behavior for configurations with three and four lines. Specifically, at approximately 5\,GHz, the three-line configuration exhibits an eigenvalue of approximately 8, whereas the four-line configuration shows an eigenvalue of approximately 12. Consequently, normalization becomes necessary to enable meaningful comparisons between line sets of different counts and lengths. This normalization approach is discussed in detail in Section~\ref{sec:2c}.

It is worth noting that even when using multiple lines to expand the calibration frequency range, the eigenvalue in multiline TRL calibration still exhibits a cyclic behavior in the lossless case. This occurs particularly when the lengths of lines are harmonically related, as depicted in Fig.~\ref{fig:2.3}. This periodic behavior is analogous to that observed in classical TRL, enabling multiple usable frequency bands across the spectrum. For lossy lines, the eigenvalue $\lambda$ never reaches zero except at DC frequency.

\subsection{Defining an Effective Phase for Multiline TRL}
\label{sec:2c}
We now define an effective phase for multiline TRL. Unlike classical TRL, associating the phase with the multiline TRL is challenging due to the multiple line pairs summed in the eigenvalue expression of \eqref{eq:2.16}. Furthermore, since the eigenvalue in multiline TRL increases with the number of lines, direct comparison with classical TRL or quantifying which set of lines performs better becomes difficult. Therefore, we need a normalization that scales properly with the number of lines.

Introducing such normalization to the multiline TRL eigenvalue does not affect the weighted eigendecomposition in \eqref{eq:2.14}, as all eigenvalues are scaled by the same factor. The sensitivity of the eigenvectors remains unaffected as long as this normalization is non-zero. Using this normalized eigenvalue, we can define an effective phase equivalent to that of classical TRL.

It is important to note that the phase concept in TRL is just one established metric for classical TRL that we use for historical consistency. Additionally, when two sets of line lengths yield the same phase response but one set uses more lines, the set with more lines naturally performs better in the presence of measurement noise and perturbations. This improved performance results from having more measurements.  Additionally, the eigenvector sensitivity is inversely proportional to the eigengap \cite{Wilkinson1978}. This is further discussed in Section~\ref{sec:3a}.

As established so far, the stability of multiline TRL relies on the eigenvalue in \eqref{eq:2.16} being non-zero for all considered frequencies. While the multiline TRL formulation appears more complicated than classical TRL, it actually reduces to classical TRL in the special case of two lines. In fact, the eigenvalue $\lambda$ becomes the square of the eigengap used to define the phase in TRL; that is, taking the square root of the multiline TRL eigenvalue for two lines is equivalent to the eigengap in classical TRL:
\begin{equation}
    \left|e^{\gamma l_{12}}-e^{-\gamma l_{12}}\right| = \sqrt{\lambda_\mathrm{TRL}}, \qquad l_{12} = l_1 - l_2.
    \label{eq:2.17}
\end{equation}

From the above equation, it is clear that the phase of a classical TRL can be defined from the eigenvalue $\lambda$ as follows:
\begin{equation}
    \phi = \arcsin\left(\frac{\sqrt{\lambda_\mathrm{TRL}}}{2}\right),
    \label{eq:2.18}
\end{equation}
where for the lossless case, \eqref{eq:2.18} is confined between $0^\circ$ and $90^\circ$. For the lossy case, any value exceeding the limits of the arcsine function by convention is truncated to $90^\circ$ \cite{Marks1991}.

For two lines, as in \eqref{eq:2.17} and \eqref{eq:2.18}, it is straightforward to define a phase, since there is only one line pair. However, when more than two lines are used, the eigenvalue increases, even in the lossless case, simply because there are more line pairs. However, this does not necessarily mean that the bandwidth of the multiline TRL is improved. As highlighted earlier, this means that the sensitivity of the eigenvectors improves against noise and perturbations in the measurements. For example, suppose we have three lines, where two are identical, i.e., $\bs{l} = \{l_1, l_2, l_2\}$. This is no different from classical TRL, but now there are three pairs, one of which has zero length difference:
\begin{equation}
    \lambda = \left|e^{\gamma l_{12}}-e^{-\gamma l_{12}}\right|^2 + \left|e^{\gamma l_{12}}-e^{-\gamma l_{12}}\right|^2 + \underbrace{\left|e^{\gamma l_{22}}-e^{-\gamma l_{22}}\right|^2}_{=0}
    \label{eq:2.19}
\end{equation}

To define a phase for this example, which should match the classical TRL case in \eqref{eq:2.17}, we solve for the eigengap as follows:
\begin{equation}
    \left|e^{\gamma l_{12}}-e^{-\gamma l_{12}}\right| = \sqrt{\lambda/2}, \quad \text{or } \, \phi = \arcsin\left(\frac{\sqrt{\lambda/2}}{2}\right)
    \label{eq:2.20}
\end{equation}

In general, for multiple lines of different lengths, the eigenvalue summation in \eqref{eq:2.16} involves all line pairs, not just one as in the repeated line example. To accommodate the general case, we need an average metric for all line pairs. A natural choice is the root mean square (RMS), since the eigenvalue sums over squared terms, and we need to take the square root to use it in the $\arcsin$ function to measure phase. To obtain the average, we scale by the number of line pairs before taking the square root:
\begin{equation}
    \phi = \arcsin\left(\frac{\sqrt{\lambda/C(N,2)}}{2}\right)
    \label{eq:2.21}
\end{equation}
where $C(N,2) = N(N-1)/2$ is the number of unique line pairs.

While \eqref{eq:2.21} meets our needs in many cases, it does not adequately handle repeated line pairs. For example, in the previous example of three lines with one line repeated, the denominator under the square root should be 2, but \eqref{eq:2.21} would give 3, which is incorrect. Clearly, we need to modify the scaling to account for repeated lines and for line pairs with a very small eigengap.

This motivates a normalization that also depends on the line lengths. We start by rewriting the eigenvalue expression in \eqref{eq:2.16} as a weighted sum:
\begin{equation}
    \lambda = \sum\limits_{\begin{gathered}\\[-18pt]
            \scriptstyle i=1\\[-8pt]\scriptstyle i<j\leq N\end{gathered}}^{N-1}\left|e^{\gamma l_{ij}}-e^{-\gamma l_{ij}}\right|^2 = \sum\limits_{\begin{gathered}\\[-18pt]
            \scriptstyle i=1\\[-8pt]\scriptstyle i<j\leq N\end{gathered}}^{N-1}|w_{ij}|\cdot|w_{ij}|
    \label{eq:2.22}
\end{equation}
where $|w_{ij}| = \left|e^{\gamma l_{ij}}-e^{-\gamma l_{ij}}\right|$ is the eigengap of each line pair. Since we are interested in an average metric for $|w_{ij}|$, we should divide the sum by the sum of the $|w_{ij}|$ terms. Thus, a normalized eigenvalue can be written as follows:
\begin{equation}
        \kappa = \frac{\lambda}{\sum\limits_{\begin{gathered}\\[-18pt]
                    \scriptstyle i=1\\[-8pt]\scriptstyle i<j\leq N\end{gathered}}^{N-1}|w_{ij}|} = \frac{\lambda}{\left\|\vc{\bs{W}}\right\|_1/2} = \frac{\left\|\vc{\bs{W}}\right\|_2^2}{\left\|\vc{\bs{W}}\right\|_1}
        \label{eq:2.23}
\end{equation}
where $\left\|\vc{\bs{W}}\right\|_2^2 = \left\|\bs{W}\right\|_\mathrm{F}^2$ is the squared L2-norm (Frobenius norm) of the vectorized $\bs{W}$, and $\left\|\vc{\bs{W}}\right\|_1$ is the L1-norm (sum-norm or Manhattan norm) \cite{Hogben2014}.

This new definition of normalized eigenvalue in \eqref{eq:2.23} addresses most of our requirements. For two lines, it reduces to the eigengap of classical TRL:
\begin{equation}
    \kappa(l_1, l_2) = \left|e^{\gamma l_{12}}-e^{-\gamma l_{12}}\right|
    \label{eq:2.24}
\end{equation}

For the repeated line example in \eqref{eq:2.19}, this normalization also reduces to the classical TRL case:
\begin{equation}
    \kappa(l_1, l_2, l_2) = \kappa(l_1, l_2) = \left|e^{\gamma l_{12}}-e^{-\gamma l_{12}}\right|
    \label{eq:2.25}
\end{equation}

However, in general, this normalization does not make the eigenvalue invariant to repeated lines when there are more than two different lines, e.g.,
\begin{equation}
    \kappa(l_1, l_2, l_3, l_3) \neq \kappa(l_1, l_2, l_3)
    \label{eq:2.26}
\end{equation}

While this may seem like a limitation, it is actually not, as the normalization in \eqref{eq:2.23} provides a good average metric for the eigengap, being bounded by the minimum and maximum eigengap of all line pairs due to this self-weighting \cite{Lann2006}. Hence, $\kappa$ is bounded as follows:
\begin{equation}
    |w_{ij}|_{\min} \leq \kappa \leq |w_{ij}|_{\max}, \quad \forall i,j \leq N
    \label{eq:2.27}
\end{equation}

\begin{figure}[th!]
    \centering
    \includegraphics[width=1\linewidth]{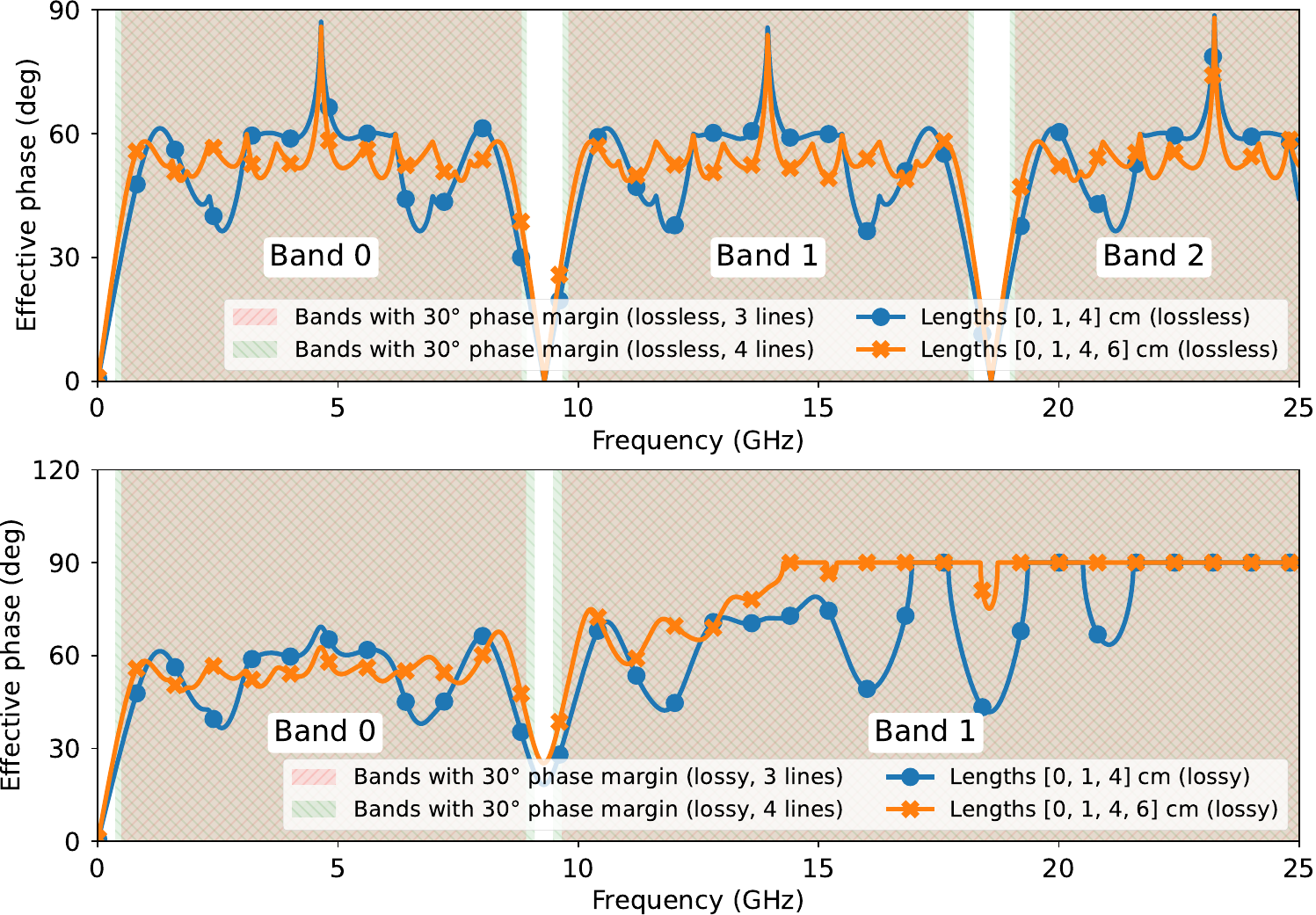}
     \caption{Comparison of effective phase for three and four lines configurations under both lossless $\epsilon_\mathrm{r,eff} = 2.6-0j$ and lossy $\epsilon_\mathrm{r,eff} = 2.6(1-0.06j)$ conditions across the 0-25\,GHz frequency range.}
    \label{fig:2.4}
\end{figure}

Therefore, $\kappa$ is always bounded by the best line pair at any frequency point. In fact, due to self-weighting, it is naturally biased towards the better line pairs. The worst case is when there are line pairs with zero length difference, $|w_{ij}|_{\min}=0$, and at best for the lossless case, $|w_{ij}|_{\max} = 2$ at quarter-wavelength. Fig.~\ref{fig:2.4} illustrates this normalized eigenvalue using the same example from Fig.~\ref{fig:2.3} for both three and four lines, where the normalized eigenvalue is plotted as a function of frequency.

To compute an effective phase, we simply use the arcsine function as in \eqref{eq:2.18}, and substituting $\kappa$ in it:
\begin{equation}
    \phi = \arcsin\left(\frac{\kappa}{2}\right)
    \label{eq:2.28}
\end{equation}

A bandwidth can be defined, as in classical TRL, by requiring a phase margin $\phi > \varphi$. The example in Fig.~\ref{fig:2.4} shows the $\varphi = 30^\circ$ phase margin bands for the lossless case for both the three- and four-line examples, which is equivalent to $\kappa \geq 1$. 

\begin{figure}[th!]
    \centering
    \includegraphics[width=1\linewidth]{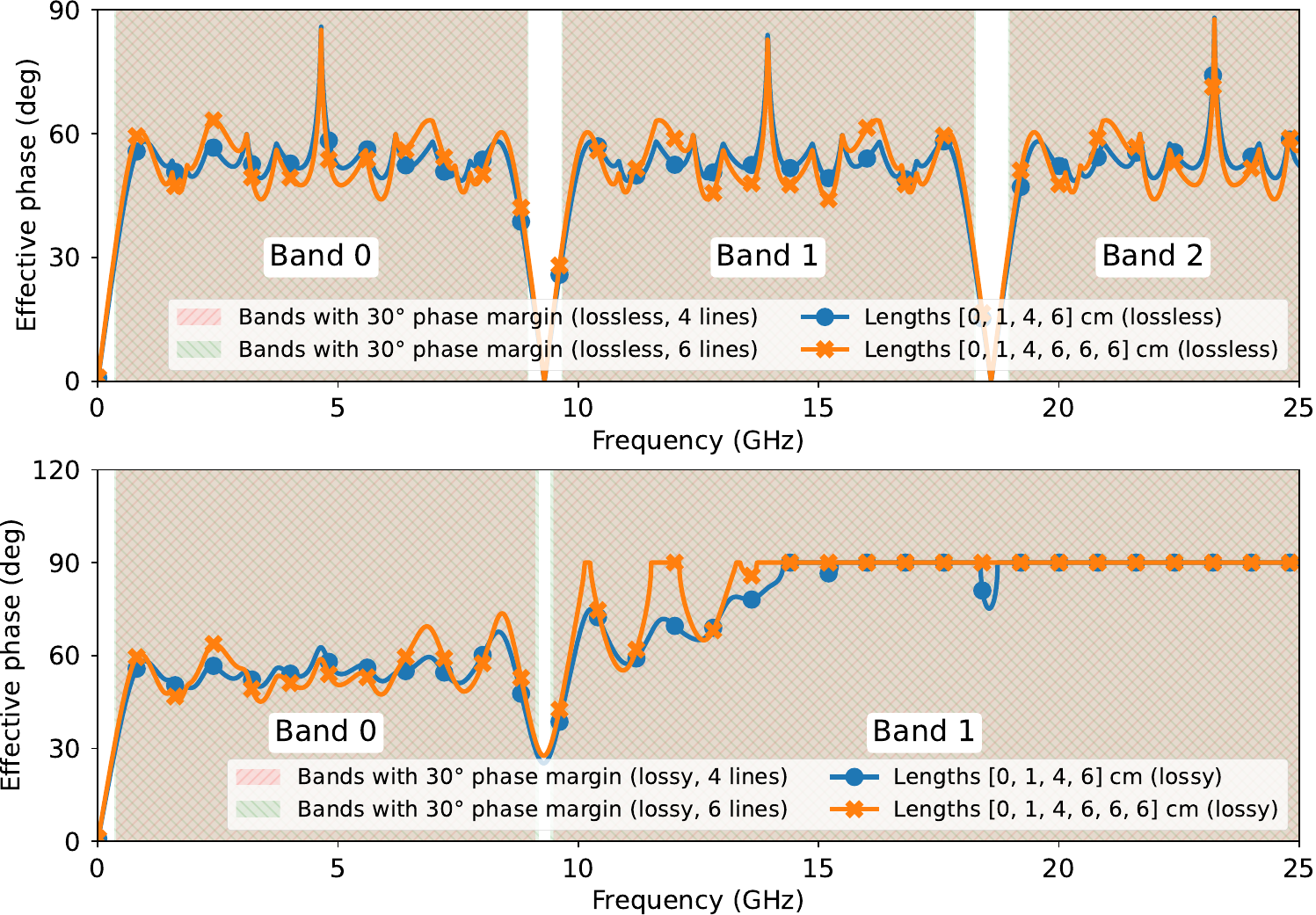}
    \caption{Comparison of effective phase between four and six-line configurations, where in the six-line configuration, the fourth line is repeated three times. Results are shown for both lossless $\epsilon_\mathrm{r,eff} = 2.6-0j$ and lossy $\epsilon_\mathrm{r,eff} = 2.6(1-0.06j)$ conditions across the 0-25\,GHz frequency range.}
    \label{fig:2.5}
\end{figure}

While the normalized eigenvalue $\kappa$ is not strictly invariant to repeated lines, it remains a robust metric for comparing sets of lines, even when repeated lines are present. For example, Fig.~\ref{fig:2.5} shows the normalized eigenvalue for multiline TRL calibration using four lines \{0, 1, 4, 6\}\,cm and six lines \{0, 1, 4, 6, 6, 6\}\,cm. The only difference is that the last line in the six-line example is repeated, i.e., $l_6 = l_5 = 6$\,cm. The figure demonstrates that the normalized eigenvalue for both sets is very similar. The only way to mitigate the effect of repeated lines is to provide that information as part of the weighted eigendecomposition by modifying the weighting matrix, for example, using the scaled version described in Appendix~\ref{anx:A}.

It should be noted that, due to averaging, the maximum phase of $90^\circ$ seen in classical TRL is not achieved here. This would require all line pairs to simultaneously have a phase of $90^\circ$, which can only occur at discrete frequencies. This does not imply reduced performance; rather, it reflects how all line pairs contribute, each scaled by their own weights, and how measurement noise tends to average out when all lines contribute. 

Alternatively, the weighting in multiline TRL could be modified to bias the result toward the line pairs with the higher phase by incorporating a scaling matrix into the weighting matrix, as described in Appendix~\ref{anx:A}. This approach would progressively discard line pairs with low phase contribution as the scaling increases, resulting in an outcome that is more biased toward line pairs with higher phase. 

The default weighting choice, as in \cite{Hatab2022,Hatab2023}, is the L$1$-norm, where each line pair is weighted by its own eigengap. Increasing the exponent on the eigengap leads to L$m$-norm weighting of the line pairs, as discussed in Appendix~\ref{anx:A}. In the extreme case, as the norm approaches $\infty$, the weighting acts as L$\infty$-norm, effectively becoming selective TRL, where at each frequency the line pair with maximum phase is chosen.

\section{Procedures for Computing Line Lengths for Multiline TRL Calibration}
\label{sec:3}

In this section, we cover three topics. In a first step, we present the brute-force approach for finding optimal line lengths through nonlinear constrained optimization. The second method is a simpler approach using sparse rulers, which, while not optimal, provides a convenient way to quickly compute line lengths. The last topic addresses determining the required number of lines for a given frequency range.

\subsection{The Brute-Force Optimization Method}
\label{sec:3a}

As elaborated in Subsections~\ref{sec:2b} and \ref{sec:2c}, multiline TRL calibration will only have a solution for the error terms if the eigenvalue $\lambda$ is non-zero across all considered frequencies. When selecting line lengths, our main concern is the sensitivity of the eigenvectors, as they contain the error terms we are solving for. According to eigenvector perturbation theory \cite{Wilkinson1978}, the Jacobian of the eigenvectors is inversely proportional to the eigenvalue difference (i.e., the eigengap). For example, given the matrix system $\bs{F}$ from \eqref{eq:2.15} and its left and right eigendecompositions $\bs{F} = \bs{U}\Lambda\bs{U}^{-1}$ and $\bs{F}^T = \bs{V}\Lambda\bs{V}^{-1}$, respectively, where $\bs{V} = \bs{U}^{-T}$, the Jacobian of the $i$th eigenvector is given by \cite{Magnus1985,Brewer1978}:
\begin{equation} 
    \bs{J}_{\bs{u}_i} = (\bs{F} - \mu_i\bs{I})^{+}\left(\frac{1}{\bs{v}_i^T\bs{u}_i}(\bs{u}_i^T\otimes\bs{u}_i\bs{v}_i^T) - (\bs{u}_i^T\otimes\bs{I})\right)\bs{J}_{\bs{F}}
    \label{eq:3.1} 
\end{equation}
where $\mu_i$ is the $i$th eigenvalue, and $\bs{J}_{\bs{u}_i}$ and $\bs{J}_{\bs{F}}$ are the Jacobians of the $i$th eigenvector $\bs{u}_i$ and the matrix $\bs{F}$, respectively. The symbol $()^+$ denotes the pseudo-inverse and $\otimes$ is the Kronecker product.

The pseudo-inverse in \eqref{eq:3.1} indicates that the eigenvector sensitivity is inversely proportional to the eigengap. As presented in \eqref{eq:2.15}, we can factor $\lambda$ out of the equation, leading to the fact that the corresponding Jacobian of the eigenvector is inversely proportional to only $\lambda$, i.e.,
\begin{equation} 
    \bs{J}_{\bs{u}} \propto \frac{1}{\lambda}\bs{J}_{\bs{F}}
    \label{eq:3.2} 
\end{equation}

Therefore, to minimize the sensitivity of the eigenvector solutions, we need to maximize the eigenvalue $\lambda$ across all considered frequencies. This can be formulated as an optimization problem using the following loss function:
\begin{equation}
    \bs{l}_\mathrm{opt} = \argmin{\bs{l}}\,\frac{1}{2}\left( \maxlim{f \in [f_\mathrm{min}, f_\mathrm{max}]}{-\lambda(\bs{l},f)} - \frac{1}{M}\kern-5pt\sum_{f=f_\mathrm{min}}^{f_\mathrm{max}}\kern-5pt\lambda(\bs{l},f)\right)
    \label{eq:3.3}
\end{equation}
where $\lambda$ is a function of both frequency and line lengths, with other parameters such as relative permittivity being implicit. $f_\mathrm{min}$ and $f_\mathrm{max}$ are the frequency bounds over which the optimization is performed, and $M$ is the number of frequency points between $f_\mathrm{min}$ and $f_\mathrm{max}$. The negative sign is included to convert the maximization problem into a minimization problem. The loss function in \eqref{eq:3.3} consists of two components: the first is a min-max term that ensures the worst-case frequency point is optimized, while the second is the mean of the eigenvalue, which promotes a flat frequency response. In other words, the line lengths are selected to ensure the eigenvalue is maximized while remaining as uniform as possible across all frequency points.

Without constraints, the optimization formulation in \eqref{eq:3.3} is ill-posed, as any common offset in the line lengths would also be a valid solution. To address this, we need to anchor the first line to be the thru and set it to zero by convention. The second degree of freedom is the maximum length. If unconstrained, the optimization may yield nonphysical line lengths, especially to satisfy the low-frequency requirement. Therefore, both the thru and maximum length must be predefined. The optimization problem can be reformulated as a constrained minimization, where the first and last lines are predefined, i.e., $l_1=0$ and $l_N=l_\mathrm{max}$, and the remaining lines are optimized:
\begin{equation}
    \begin{aligned}
        \bs{l}_\mathrm{opt} &= \argmin{\bs{l}}\,\frac{1}{2}\left( \maxlim{f \in [f_\mathrm{min}, f_\mathrm{max}]}{-\lambda(\bs{l},f)} - \frac{1}{M}\kern-6pt\sum_{f=f_\mathrm{min}}^{f_\mathrm{max}}\kern-5pt\lambda(\bs{l},f)\right) \\
        &\text{subject to} \quad  \bs{a} \leq \bs{C}\bs{l} \leq \bs{b}
    \end{aligned}
    \label{eq:3.4}
\end{equation}
where $\bs{C}$ is the constraint matrix encoding the linear constraints on the line lengths to ensure each line is unique, ordered, and always smaller than $l_\mathrm{max}$. Vectors $\bs{a}$ and $\bs{b}$ are the lower and upper bounds, respectively.

The linear constraint has the following structure:
\begin{equation}
    \underbrace{\begin{bmatrix}
            0 \\
            -l_\mathrm{max} \\
            -l_\mathrm{max} \\
            \vdots \\
            -l_\mathrm{max} \\
            l_\mathrm{max}
    \end{bmatrix}}_{\bs{a}: (N+1)\times 1} \leq \underbrace{\begin{bmatrix}
        -1 & 0 & 0 & \cdots  & 0 \\
        1 & -1 & 0 & \cdots  & 0 \\
        0 & 1 & -1 & \cdots  & 0 \\
        \vdots & \vdots & \ddots & \ddots  & \vdots \\
        0 & 0 & \cdots & 1  & -1\\
        0 & 0 & \cdots &  0 & 1
    \end{bmatrix}}_{\bs{C}: (N+1)\times N}\underbrace{\begin{bmatrix}
        l_1 \\
        l_2 \\
        l_3 \\
        \vdots \\
        l_{N} 
    \end{bmatrix}}_{\bs{l}: N\times 1} \leq \underbrace{\begin{bmatrix}
        0 \\
        -l_\mathrm{min} \\
        -l_\mathrm{min} \\
        \vdots \\
        -l_\mathrm{min} \\
        l_\mathrm{max}
        \end{bmatrix}}_{\bs{b}: (N+1)\times 1}
    \label{eq:3.5}
\end{equation}
where $l_\mathrm{min}$ and $l_\mathrm{max}$ are the minimum and maximum allowed length differences. Note that the first and last entries enforce the first line to be zero length and the last line to have the maximum length. Generally, the choice for $l_\mathrm{max}$ is dictated by the minimum required frequency, which can be calculated from \eqref{eq:2.7a}, whereas $l_\mathrm{min}$ is dictated by the maximum desired frequency, but is also restricted by the minimum resolution of length that can be manufactured.

\begin{figure}[th!]
    \centering
    \includegraphics[width=1\linewidth]{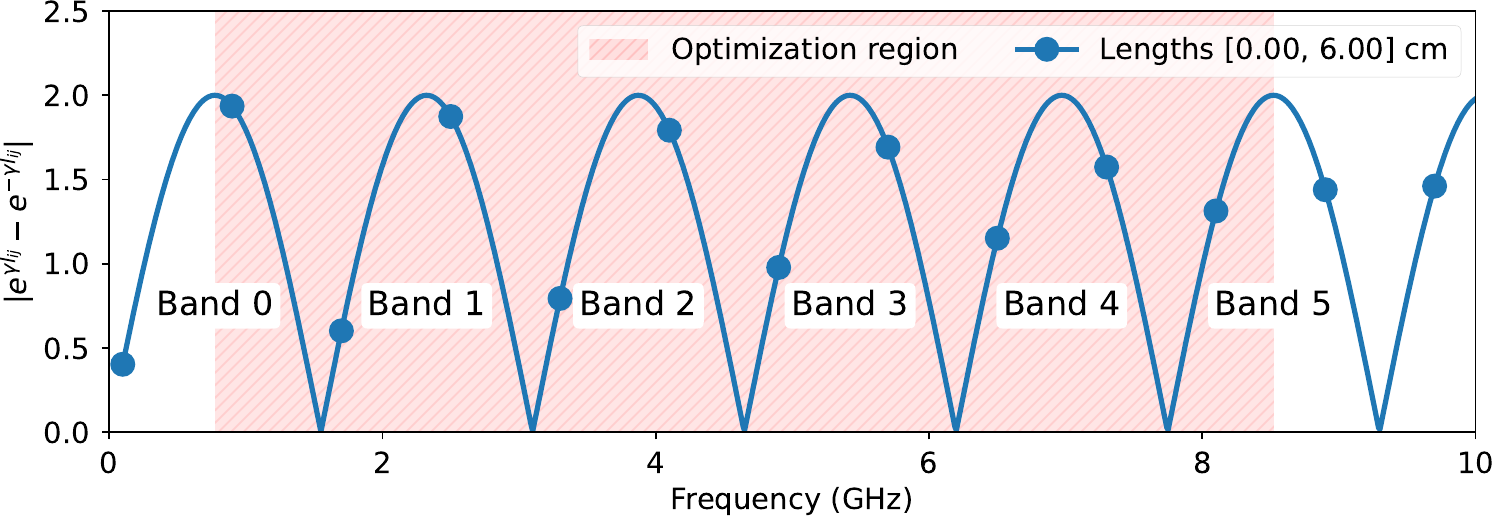}
    \caption{Illustration of the optimization frequency range determined from the TRL eigengap curve for the longest line. In this example, $l_\mathrm{max} = 6\,\mathrm{cm}$ and $\epsilon_\mathrm{r,eff} = 2.6-0j$, which leads to a minimum frequency anchored at $f_\mathrm{min} = 0.775\,\mathrm{GHz}$ (band-0). The maximum frequency was chosen at $f_\mathrm{max} = 8.521\,\mathrm{GHz}$ (band-5).}
    \label{fig:3.1}
\end{figure}

To select $f_\mathrm{min}$ and $f_\mathrm{max}$ for the optimization, we use the TRL formulation and choose the frequencies closest to the peaks in the eigengap curve for the longest line length relative to the thru, corresponding to a $90^\circ$ phase as given by \eqref{eq:2.6}. For example, Fig.~\ref{fig:3.1} shows the TRL eigengap for $l_\mathrm{max} = 6\,\mathrm{cm}$, where $f_\mathrm{min}$ is chosen at the quarter-wavelength of band-0 and $f_\mathrm{max}$ at the quarter-wavelength of band-5. We avoid selecting $f_\mathrm{min}$ at the desired phase margin (e.g., $\phi=30^\circ$) because the tail of the multiline TRL eigenvalue will limit the optimization; the lower end cannot be maximized beyond the response of the longest line. The choice of $f_\mathrm{max}$ is user-defined, but it affects the minimum possible line length and the required number of lines. In general, a higher $f_\mathrm{max}$ requires shorter lines and a greater number of lines.

While the minimization formulation in \eqref{eq:3.4} is suitable for obtaining unique, optimal line lengths, this constrained problem can yield solutions that are sensitive to length perturbations, e.g., due to manufacturing and measurement uncertainties. Therefore, it can be beneficial to seek line length solutions that remain near optimal even under known perturbations. To this end, we regularize the objective function in \eqref{eq:3.4} to also minimize the standard deviation of the eigenvalue:
\begin{equation}
    \begin{aligned}
        \bs{l}_\mathrm{opt}& = \argmin{\bs{l}}
         \left(\kern-2pt\frac{1}{2}\kern-2pt\left(\kern-1pt\maxlim{f \in [f_\mathrm{min}, f_\mathrm{max}]}{-\lambda(\bs{l},f)} - \frac{1}{M}\kern-6pt\sum_{f=f_\mathrm{min}}^{f_\mathrm{max}}\kern-5pt\lambda(\bs{l},f)\kern-2pt\right)\right.\\
        &\left.+ \sqrt{\frac{1}{M}\sum_{f=f_\mathrm{min}}^{f_\mathrm{max}} \kern-4pt\bs{J}_{\lambda}(\bs{l},f)\bs{\Sigma}\bs{J}^T_{\lambda}(\bs{l},f)}\right) \\
        &\text{subject to} \quad  \bs{a} \leq \bs{C}\bs{l} \leq \bs{b}
    \end{aligned}
    \label{eq:3.6}
\end{equation}
where $\bs{J}_{\lambda}(\bs{l},f)$ is the Jacobian of $\lambda$ with respect to the lengths $\bs{l}$, while $\bs{\Sigma}$ is the covariance matrix of the line lengths, which in most cases can be simplified as a scalar, as each line would experience the same perturbation and are often uncorrelated. The details on the calculation of Jacobian $\bs{J}_{\lambda}(\bs{l},f)$ are provided in Appendix~\ref{anx:B}.

To solve the optimization problem for either \eqref{eq:3.4} or \eqref{eq:3.6}, an appropriate optimization procedure is required. Although it may seem that any optimization method would suffice, the loss functions in \eqref{eq:3.4} and \eqref{eq:3.6} are not well-behaved; they exhibit multiple local minima due to the complex exponential terms. As a result, gradient-based optimization methods often fail, becoming trapped in local minima.

\begin{figure}[th!]
    \centering
    \subfloat[]{\includegraphics[width=0.49\linewidth]{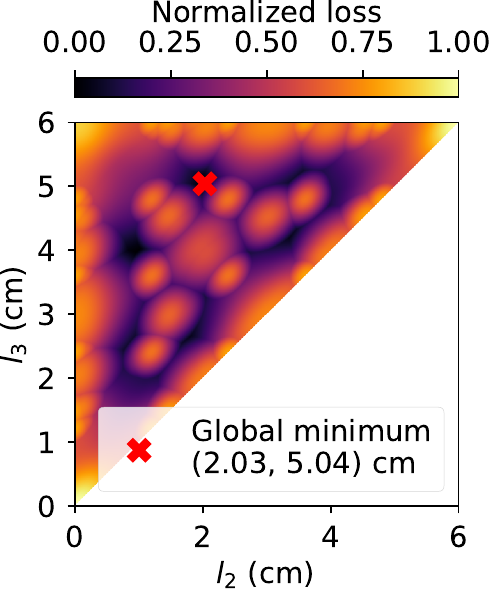}}
    \hfill
    \subfloat[]{\includegraphics[width=0.49\linewidth]{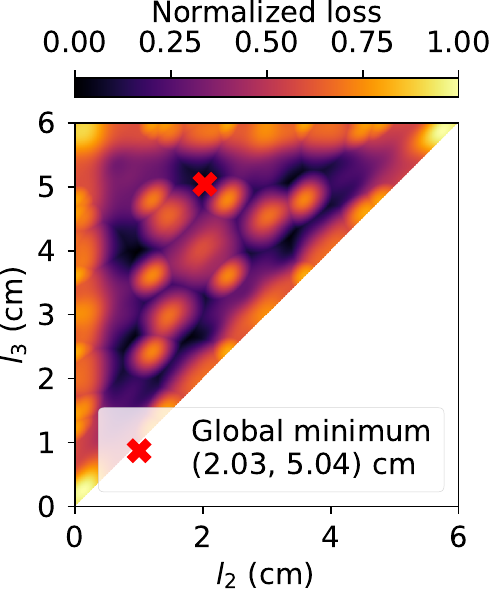}}
    \caption{Grid search visualization for optimizing two line lengths in a four-line system. (a) Loss function landscape using the standard formulation in \eqref{eq:3.4}, and (b) loss function with length uncertainty regularization from \eqref{eq:3.6}, both computed with $\epsilon_\mathrm{r,eff} = 2.6-0j$.}
    \label{fig:3.2}
\end{figure}

To illustrate this concept, consider a system with four lines, where the first and last are anchored at $0\,\mathrm{cm}$ and $6\,\mathrm{cm}$, respectively, with $\epsilon_\mathrm{r,eff} = 2.6-0j$. Fig.~\ref{fig:3.2} displays the loss function landscape when searching for optimal values of $l_2$ and $l_3$ using the two loss functions defined in \eqref{eq:3.4} and \eqref{eq:3.6}. Due to the symmetry of the search space, we only examine the upper triangular region, which ensures unique line lengths as enforced by the linear constraints. For the regularized loss function in \eqref{eq:3.6}, in this example we assumed a length uncertainty of $0.2\,\mathrm{cm}$ for each line, resulting in a covariance matrix $\bs{\Sigma} = (0.2\,\mathrm{cm})^2\bs{I}$.

As shown in Fig.~\ref{fig:3.2}, both loss functions exhibit multiple local minima, making gradient-based optimization challenging. Therefore, we recommend solving such problems using global optimization methods, such as the differential evolution (DE) algorithm \cite{Storn1997}. The DE method is used in this work due to its robustness as a global optimizer and its ability to parallelize function evaluations across multiple CPU cores; however, it is not the only applicable solver. Other global optimization methods could also be employed. It is also worth noting that, in Fig.~\ref{fig:3.2}, both loss functions yielded the same global minimum. However, this is not necessarily the case in general, especially when considering more complex problems involving multiple lines. Typically, the regularized loss function with length uncertainty produces a more spread distribution of peaks and valleys in the loss function landscape.

\begin{figure}[th!]
    \centering
    \includegraphics[width=1\linewidth]{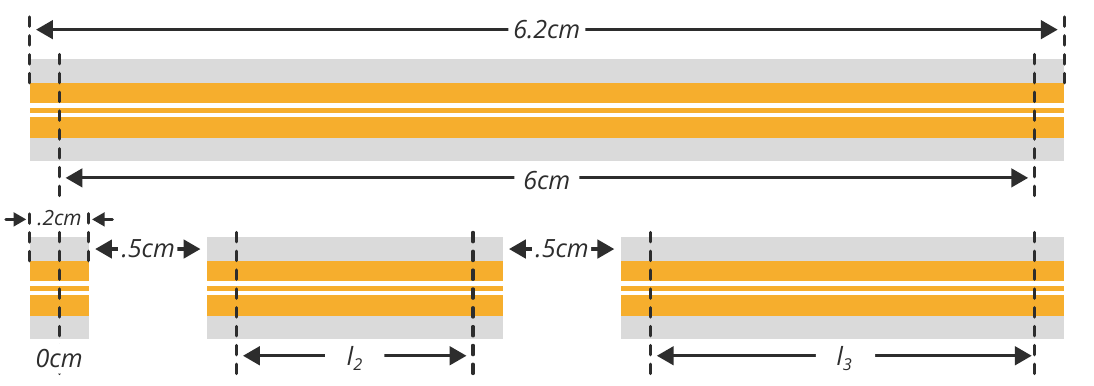}
    \caption{Illustration of four CPW lines arranged in two rows to optimize space utilization for multiline TRL calibration.}
    \label{fig:3.3}
\end{figure}

Additional regularization or constraints can always be introduced to further restrict the solution space. For example, we may wish to constrain the solution so that the line lengths fit within a predefined physical area. This is particularly relevant for on-wafer applications, where available space is often limited. Typically, the maximum allowable lateral space is assigned to $l_\mathrm{max}$, with the remaining lines arranged in a second row, and so on. Fig.~\ref{fig:3.3} illustrates an example in which four coplanar waveguide (CPW) lines are fitted into two rows: the first row is fully occupied by the longest line, while the second row contains the remaining lines. The objective is to determine the lengths of the other lines such that the total length of each row and the spacing between lines are constant.

To achieve the constraint shown in Fig.~\ref{fig:3.3}, we incorporate an additional linear equality constraint on the lengths, such that the sum of lengths must equal some constraint. The optimization problem is extended from \eqref{eq:3.6} to become:
\begin{equation}
    \begin{aligned}
        \bs{l}_\mathrm{opt} &= \argmin{\bs{l}}
        \left(\kern-2pt\frac{1}{2}\kern-2pt\left(\kern-1pt\maxlim{f \in [f_\mathrm{min}, f_\mathrm{max}]}{-\lambda(\bs{l},f)} - \frac{1}{M}\kern-6pt\sum_{f=f_\mathrm{min}}^{f_\mathrm{max}}\kern-5pt\lambda(\bs{l},f)\kern-2pt\right)\right.\\
        &+ \sqrt{\frac{1}{M}\sum_{f=f_\mathrm{min}}^{f_\mathrm{max}} \kern-4pt\bs{J}_{\lambda}(\bs{l},f)\bs{\Sigma}\bs{J}^T_{\lambda}(\bs{l},f)}
        \Bigg) \\
        &\text{subject to} \quad  \bs{a}_1 \leq \bs{C}_1\bs{l} \leq \bs{b}_1 \quad\&\quad \bs{C}_2\bs{l} = \bs{b}_2
    \end{aligned}
    \label{eq:3.7}
\end{equation}
where $\bs{C}_2\bs{l} = \bs{b}_2$ is the additional linear constraint. For the example in Fig.~\ref{fig:3.3}, the linear constraint matrix $\bs{C}_2$ and equality bound $\bs{b}_2$ will have the following structure:
\begin{equation}
    \underbrace{\begin{bmatrix}
            0 & 0 & 0 &  1 \\
            1 & 1 & 1 &  0 
    \end{bmatrix}}_{\bs{C}_2}\underbrace{\begin{bmatrix}
            l_1 \\
            l_2 \\
            l_3 \\
            l_4 
    \end{bmatrix}}_{\bs{l}} = \underbrace{\begin{bmatrix}
            6.2\,\mathrm{cm}-0.2\,\mathrm{cm} \\
            6.2\,\mathrm{cm}-1\,\mathrm{cm}-0.6\,\mathrm{cm}
    \end{bmatrix}}_{\bs{b}_2}
    \label{eq:3.8}
\end{equation}

Each additional linear constraint increases the complexity of the optimization problem by expanding the solution space and introducing more parameters~\cite{Lampinen2002}. As a result, such problems may require significantly more time to converge.

We conclude this subsection with a summary of the optimization procedure provided in Algorithm~\ref{alg:1}.
\begin{algorithm}[ht!]
\caption{Computing line lengths for multiline TRL calibration based on constrained optimization.}
\label{alg:1}
\begin{algorithmic}[1]
    \REQUIRE Frequency bounds $f_\mathrm{min}$, $f_\mathrm{max}$, complex relative equivalent permittivity $\epsilon_\mathrm{r,eff}$, and the thru length defined to have zero length by convention, i.e., $l_1=0$.
    \STATE If not provided, compute $l_\mathrm{max}$ from $f_\mathrm{min}$ and desired minimum phase margin $\phi$ based on \eqref{eq:2.7a}.
    \STATE If not provided, determine the number of lines based on the procedure discussed in Subsection~\ref{sec:3c}.
    \STATE Construct constraint matrix $\bs{C}$ and bounds $\bs{a}$, $\bs{b}$ in \eqref{eq:3.5}. If $l_\mathrm{min}$ is not provided, set it to zero. Optional to incorporate further constraints, e.g., \eqref{eq:3.8}. 
    \STATE Define the actual $f_\mathrm{min}$ and $f_\mathrm{max}$ used in the optimization, which are based on the lowest and highest TRL bands from the longest line that are closest to and include the target frequency range using \eqref{eq:2.6} while setting $\phi=90^\circ$.
    \STATE Define the loss function, e.g., any of \eqref{eq:3.4}, \eqref{eq:3.6}, or \eqref{eq:3.7}.
    \STATE Solve for optimal $l_2, \ldots, l_{N-1}$ via a global optimization procedure, e.g., using DE method \cite{Storn1997}.
    \RETURN Optimized line lengths $\bs{l}_\mathrm{opt}$
\end{algorithmic}
\end{algorithm}

\subsection{Simplified Procedure Using Sparse Rulers}
\label{sec:3b}

\begin{figure}[th!]
    \centering
    \includegraphics[width=1\linewidth]{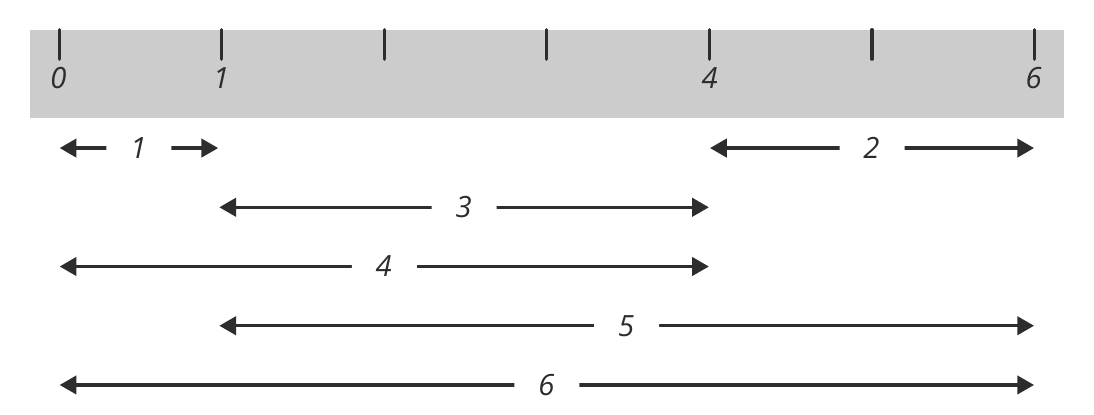}
    \caption{Illustration of a sparse ruler using the set $\{0,1,4,6\}$, showing how all integer distances from 0 to 6 can be measured using only these four marks.}
    \label{fig:3.4}
\end{figure}

A sparse ruler is a set of marks (numbers) such that every integer distance up to the maximum can be measured as the difference between two marks in the set~\cite{Linebarger1993}. For example, the set $\{0,1,4,6\}$ is a sparse ruler because all integer values from 0 to 6 can be obtained as differences between pairs of marks: $\{0,1,2,3,4,5,6\}$. This concept is illustrated in Fig.~\ref{fig:3.4}, where a ruler with a subset of marks allows reconstruction of all intermediate positions by taking pairwise differences.

Sparse rulers are widely used in various applications, such as sensor array design and multi-input, multi-output (MIMO) systems~\cite{Xiao2023,Romero2016}. Their relevance to multiline TRL calibration stems from the fact that the multiline TRL calibration fundamentally depends on the differences between line lengths. As expressed in \eqref{eq:2.16}, the eigenvalue $\lambda$ is composed of the sum over all line pair differences. In the lossless case, this sum is equivalent to a Fourier series in terms of the line differences, which can be expressed as:
\begin{equation}
    \begin{aligned}
        \lambda \stackrel{\gamma = j\beta}{=}\kern-4pt \sum\limits_{\begin{gathered}\\[-18pt]
            \scriptstyle i=1\\[-8pt]\scriptstyle i<j\leq N\end{gathered}}^{N-1}\kern-6pt\left(e^{j\beta l_{ij}}-e^{-j\beta l_{ij}}\right)^2 &= -4\kern-6pt \sum\limits_{\begin{gathered}\\[-18pt]
            \scriptstyle i=1\\[-8pt]\scriptstyle i<j\leq N\end{gathered}}^{N-1}\kern-6pt\sin^2\beta l_{ij}\\
            &= 2\kern-6pt\sum\limits_{\begin{gathered}\\[-18pt]
                \scriptstyle i=1\\[-8pt]\scriptstyle i<j\leq N\end{gathered}}^{N-1}\kern-6pt(\cos2\beta l_{ij} - 1)
    \end{aligned}
    \label{eq:3.9}
\end{equation}

While \eqref{eq:3.9} is a sum of cosines and a constant term, it becomes a true Fourier series when the line pair differences are integer multiples of a fundamental length. This occurs when $l_{ij} = k_{ij}l_0$, or, in other words, when the lengths are harmonically related. In this case, the eigenvalue simplifies to:
\begin{equation}
    \lambda = 2\kern-6pt\sum\limits_{\begin{gathered}\\[-18pt]
            \scriptstyle i=1\\[-8pt]\scriptstyle i<j\leq N\end{gathered}}^{N-1}\kern-6pt(\cos2\beta l_0k_{ij} - 1), \quad k_{ij} = k_i-k_j \in \mathbb{N}^+
    \label{eq:3.10}
\end{equation}

If the set of $k_{ij}$ covers all integers up to a maximum without gaps, then \eqref{eq:3.10} forms a complete Fourier series. This is the key connection between sparse rulers and multiline TRL calibration: when the line lengths are chosen according to a sparse ruler, all required differences are covered. The line lengths can then be expressed as:
\begin{equation}
    \bs{l}_\mathrm{ruler} = \text{Sparse Ruler Set} \times l_0
    \label{eq:3.11}
\end{equation}
where $l_0$ is the smallest step size, i.e., the minimum length difference.

\begin{figure}[th!]
    \centering
    \includegraphics[width=1\linewidth]{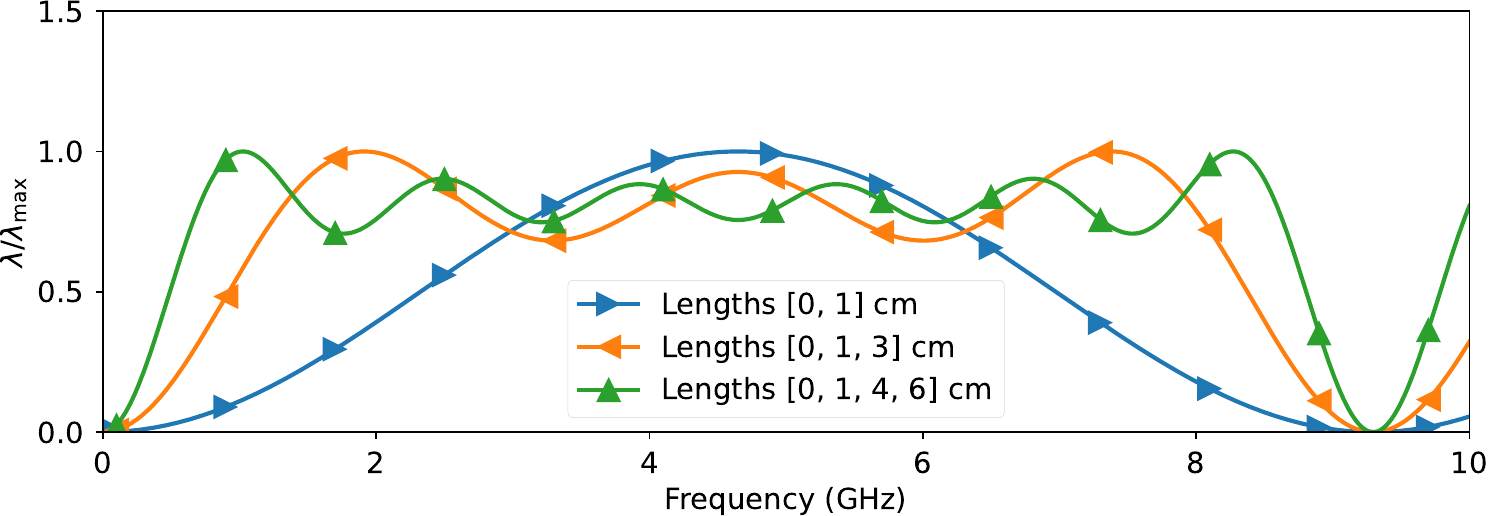}
    \caption{Comparison of the multiline TRL eigenvalue responses showing how bandwidth increases as larger sparse ruler sets are used ($\{0,1\}$, $\{0,1,3\}$, and $\{0,1,4,6\}$).}
    \label{fig:3.5}
\end{figure}

The significance of this Fourier series structure is that the sum of harmonically related sinusoids approaches a square wave as more terms are added. Consequently, the eigenvalue response becomes broader and flatter with an increasing number of line pairs. For example, consider the sets $\{0,1\}$, $\{0,1,3\}$, and $\{0,1,4,6\}$ with $l_0 = 1\,\mathrm{cm}$ and $\epsilon_\mathrm{r,eff} = 2.6-0j$. As shown in Fig.~\ref{fig:3.5}, increasing the number of terms widens the eigenvalue's bandwidth and makes the response more square-like. In Fig.~\ref{fig:3.5}, $\lambda$ is normalized to its maximum value for visualization purposes.

In the example shown in Fig.~\ref{fig:3.5}, the sparse ruler sets are known as ``perfect rulers,'' meaning their pairwise differences cover all integers up to the maximum value without redundancy. Only a few perfect rulers exist, such as $\{0,1\}$, $\{0,1,3\}$, and $\{0,1,4,6\}$~\cite{Leech1956}. Other types include Wichmann rulers~\cite{Wichmann1963}, which cover all differences but with redundancy, and Golomb rulers, which guarantee unique pairwise differences~\cite{Sidon1932,Meyer2009}. When Golomb rulers cover all possible differences, they are by definition perfect rulers.

An additional distinction concerns ``optimal'' sparse rulers. An optimal sparse ruler of order~$n$ is one that uses the fewest marks (i.e., $n$ marks) to measure all integer distances up to its length, meaning no ruler of the same order has a shorter maximum mark. Most known optimal sparse rulers can be constructed systematically using Wichmann's algorithm~\cite{Wichmann1963}. However, for certain orders, the shortest known sparse rulers cannot be generated by Wichmann's construction; these are referred to as non-Wichmann optimal rulers. A comprehensive listing and discussion of both types is provided in~\cite{Luschny2013}.

\begin{figure}[th!]
    \centering
    \includegraphics[width=1\linewidth]{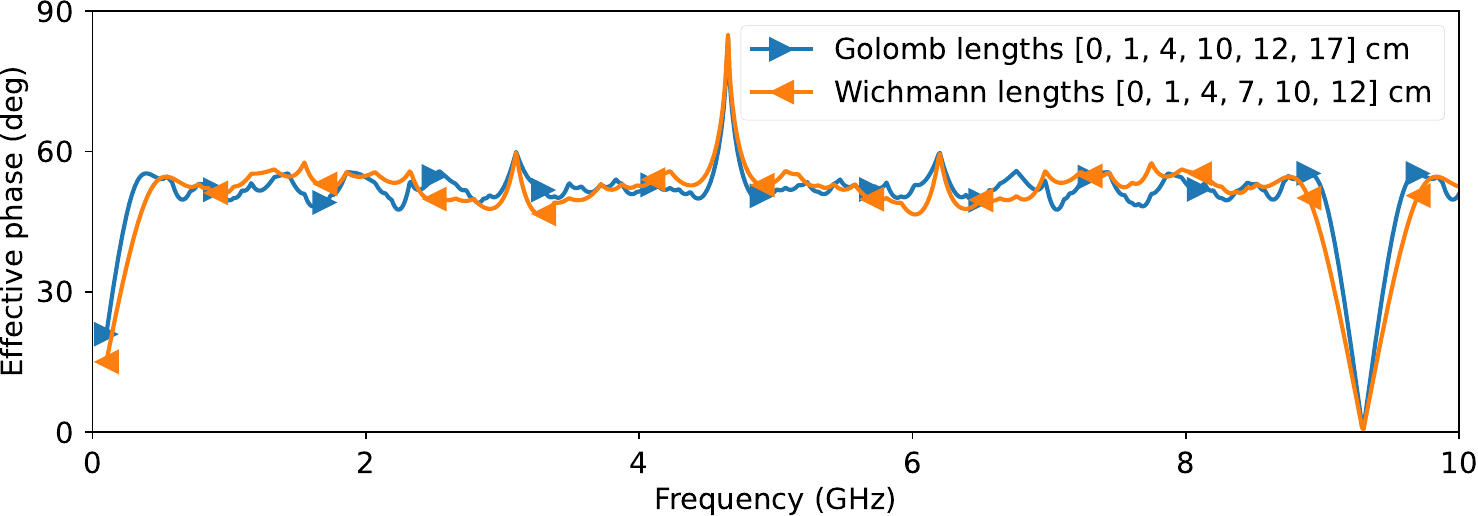}
    \caption{Comparison of multiline TRL effective phase responses using six lines configurations based on Wichmann and Golomb sparse rulers.}
    \label{fig:3.6}
\end{figure}

For multiline TRL calibration, there is no strict requirement that the eigenvalue cover all possible differences. In practice, Golomb rulers are often preferred because they avoid redundancy, resulting in a broader bandwidth for a given number of lines compared to Wichmann rulers. Fig.~\ref{fig:3.6} compares the effective phase responses for six lines chosen according to Wichmann and Golomb rulers, with $l_0 = 1\,\mathrm{cm}$ and $\epsilon_\mathrm{r,eff} = 2.6-0j$. The Golomb set achieves a broader bandwidth by including a longer line, while both sets provide similar effective phase coverage.

Because sparse rulers generate harmonically related line pairs, the frequency response of the multiline TRL eigenvalue exhibits cyclic banded behavior similar to classical TRL calibration. The minimum and maximum frequencies for each band can be computed using the classical TRL equations from \eqref{eq:2.6}. The lower frequency edge is determined by the longest line, while the upper frequency edge is determined by the smallest length difference, $l_0$.

Golomb rulers have proven more suitable for multiline TRL calibration in practice because they provide wider bandwidth for a given number of lines, as demonstrated with successful practical multiline TRL kits operating up to 150\,GHz~\cite{Hatab2023a, Hatab2023b, Arsanjani2024, Arsanjani2024a}. While the sparse ruler approach offers convenience and simplicity---requiring only the application of TRL equations to determine the frequency range---it lacks an explicit optimality metric, unlike the optimization-based approach discussed in Subsection~\ref{sec:3a}. This limitation makes it difficult to incorporate loss or frequency-dependent permittivity into the sparse ruler method. Nevertheless, sparse rulers generally perform well in practical applications, even with moderate parameter variations.

The procedure for computing multiline TRL line lengths using sparse rulers is summarized in Algorithm~\ref{alg:2}.
\begin{algorithm}[ht!]
    \caption{Computing line lengths for multiline TRL calibration based on sparse rulers.}
    \label{alg:2}
    \begin{algorithmic}[1]
        \REQUIRE Frequency bounds $f_\mathrm{min}$, $f_\mathrm{max}$, average real part of the complex relative effective permittivity $\epsilon_\mathrm{r,eff}$, and the thru length defined to be zero, i.e., $l_1=0$.
        \STATE Compute the minimum line length $l_0$ from $f_\mathrm{max}$ and the desired band of operation using \eqref{eq:2.7b}.
        \STATE If not provided, determine the number of lines using the procedure in Subsection~\ref{sec:3c}.
        \STATE Select a sparse ruler set (e.g., Golomb or Wichmann) based on the number of lines.
        \RETURN Multiply $l_0$ by the sparse ruler set to obtain the line lengths.
    \end{algorithmic}
\end{algorithm}

\subsection{Determining the Recommended Number of Lines}
\label{sec:3c}
In the lossless case, the general rule is that the longest line determines the lowest frequency, while the shortest line difference sets the highest frequency limit. These limits can be calculated using the classical TRL equations from \eqref{eq:2.6}. The purpose of having multiple lines is to fill the gap between the longest and shortest length differences and ensure there are no gaps or nulls in the frequency response of the eigenvalue.

For the lossless case, it is possible to develop a procedure to determine the recommended number of lines. However, for the lossy case, this is not straightforward. In lossy media, the relationship changes: longer lines experience higher loss and thus exhibit larger effective phase at higher frequencies. This means the required number of lines depends on the loss characteristics of the transmission medium. In extreme cases, a single long lossy line alongside a thru standard could be sufficient if the loss is sufficient to establish an effective phase above the desired phase margin. Alternatively, for lossy lines, one can perform the optimization procedure discussed in Subsection~\ref{sec:3a} and incrementally increase the number of lines until the desired effective phase response for the eigenvalue is achieved.

While there is no straightforward way to determine the number of lines for the lossy case beyond the approaches mentioned above, for the lossless case, we can establish a recommended number of lines that ensures sufficient distribution to allow a flat effective phase across the desired frequency range.

\begin{figure}[th!]
    \centering
    \includegraphics[width=1\linewidth]{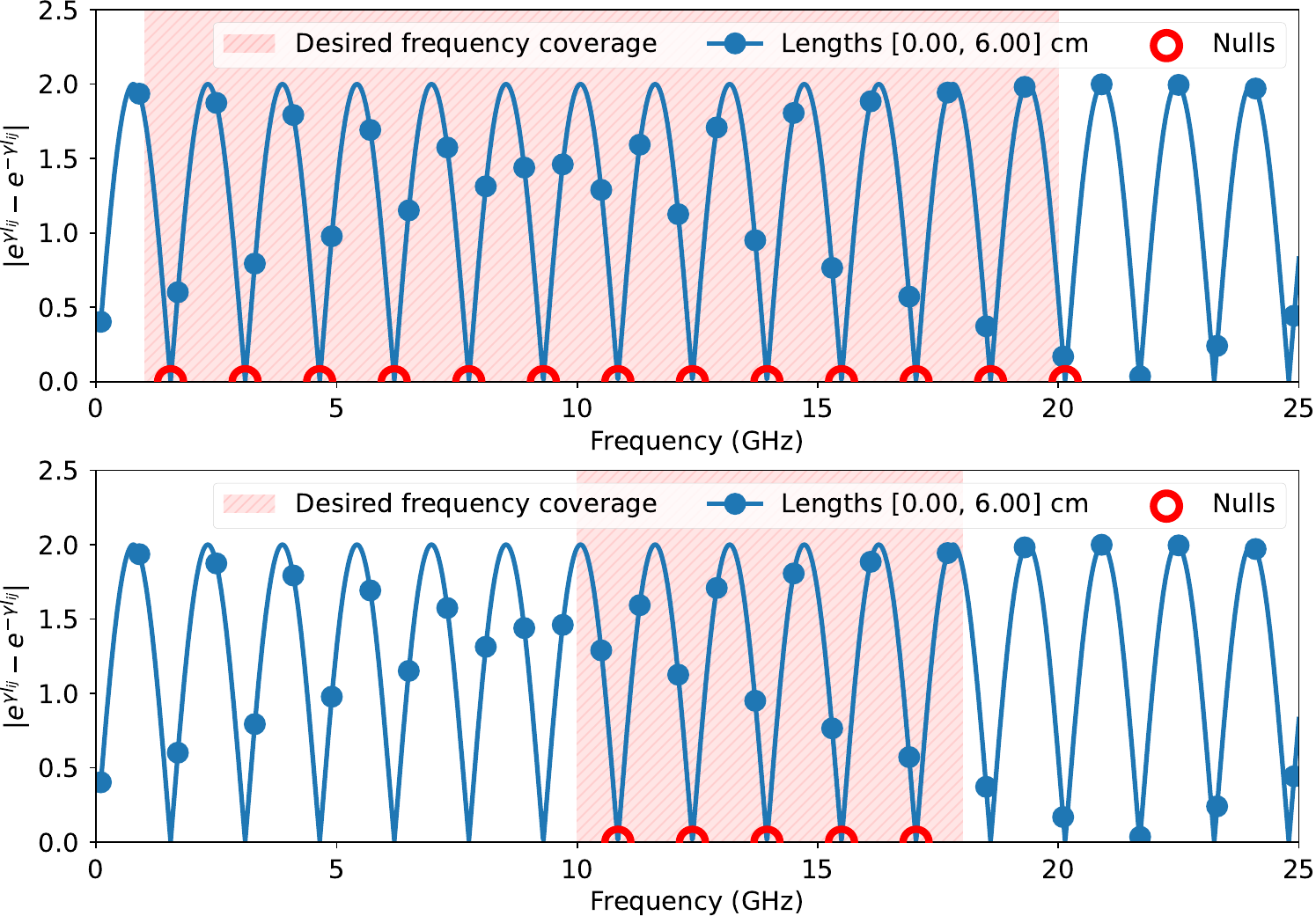}
    \caption{Illustration of eigenvalue nulls for the longest line within the desired frequency range. The upper plot demonstrates full coverage from DC to $f_\mathrm{max}$, while the lower plot shows a limited bandwidth case between $f_\mathrm{min}$ and $f_\mathrm{max}$. In this example, $l_\mathrm{max} = 6\,\mathrm{cm}$ and $\epsilon_\mathrm{r,eff} = 2.6-0j$.}
    \label{fig:3.7}
\end{figure}

Recall that the purpose of using multiple lines is to spread the phase response of each line pair across frequency, creating an effectively constant phase. Without sufficient line lengths between the thru and the longest lines, the nulls of the longest line become the limiting factor of the calibration kit. Therefore, we need enough line pairs to compensate for these nulls across the desired frequency range. For illustration, Fig.~\ref{fig:3.7} shows the eigengap curve of a single long line relative to the thru, highlighting the nulls in the desired frequency range.

In general, each unique line pair shorter than the maximum length difference will have its peak (i.e., $90^\circ$) at at least one null of the longest line. Hence, counting the number of nulls in the analyzed frequency range indicates how many line pairs we need. To obtain the recommended line lengths, we first compute the maximum number of line pairs required to cover all nulls of the longest line across the solution frequency range, and from that, derive the appropriate number of lines based on the number of pairs.

To compute the number of nulls in the longest line (i.e., the number of bands), we can use the classical TRL equation for all bands from DC to the solution's maximum frequency. This gives us the maximum number of unique pairs needed to cover all frequencies from DC to $f_\mathrm{max}$:
\begin{equation}
    M_{\text{max}} = \left\lceil \frac{2l_{\text{max}} f_{\text{max}} \sqrt{\epsilon_\mathrm{r,eff}^\prime}}{c_0} - 1 + \frac{\varphi}{180} \right\rceil + 1
    \label{eq:3.12}
\end{equation}

Typically, our lowest desired frequency is above DC, so the actual number of nulls in the eigenvalue of the longest line that are relevant is a function of the bandwidth. We can calculate this by reusing \eqref{eq:3.12} and replacing $f_\mathrm{max}$ with the frequency difference, i.e., the bandwidth:
\begin{equation}
    M_{\text{min}} = \left\lceil \frac{2l_{\text{max}}(f_{\text{max}} - f_{\text{min}})\sqrt{\epsilon_\mathrm{r,eff}^\prime}}{c_0} - 1 + \frac{\varphi}{180} \right\rceil + 1
    \label{eq:3.13}
\end{equation}

Note that in both \eqref{eq:3.12} and \eqref{eq:3.13} we round up to the nearest integer to accommodate the phase margin condition, as we want to be one pair above the phase margin limit.

Either \eqref{eq:3.12} or \eqref{eq:3.13} can be used to determine the number of pairs, depending on the desired frequency range and number of lines. When using sparse rulers to generate line lengths, we might not be able to use $M_\mathrm{min}$ directly, as it may not satisfy the requirement of a cyclic band structure with harmonically related line lengths. Therefore, we may need to increase $M_\mathrm{min}$ until we obtain the closest value that meets the band requirement, which is equivalent to searching for the closest integer that equally divides $f_\mathrm{max}$, since harmonically related lengths must deliver equal bandwidths. Thus, the proper number of line pairs is given by:
\begin{equation}
    \begin{aligned}
        M =& \min \{\, m \in \{M_{\text{min}},\ldots,M_{\text{max}}\} \}, \\
        &\text{s.t. } M_{\text{max}} \bmod m = 0 \,
    \end{aligned}
    \label{eq:3.14}
\end{equation}

Finally, the recommended number of lines is computed by inverting the number of pairs equation $M = N(N-1)/2$, which gives:
\begin{equation}
    N = \left\lfloor \frac{1 + \sqrt{1 + 8M}}{2} \right\rceil
        \label{eq:3.15}
\end{equation}

In \eqref{eq:3.15}, we choose to round to the nearest integer, because in all of \eqref{eq:3.12}--\eqref{eq:3.14}, we are always overestimating the number of pairs. The number of lines determined here remains an approximation; using one more or one less based on rounding is also valid. Determining the truly optimal number of lines ultimately depends on the chosen optimality metric. For example, if the flatness of the eigenvalue is not critical and some dips are acceptable as long as they remain above the specified phase margin, then fewer lines may suffice.

\begin{figure}[th!]
    \centering
    \includegraphics[width=1\linewidth]{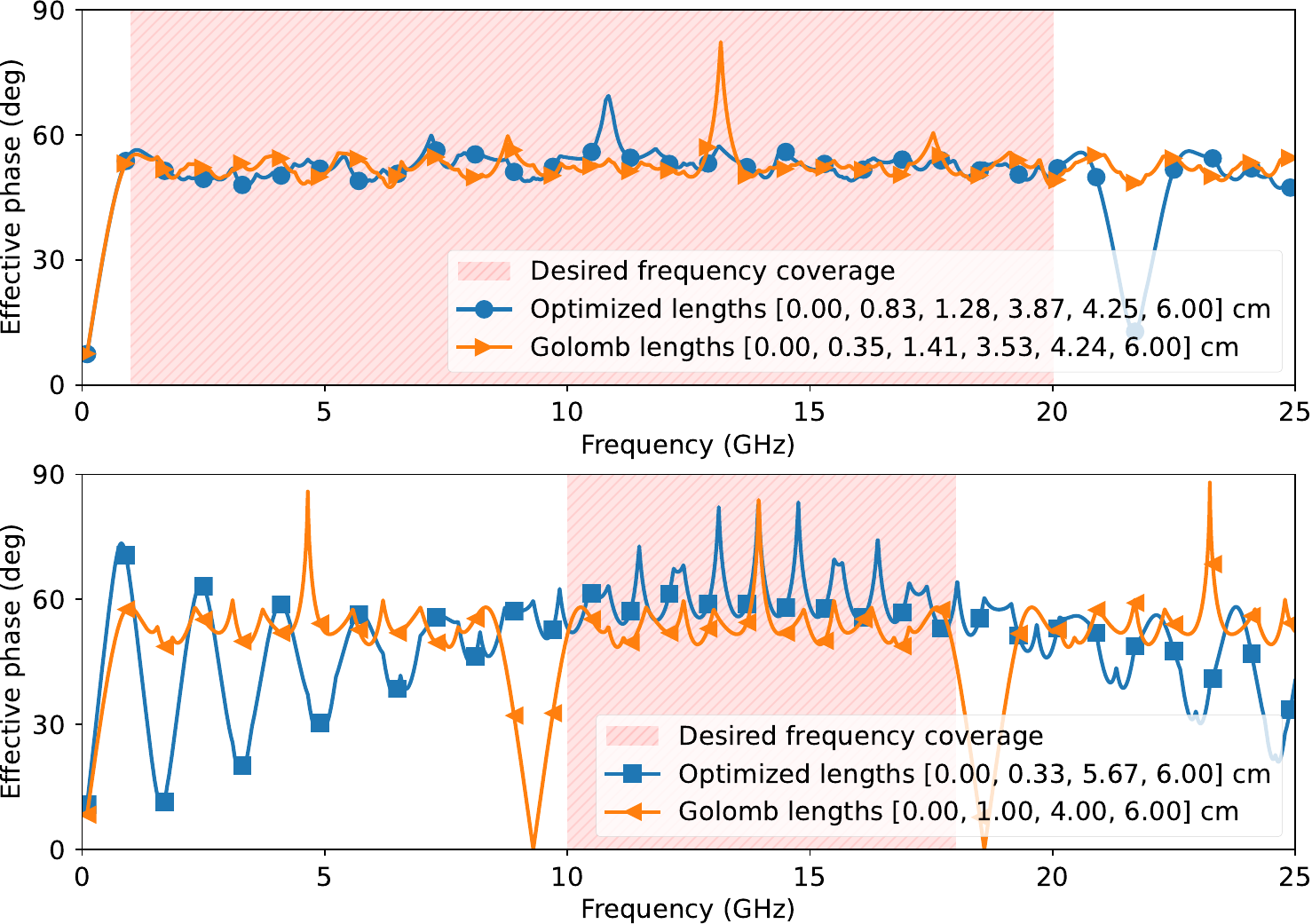}
    \caption{Effective phase responses for different line sets based on the example in Fig.~\ref{fig:3.7}. The upper plot demonstrates full frequency coverage from DC to $f_\mathrm{max}$, while the lower plot shows a limited bandwidth case operating between $f_\mathrm{min}$ and $f_\mathrm{max}$. For both cases, $l_\mathrm{max} = 6\,\mathrm{cm}$ and $\epsilon_\mathrm{r,eff} = 2.6-0j$.}
    \label{fig:3.8}
\end{figure}

To illustrate this, Fig.~\ref{fig:3.8} shows the equivalent phase for different sets of lines, computed using the number of pairs from \eqref{eq:3.12} and \eqref{eq:3.14}, and applying both the optimization and sparse ruler approaches to determine the lengths. In both plots in Fig.~\ref{fig:3.8}, the longest line is fixed, and the number of lines is calculated based on the required number of pairs for the coverage bandwidth. In the first plot, \eqref{eq:3.12} is used to calculate full-bandwidth coverage from DC, whereas in the second plot, \eqref{eq:3.14} is used for the limited-bandwidth example.

\section{Experimental Validation and Sensitivity Analysis}
\label{sec:4}

In this section, we validate the proposed line length computation methods through experimental measurements and Monte Carlo (MC) analysis. The first subsection revisits four multiline TRL calibration kits from prior measurement campaigns \cite{Hatab2023a,Hatab2023c,Hatab2023b,Schafsteller2024,Arsanjani2024}, all of which employed the sparse ruler method for selecting line lengths. Among the approaches proposed in this article, the sparse ruler method is the most convenient and straightforward, making it accessible even to users with limited VNA experience. We have applied this approach in numerous measurement campaigns covering frequencies up to $150\,\mathrm{GHz}$ on a probe station, spanning microstrip, grounded coplanar waveguide (GCPW), and stripline structures on various PCB materials \cite{Hatab2023b,Arsanjani2024,Arsanjani2024a,Hatab2023a,Hatab2023c,Hatab2026,Schafsteller2024}, as well as connectorized setups \cite{Hatab2023d,Hatab2024}. In those studies, line length selection was not the primary focus; rather, each publication addressed a different topic, such as device design and measurement--simulation correlation \cite{Hatab2023b,Arsanjani2024,Arsanjani2024a}, calibration algorithm development \cite{Hatab2023a,Hatab2023c,Hatab2026}, or material characterization \cite{Schafsteller2024}. Here, we revisit four representative kits covering different transmission line structures and substrate materials, and examine the effects of adding or removing lines from the calibration set.

The second subsection takes a complementary approach: we compare the sparse ruler and optimization methods against the line lengths of a commercial impedance standard substrate (ISS) commonly used by on-wafer probing vendors, quantifying the differences through MC analysis. Because the multiline TRL algorithm in \cite{Hatab2022,Hatab2023} inherently uses all available line pairs and automatically down-weights singular combinations through its eigenvalue formulation, it is difficult to distinguish between well-chosen length sets from measurement alone, as the algorithm produces robust results in all cases. To enable a controlled comparison, we therefore employ a MC analysis based on a CPW model validated against actual ISS measurements. This MC framework, originally demonstrated in \cite{Hatab2023}, is adapted here to incorporate lengths computed using both the optimization and sparse ruler approaches.

\subsection{PCB Measurements}
\label{sec:4a}

\begin{figure}[th!]
	\centering
	\subfloat[]{\includegraphics[width=0.49\linewidth]{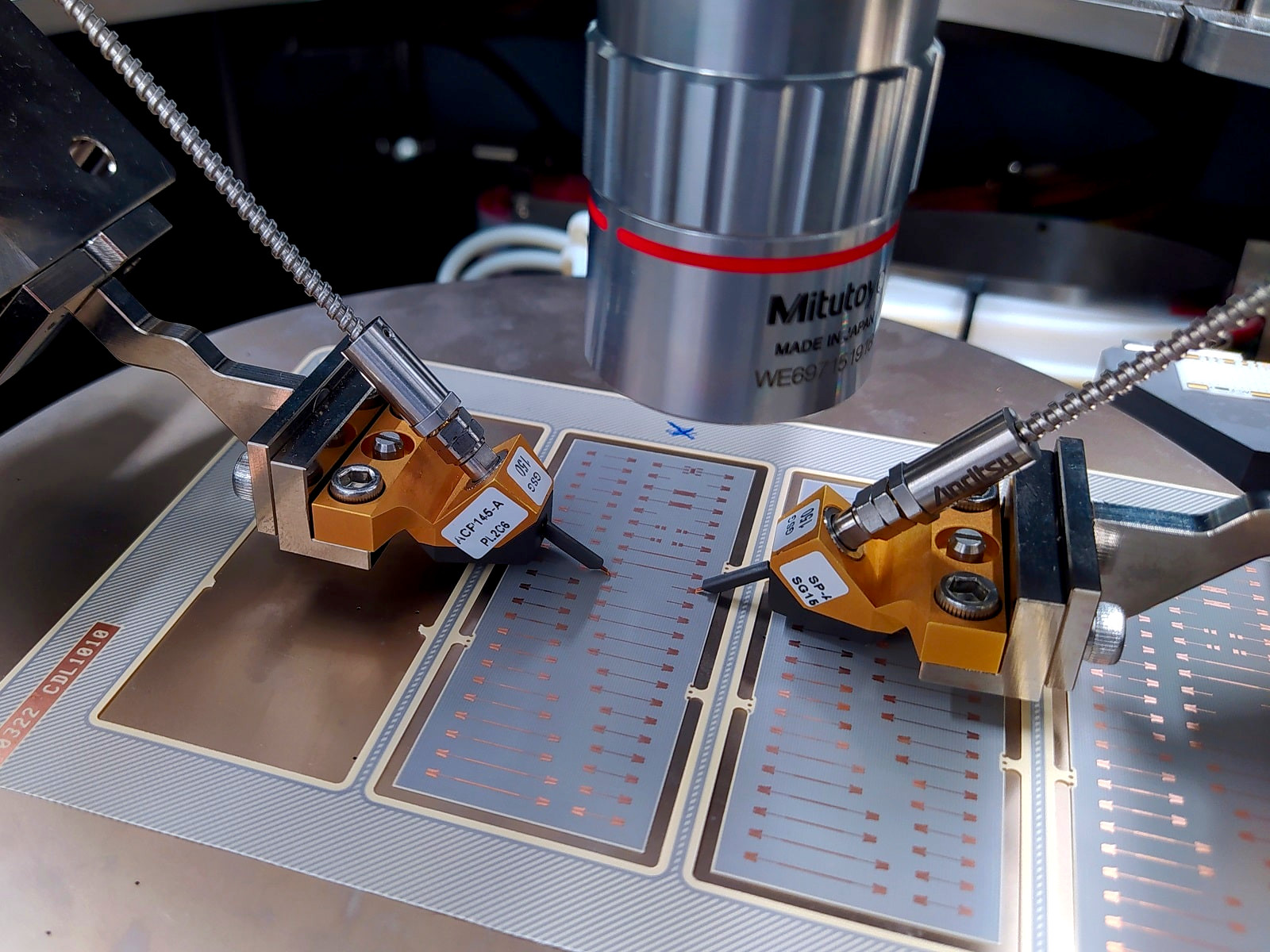}}
	\hfill
	\subfloat[]{\includegraphics[width=0.49\linewidth]{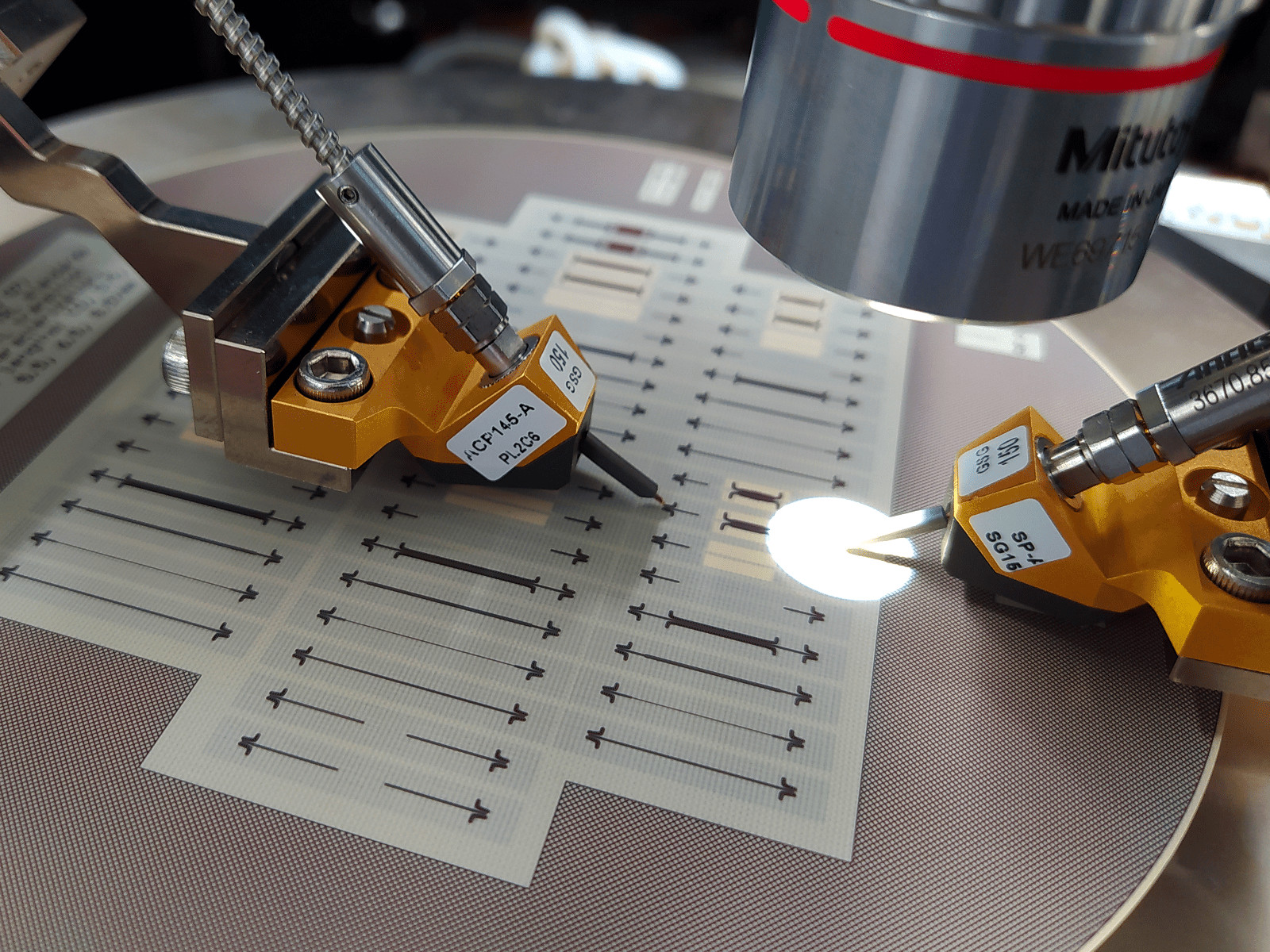}}\\
	\subfloat[]{\includegraphics[width=0.49\linewidth]{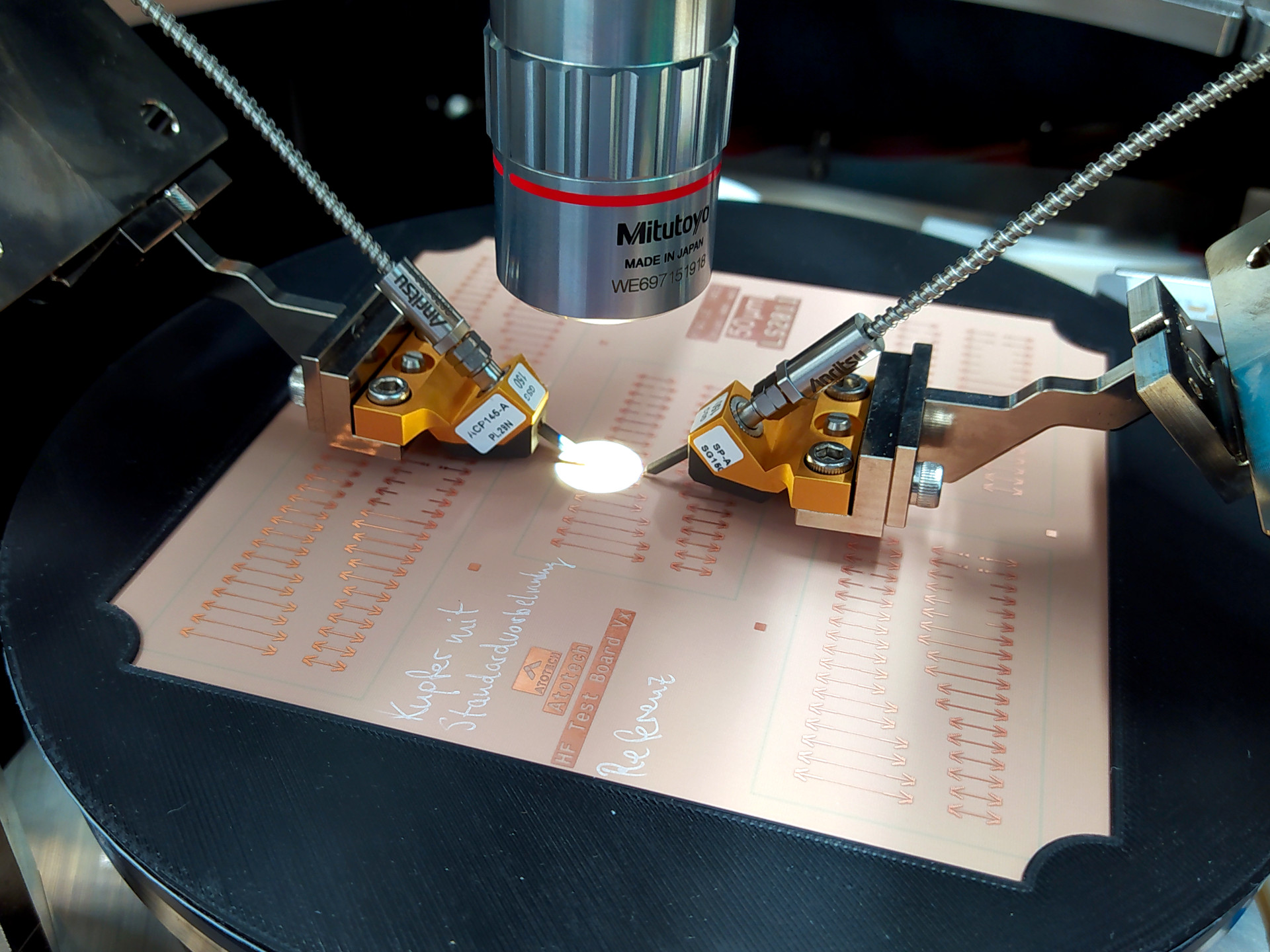}}
	\hfill
	\subfloat[]{\includegraphics[width=0.49\linewidth]{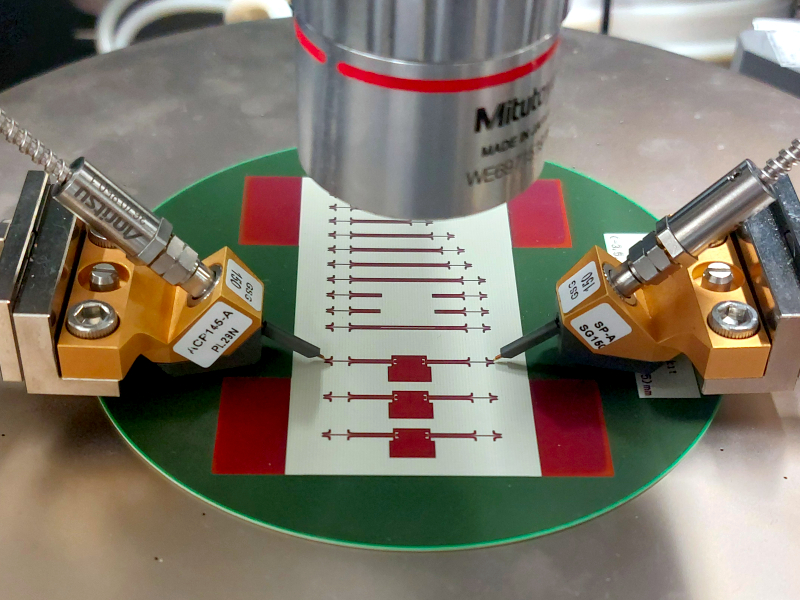}}
	\caption{Measurement setups for the four PCB calibration kits. (a)~Kit~1: microstrip on Megtron~7, (b)~Kit~2: stripline on Tachyon~100G, (c)~Kit~3: microstrip on Astra~MT77, and (d)~Kit~4: GCPW on Megtron~7. All measurements use FormFactor ACP $150\,\mu\mathrm{m}$ pitch GSG probes with an Anritsu VectorStar VNA with millimeter-wave frequency extension modules up to $150\,\mathrm{GHz}$ and a FormFactor SUMMIT200 probe station.}
	\label{fig:4.1}
\end{figure}

\begin{figure}[th!]
	\centering
	\subfloat[]{\includegraphics[width=0.5\linewidth]{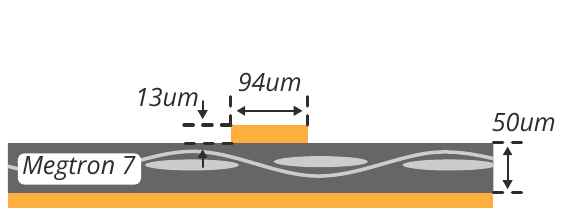}}
	\hfill
	\subfloat[]{\includegraphics[width=0.5\linewidth]{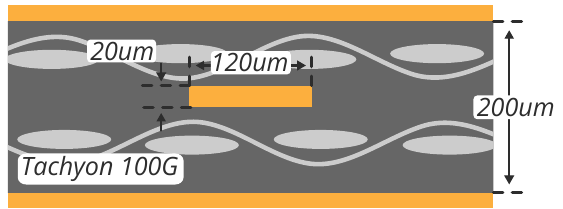}}\\
	\subfloat[]{\includegraphics[width=0.5\linewidth]{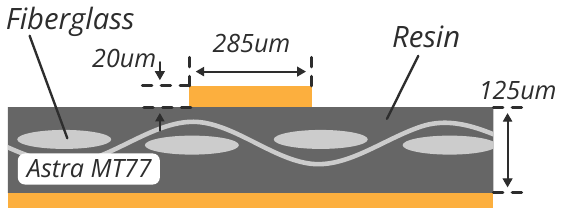}}
	\hfill
	\subfloat[]{\includegraphics[width=0.5\linewidth]{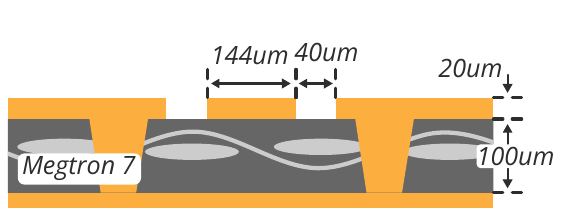}}
	\caption{Illustration of transmission line cross-section stackup and geometry for each kit. (a)~Kit~1: microstrip on $50\,\mu\mathrm{m}$ Megtron~7, (b)~Kit~2: stripline on $100\,\mu\mathrm{m}$ (each side) Tachyon~100G, (c)~Kit~3: microstrip on $125\,\mu\mathrm{m}$ Astra~MT77, and (d)~Kit~4: GCPW on $100\,\mu\mathrm{m}$ Megtron~7. These sketches are illustrative and are not to scale.}
	\label{fig:4.2}
\end{figure}

We consider four multiline TRL calibration kits from prior measurement campaigns, summarized in Table~\ref{tab:4.1}. Each kit was designed for a different application; the referenced publications discuss the respective use cases in detail. In all cases, the target frequency coverage was $2\,\mathrm{GHz}$ to $150\,\mathrm{GHz}$, although measurements were performed down to $1\,\mathrm{GHz}$. All measurements were conducted on the same hardware platform: FormFactor ACP $150\,\mu\mathrm{m}$-pitch GSG probes with PC~0.8\,mm connectors, connected to an Anritsu VectorStar VNA with millimeter-wave frequency extension modules up to $150\,\mathrm{GHz}$, used together with a FormFactor SUMMIT200 probe station. Photographs of the measurement setups are shown in Fig.~\ref{fig:4.1}, and the corresponding transmission line cross-section geometries are illustrated in Fig.~\ref{fig:4.2}. For kits~1 and~3, the line lengths were derived from the optimal sparse ruler $\{0, 1, 2, 6, 10, 13\}$, which is a non-Wichmann optimal ruler \cite{Luschny2013}. For kits~2 and~4, the optimal Golomb ruler $\{0, 1, 8, 11, 13, 17\}$ was used \cite{Meyer2009}. In both cases, the minimum length spacing was set to $0.5\,\mathrm{mm}$, ensuring coverage up to $150\,\mathrm{GHz}$ via \eqref{eq:2.7b} with margin for the expected variation in relative permittivity of the PCB substrate.
\begin{table}[ht!]
	\centering
	\caption{Summary of the four multiline TRL PCB calibration kits from prior publications, all designed using the sparse ruler method. All PCBs use non-plated copper traces.}
	\label{tab:4.1}
	\resizebox{\columnwidth}{!}{%
		\begin{tabular}{cccccc}
			\toprule
			Kit &
			Reference &
			Structure &
			\begin{tabular}[c]{@{}c@{}}Substrate \\ Material\end{tabular} &
			\begin{tabular}[c]{@{}c@{}}Substrate \\ Thickness \\ ($\mu$m)\end{tabular} &
			\begin{tabular}[c]{@{}c@{}}Lengths \\ (mm)\end{tabular} \\
			\midrule
			1 & \cite{Hatab2023a,Hatab2023c} &
			Microstrip &
			\begin{tabular}[c]{@{}c@{}}Panasonic \\ Megtron~7\cite{panasonic_megtron7}\end{tabular} &
			$50$ &
			\begin{tabular}[c]{@{}c@{}}$0, 0.5, 1,$ \\ $3, 5, 6.5$\end{tabular} \\\midrule
			2 & \cite{Hatab2023b} &
			Stripline &
			\begin{tabular}[c]{@{}c@{}}ISOLA \\ Tachyon~100G\cite{isola_tachyon100g}\end{tabular} &
			\begin{tabular}[c]{@{}c@{}}$100$ \\ (each side)\end{tabular} &
			\begin{tabular}[c]{@{}c@{}}$0, 0.5, 4,$ \\ $5.5, 6.5, 8.5$\end{tabular} \\\midrule
			3 & \cite{Schafsteller2024} &
			Microstrip &
			\begin{tabular}[c]{@{}c@{}}ISOLA \\ Astra~MT77\cite{isola_astra_mt77}\end{tabular} &
			$125$ &
			\begin{tabular}[c]{@{}c@{}}$0, 0.5, 1,$ \\ $3, 5, 6.5$\end{tabular} \\\midrule
			4 & \cite{Arsanjani2024} &
			GCPW &
			\begin{tabular}[c]{@{}c@{}}Panasonic \\ Megtron~7\cite{panasonic_megtron7}\end{tabular} &
			$100$ &
			\begin{tabular}[c]{@{}c@{}}$0, 0.5, 4,$ \\ $5.5, 6.5, 8.5$\end{tabular} \\
			\bottomrule
		\end{tabular}
	}
\end{table}

As discussed in Section~\ref{sec:3b}, the sparse ruler method computes line lengths under a lossless assumption using only the real part of the effective relative permittivity. All four substrates employ fiberglass-reinforced resin construction, making them inherently non-homogeneous. As a result, the effective relative permittivity seen by the transmission line can vary depending on the line geometry and its position relative to the glass weave \cite{Hatab2023c}. To account for this, we over-design the line lengths using the sparse ruler to accommodate worst-case permittivity scenarios. A higher-than-expected $\epsilon_\mathrm{r,eff}$ shifts $f_\mathrm{max}$ downward, whereas a lower-than-expected $\epsilon_\mathrm{r,eff}$ shifts $f_\mathrm{min}$ upward, as evident from \eqref{eq:2.6}. As shown in \cite{Hatab2023c}, PCB substrates can exhibit significant permittivity variation due to the glass weave. For simplicity, we adopt a conservative design margin of $\epsilon_\mathrm{r} \pm 0.5$, corresponding to approximately $15\%$ variation for a substrate with a nominal $\epsilon_\mathrm{r}$ of~$3$.

\begin{figure}[th!]
	\centering
	\includegraphics[width=1\linewidth]{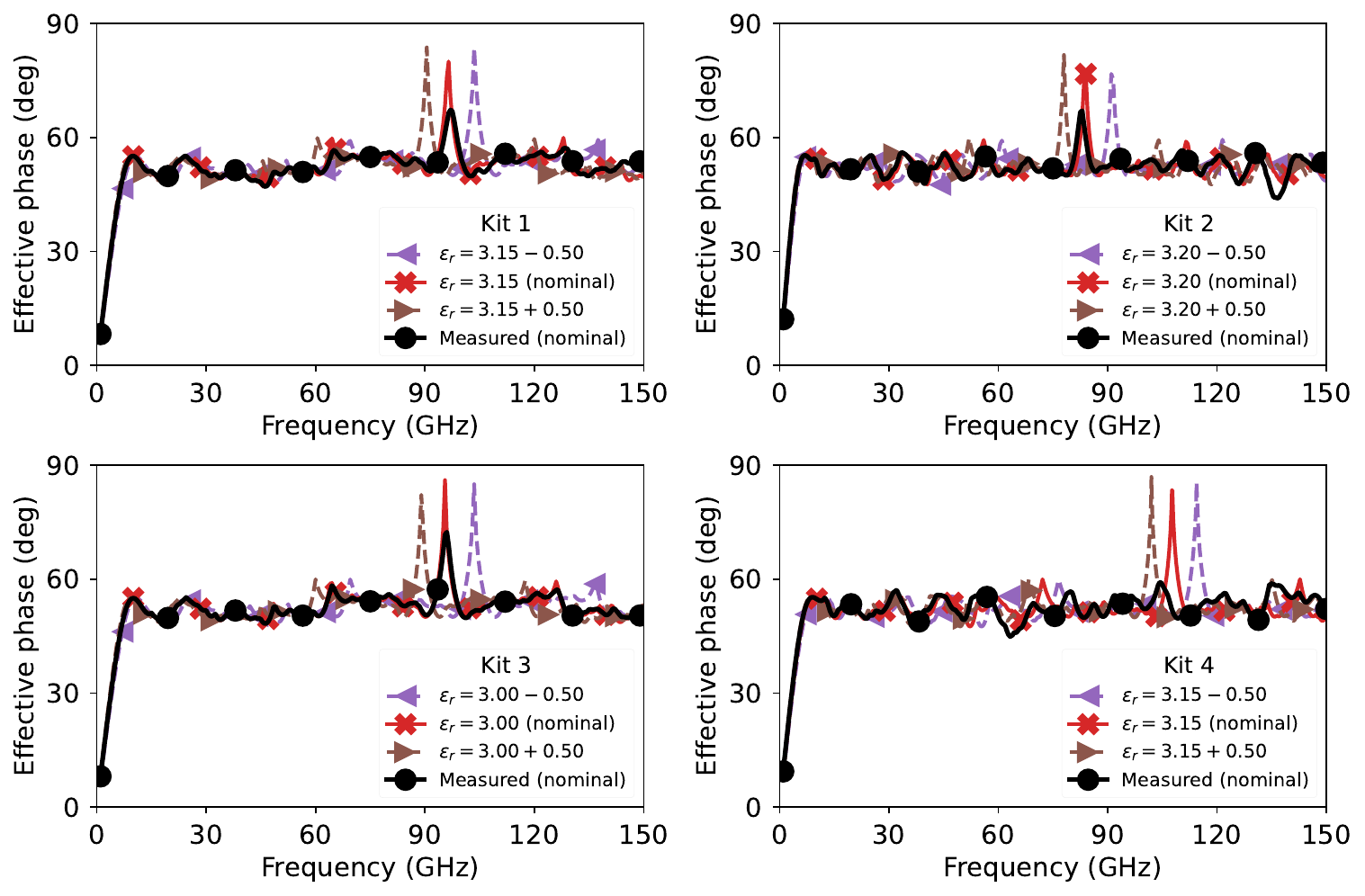}
	\caption{Comparison of the designed (theoretical) effective phase against the measured effective phase for all four calibration kits. The theoretical curves are computed from the nominal line lengths using the substrate dielectric constant $\varepsilon_\mathrm{r}$ and its variation $\varepsilon_\mathrm{r} \pm 0.5$. For microstrip (kits~1,~3) and GCPW (kit~4), the effective permittivity $\varepsilon_\mathrm{r,eff}$ is obtained via analytical models from \textit{scikit-rf} \cite{Arsenovic2022}; for stripline (kit~2), $\varepsilon_\mathrm{r,eff} = \varepsilon_\mathrm{r}$.}
	\label{fig:4.3}
\end{figure}

We first compare the designed effective phase against its measured counterpart to verify the frequency coverage of the selected line lengths. Fig.~\ref{fig:4.3} shows the theoretical effective phase, computed from the nominal line lengths and the expected effective relative permittivity using \eqref{eq:2.22}, \eqref{eq:2.23}, and \eqref{eq:2.28}, alongside the measured effective phase obtained from the calibration. The theoretical curves are evaluated at the nominal substrate dielectric constant $\varepsilon_\mathrm{r}$ from the manufacturer datasheet as well as at $\varepsilon_\mathrm{r} \pm 0.5$, representing the conservative design margin. For the microstrip kits~(1 and~3) and the GCPW kit~(4), the effective relative permittivity $\varepsilon_\mathrm{r,eff}$ is obtained from analytical transmission line models implemented in \textit{scikit-rf} \cite{Arsenovic2022}; for the stripline kit~(2), $\varepsilon_\mathrm{r,eff} = \varepsilon_\mathrm{r}$ since the conductor is fully enclosed by the substrate. As shown in Fig.~\ref{fig:4.3}, the measured effective phase falls well within the designed bounds, confirming that the sparse ruler approach provides the intended flat frequency coverage even under realistic permittivity variations.

In addition to the nominal sparse ruler sets, we examine two augmented configurations: for kit~1, two additional intermediate lines of $1\,\mathrm{mm}$ and $2\,\mathrm{mm}$ were included during the kit design; for kit~4, a line with a $-3.5\,\mathrm{mm}$ offset relative to the thru was added, extending the maximum pairwise length difference from $8.5\,\mathrm{mm}$ to $12\,\mathrm{mm}$.

To assess the robustness of the nominal sparse ruler sets, we systematically remove lines and evaluate the resulting effective phase, keeping the thru line ($0\,\mathrm{mm}$) always present so that all calibrations share the same reference plane. For the single-line removal case, each of the five non-thru lines is removed in turn, producing five reduced sets per kit. For the two-line removal case, all $C(5,2)=10$ combinations are evaluated. In each case, the effective phase is computed from the measurements of the remaining lines. The key indicator of a poor line set is the presence of localized dips in the effective phase within the calibration band. The low-frequency roll-off itself is not considered, as it is an expected consequence of reducing the maximum length difference when the longest line is removed.

\begin{figure}[th!]
	\centering
	\includegraphics[width=1\linewidth]{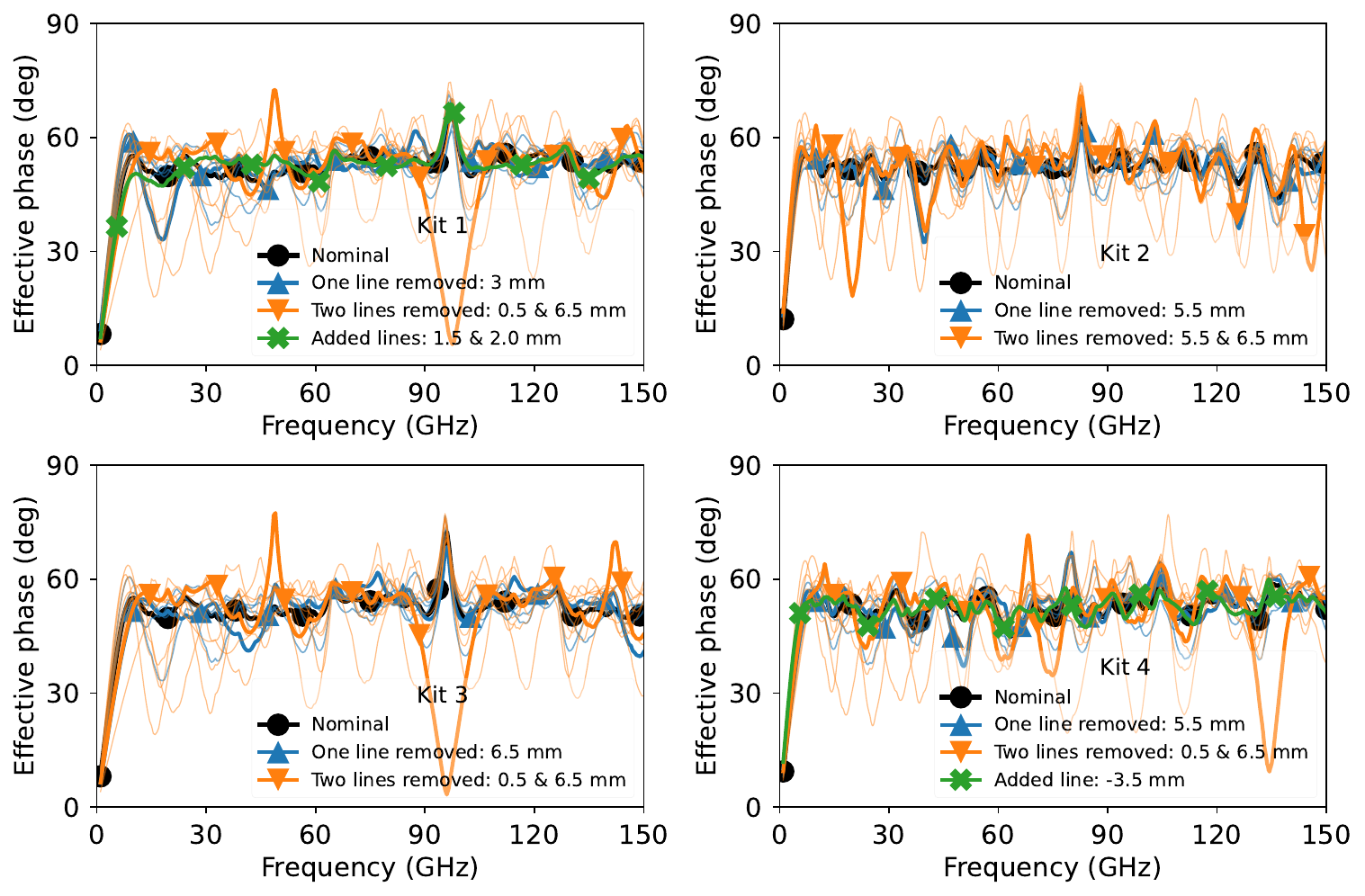}
	\caption{Comparison of measured effective phase across all four calibration kits. Shown are: the nominal sparse ruler sets (solid), augmented sets (kit~1 with additional $1\,\mathrm{mm}$ and $2\,\mathrm{mm}$ lines; kit~4 with extended maximum length difference), and all single-line and two-line removal combinations (light traces). In both removal scenarios, the worst-case combination---i.e., the one exhibiting the largest effective phase dip above the low-frequency roll-off---is highlighted with a darker, thicker trace.}
	\label{fig:4.4}
\end{figure}

Fig.~\ref{fig:4.4} compares the measured effective phase across all configurations: nominal, augmented, and reduced. For the reduced sets, all combinations are plotted with reduced opacity to convey the spread of outcomes, while the worst-case combination (i.e., the one exhibiting the largest effective phase dip at any frequency, excluding low-frequency roll-off) is highlighted with a darker, thicker trace. Removing a single line generally introduces a moderate dip in the effective phase at a specific frequency, but does not render any of the considered kits unusable. Removing two lines produces more pronounced dips, yet the effective phase remains above zero across most frequencies, even in the worst case. For kits~1 and~3, which use the same non-Wichmann sparse ruler, the worst-case two-line removal ($0.5\,\mathrm{mm}$ and $6.5\,\mathrm{mm}$) limits the usable band to just above $90\,\mathrm{GHz}$. Kits~2 and~4, which employ a Golomb sparse ruler, exhibit significant dips upon two-line removal but do not reach a null as in kits~1 and~3, indicating that the calibration still yields a usable solution, albeit with increased sensitivity in the affected frequency range.

Regarding the augmented configurations, adding lines does not produce a notable improvement in the effective phase. For kit~1, which includes two additional lines within the kit's maximum length, there is no discernible phase improvement; however, for kit~4, the added longer line improves the low-frequency roll-off compared to the nominal set, confirming that extending the maximum length difference can enhance the effective phase at low frequencies. Nevertheless, the nominal sparse ruler sets already provide a flat effective phase across the band, and the improvements from augmentation are modest in terms of phase coverage. Overall, removing any single non-thru line does not significantly degrade the calibration from a phase coverage perspective, whereas removing two lines has a more pronounced impact, but the kit remains usable, as will be demonstrated by the DUT measurement that follows. This underscores the effectiveness of the sparse ruler method in selecting line lengths that yield stable, flat frequency coverage by design.

\begin{figure}[th!]
	\centering
	\includegraphics[width=1\linewidth]{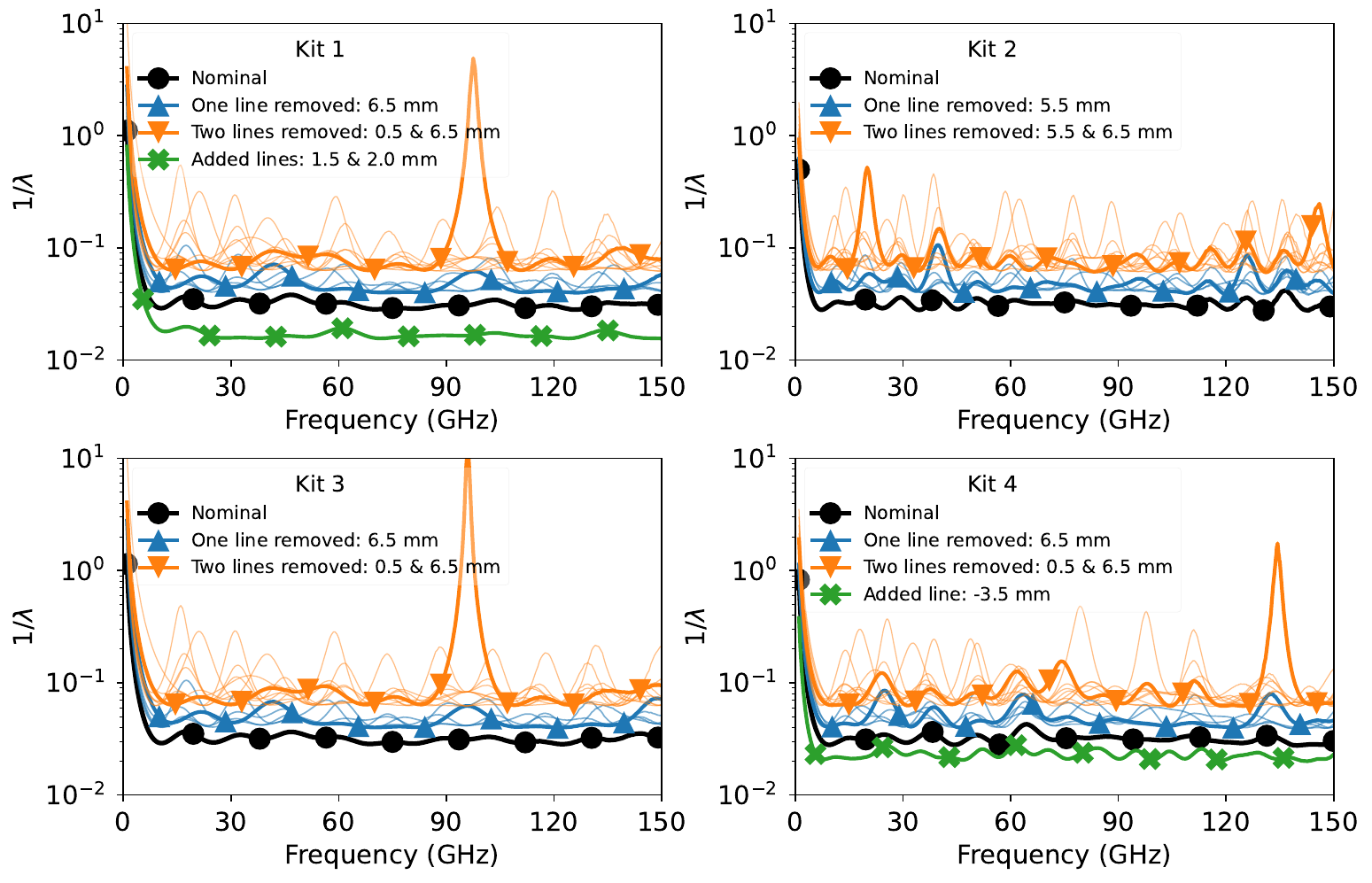}
	\caption{Measured inverse eigenvalue $1/\lambda$ versus frequency for all four calibration kits, comparing nominal, augmented, and reduced line sets. Lower values indicate reduced sensitivity of the eigenvector solution.}
	\label{fig:4.5}
\end{figure}

A complementary metric is the inverse eigenvalue $1/\lambda$, which is proportional to the sensitivity of the eigenvector solution, as discussed in Section~\ref{sec:2c}. As the number of lines increases, $\lambda$ grows and its inverse decreases, reflecting reduced statistical uncertainty. Fig.~\ref{fig:4.5} shows $1/\lambda$ for all considered configurations. As expected, incorporating additional lines further reduces $1/\lambda$, consistent with the averaging effect of more line pairs. For kit~1, adding intermediate lines does not significantly change the effective phase (which is already flat due to the sparse ruler design), but it does reduce $1/\lambda$, since additional measurements lower the statistical uncertainty. For kit~4, the additional longer line improves both the effective phase and $1/\lambda$ at the lowest frequencies.

\begin{figure}[th!]
	\centering
	\includegraphics[width=1\linewidth]{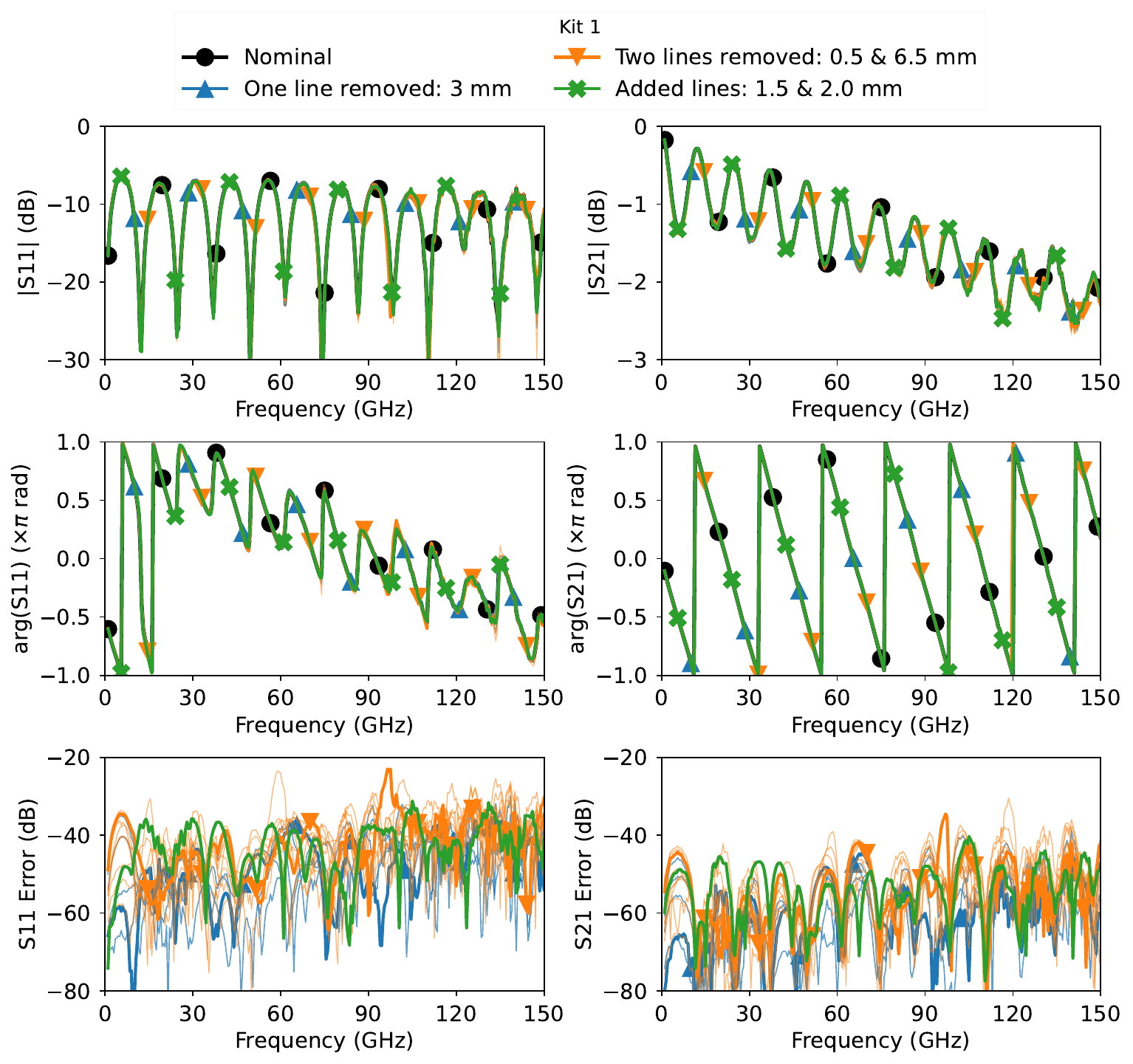}
	\caption{Calibrated S-parameters of the stepped-impedance verification line for kit~1 across all line set configurations. Left column: $S_{11}$; right column: $S_{21}$. Top row: magnitude; middle row: phase; bottom row: deviation from nominal $\mathrm{dB}(|S_{ij} - S_{ij,\mathrm{nominal}}|)$.}
	\label{fig:4.6}
\end{figure}

\begin{figure}[th!]
	\centering
	\includegraphics[width=1\linewidth]{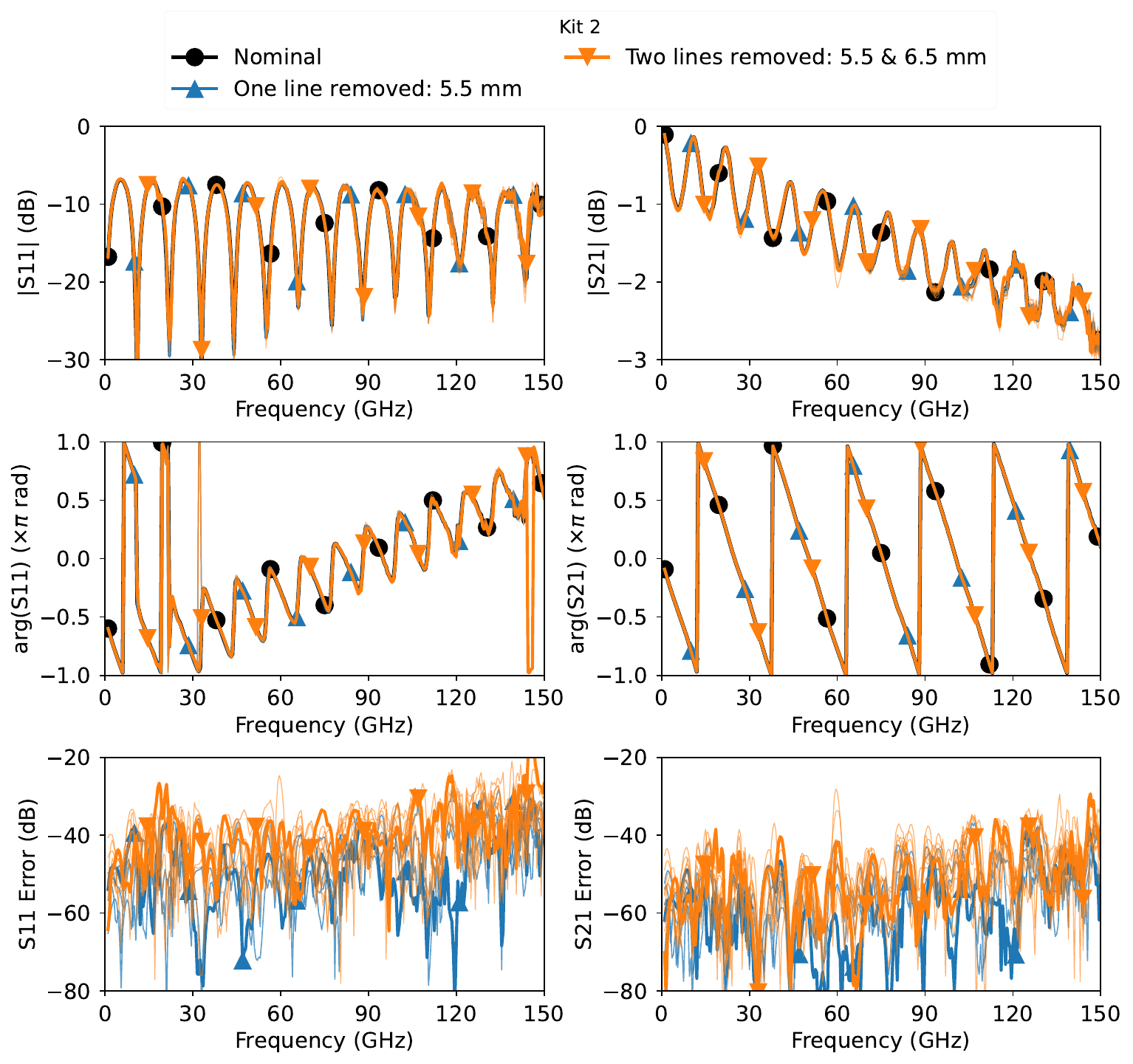}
	\caption{Calibrated S-parameters of the stepped-impedance verification line for kit~2 across all line set configurations. Left column: $S_{11}$; right column: $S_{21}$. Top row: magnitude; middle row: phase; bottom row: deviation from nominal $\mathrm{dB}(|S_{ij} - S_{ij,\mathrm{nominal}}|)$.}
	\label{fig:4.7}
\end{figure}

\begin{figure}[th!]
	\centering
	\includegraphics[width=1\linewidth]{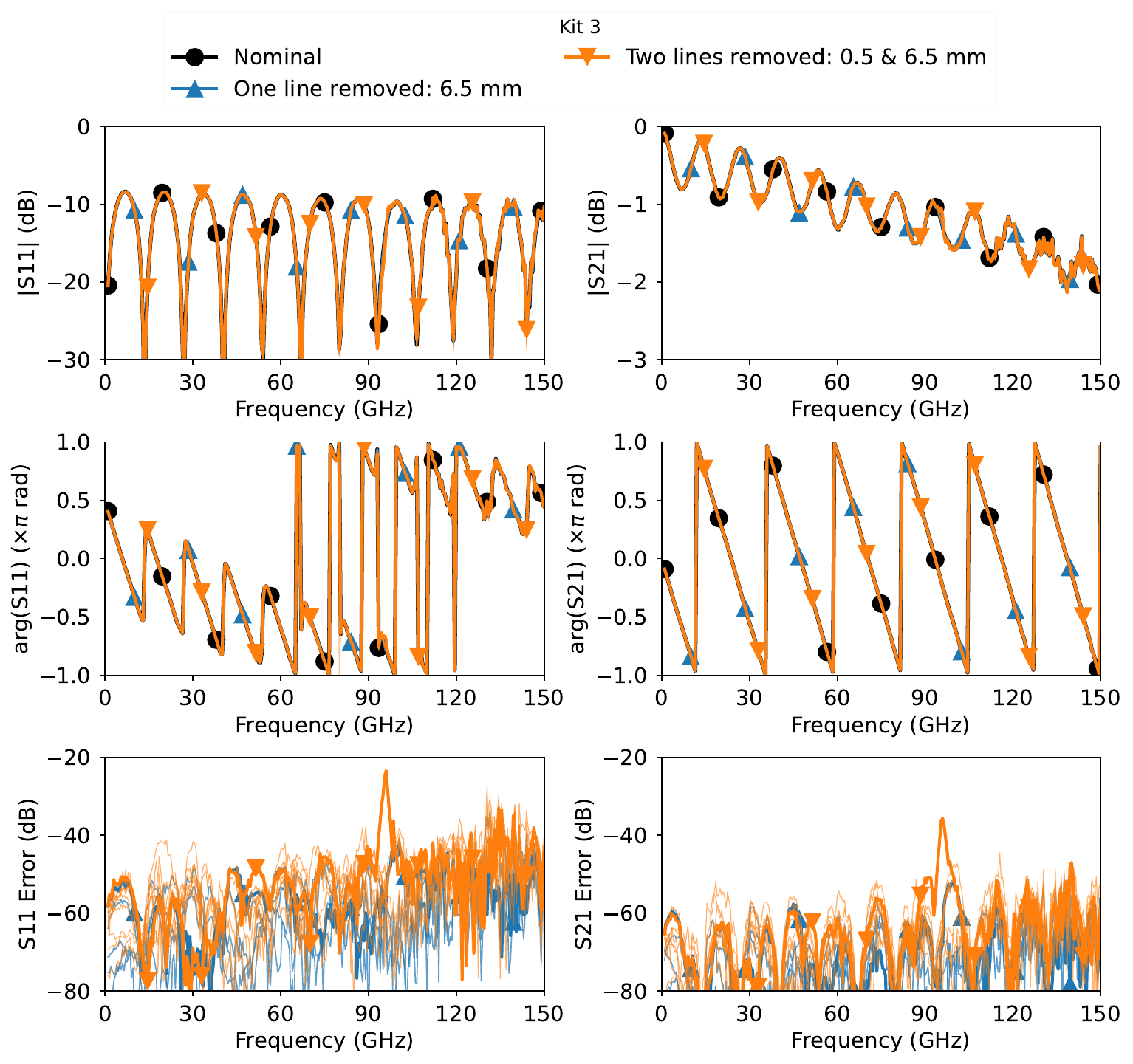}
	\caption{Calibrated S-parameters of the stepped-impedance verification line for kit~3 across all line set configurations. Left column: $S_{11}$; right column: $S_{21}$. Top row: magnitude; middle row: phase; bottom row: deviation from nominal $\mathrm{dB}(|S_{ij} - S_{ij,\mathrm{nominal}}|)$.}
	\label{fig:4.8}
\end{figure}

\begin{figure}[th!]
	\centering
	\includegraphics[width=1\linewidth]{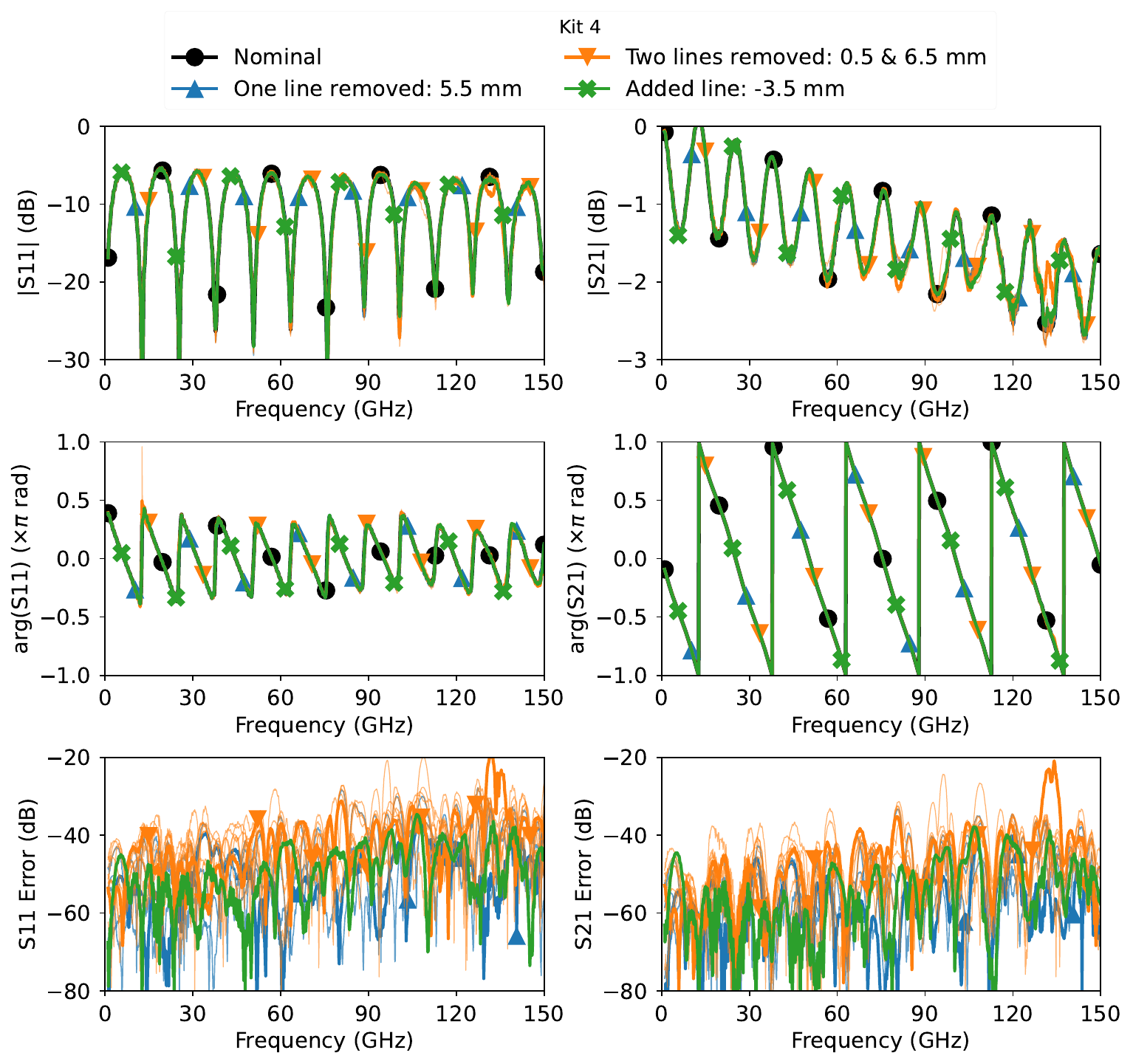}
	\caption{Calibrated S-parameters of the stepped-impedance verification line for kit~4 across all line set configurations. Left column: $S_{11}$; right column: $S_{21}$. Top row: magnitude; middle row: phase; bottom row: deviation from nominal $\mathrm{dB}(|S_{ij} - S_{ij,\mathrm{nominal}}|)$.}
	\label{fig:4.9}
\end{figure}

To assess the impact on device measurements, we compare the calibrated S-parameters of a stepped-impedance verification line across all four kits. Figs.~\ref{fig:4.6}--\ref{fig:4.9} show the magnitude and phase of $S_{11}$ and $S_{21}$, together with the deviation metric $\mathrm{dB}(|S_{ij} - S_{ij,\mathrm{nominal}}|)$ relative to the nominal sparse ruler calibration. We emphasize that this metric quantifies the deviation from the nominal sparse ruler result, not the absolute accuracy. Its purpose is to assess how much the calibrated DUT response changes when lines are added to or removed from the calibration set.
The results indicate that adding lines to the nominal set does not cause significant deviations: for kits~1 and~4 (which have augmented configurations), the maximum deviation averages around $-40\,\mathrm{dB}$ for both $S_{11}$ and $S_{21}$. This confirms that a well-designed sparse ruler set already captures the essential calibration information, and supplementing it with additional lines produces only marginal changes in the calibrated device response.
Removing a single line from the nominal set is generally acceptable, introducing only modest variation that does not substantially degrade the calibration, as observed across all kits. For kits~1 and~4, the single-line removal deviation is comparable to that of augmentation. However, removing two lines can produce noticeable deviations that correlate in frequency with the dips observed in the effective phase (Fig.~\ref{fig:4.4}) and the peaks in $1/\lambda$ (Fig.~\ref{fig:4.5}), confirming that these metrics reliably predict the frequency regions most sensitive to line removal. Even in the worst-case two-line removal scenario, the maximum deviations remain below $-20\,\mathrm{dB}$ for both $S_{11}$ and $S_{21}$ across most of the band, indicating that the calibration still yields a usable result, albeit with reduced accuracy in the affected frequency ranges. This underscores the robustness of the sparse ruler method in providing stable calibration performance, even when lines are removed.

\subsection{Measurement-Based Uncertainty Analysis}
\label{sec:4b}

\begin{figure}[!ht]
	\centering
	\includegraphics[width=.9\linewidth]{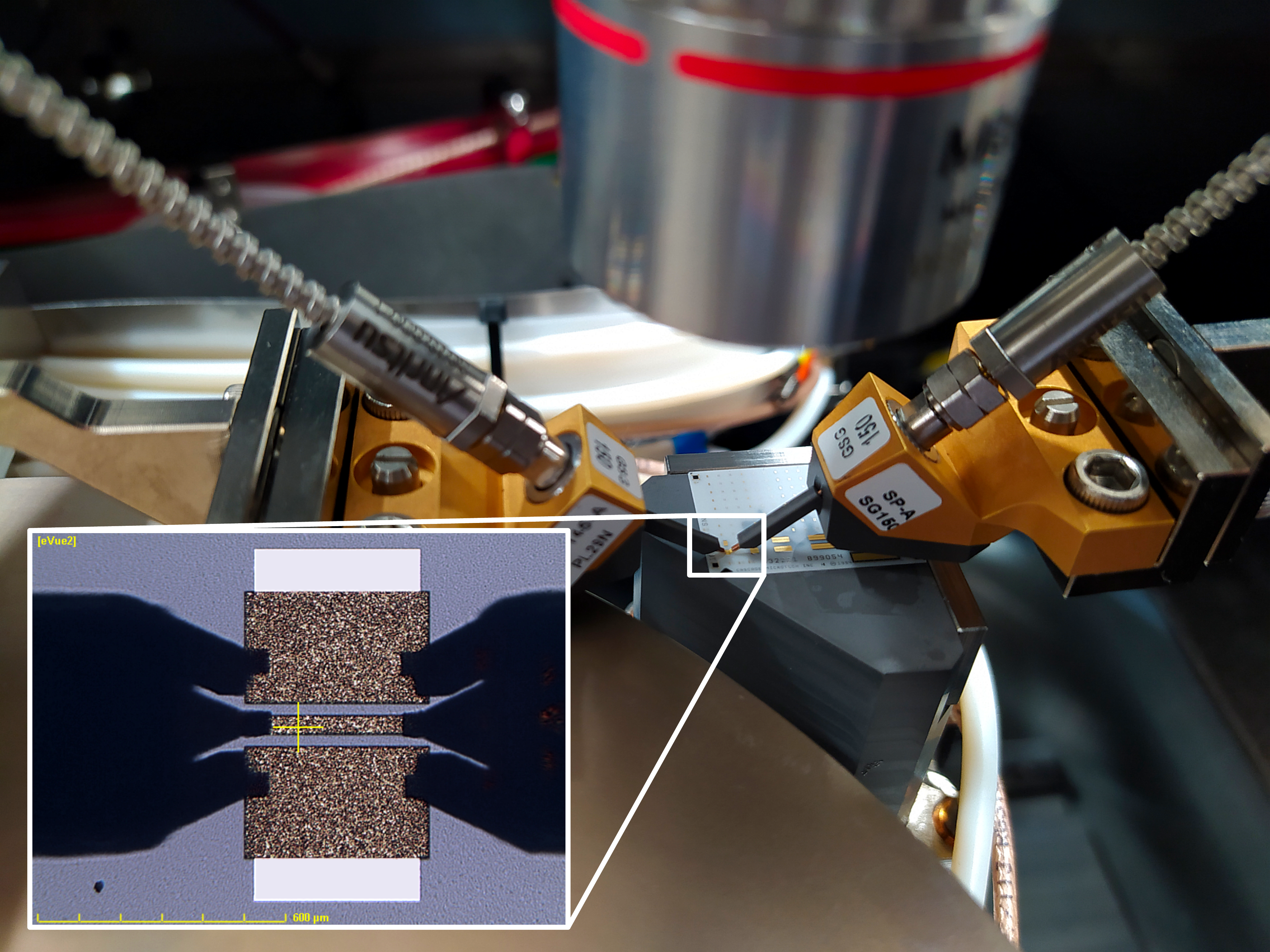}
	\caption{On-wafer measurement setup for the ISS-based CPW multiline TRL calibration, showing the FormFactor ACP $150\,\mu\mathrm{m}$ pitch GSG probes in contact with the ISS, from \cite{Hatab2023}.}
	\label{fig:4.10}
\end{figure}

\begin{figure}[!ht]
	\centering
	\includegraphics[width=0.95\linewidth]{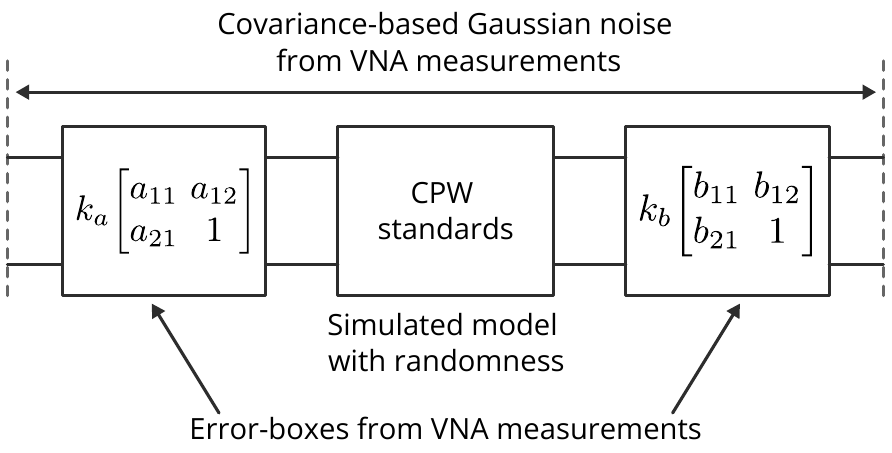}
	\caption{Overview of the MC experiment. The parameters of the calibration standards are perturbed, while correlated Gaussian noise is generated based on the sample covariance estimated from the VNA measurements.}
	\label{fig:4.11}
\end{figure}

This subsection compares the line lengths of a commercial Impedance Standard Substrates (ISS) multiline TRL calibration kit against those obtained using the optimization and sparse ruler methods proposed in this article, quantifying the differences through MC analysis. The choice of an ISS-based CPW structure is motivated by two considerations. First, it enables a direct comparison against line lengths used in an industry-standard multiline TRL calibration kit. Second, CPW on Alumina has well-established analytical models \cite{Phung2021,Schnieder2003,Heinrich1993} that have been confirmed against measurements in \cite{Hatab2023}. In contrast, the PCB transmission line structures used in Section~\ref{sec:4a} employ inhomogeneous substrates for which only nominal design models are available (e.g., the analytical models in \textit{scikit-rf} package \cite{Arsenovic2022}), making them unsuitable for a rigorous MC uncertainty study. The ISS under consideration uses CPW lines with the following lengths (relative to the first line, defined as the thru): $\{0, 0.25, 0.7, 1.6, 3.3, 5.05\}\,\mathrm{mm}$ \cite{FormFactor2023,MPICorporation2023}. The measurement setup is shown in Fig.~\ref{fig:4.10}, and an overview of the experiment based on the VNA error box model is illustrated in Fig.~\ref{fig:4.11}. The instrumentation is identical to that used in the previous subsection for the PCB measurements.

We compare the commercial ISS line lengths to an optimized set obtained from the regularized objective function in \eqref{eq:3.6}, which accounts for length uncertainty, as well as to lengths derived from the Golomb sparse ruler. For the optimizer, we assumed a length standard uncertainty of $20\,\mu\mathrm{m}$, which enters the regularization term in \eqref{eq:3.6}. To ensure a fair comparison, we constrain our solutions to match the commercial ISS: the same number of lines, the same maximum length, and quantization in steps of $50\,\mu\mathrm{m}$ (i.e., all lengths are integer multiples of $50\,\mu\mathrm{m}$). Additionally, we consider a fine-resolution optimized set, quantized in steps of $1\,\mu\mathrm{m}$ instead of $50\,\mu\mathrm{m}$, to assess whether finer quantization yields a meaningfully different solution. The resulting line lengths are: $\{0, 0.35, 0.75, 2.4, 3.85, 5.05\}\,\mathrm{mm}$ for the $50\,\mu\mathrm{m}$-step optimized set, $\{0, 1.471, 1.802, 3.93, 4.311, 5.05\}\,\mathrm{mm}$ for the $1\,\mu\mathrm{m}$-step (fine-resolution) optimized set, and $\{0, 0.3, 1.2, 2.95, 3.55, 5.05\}\,\mathrm{mm}$ for the Golomb sparse ruler set. Both optimizations were performed using the DE method from the \textit{SciPy} package in Python \cite{Virtanen2020} on an AMD Ryzen~7 6800HS processor; each ran for 2000 iterations, completing in approximately 175 seconds while consuming approximately 1.5\,GB of RAM.

\begin{figure}[th!]
    \centering
    \includegraphics[width=1\linewidth]{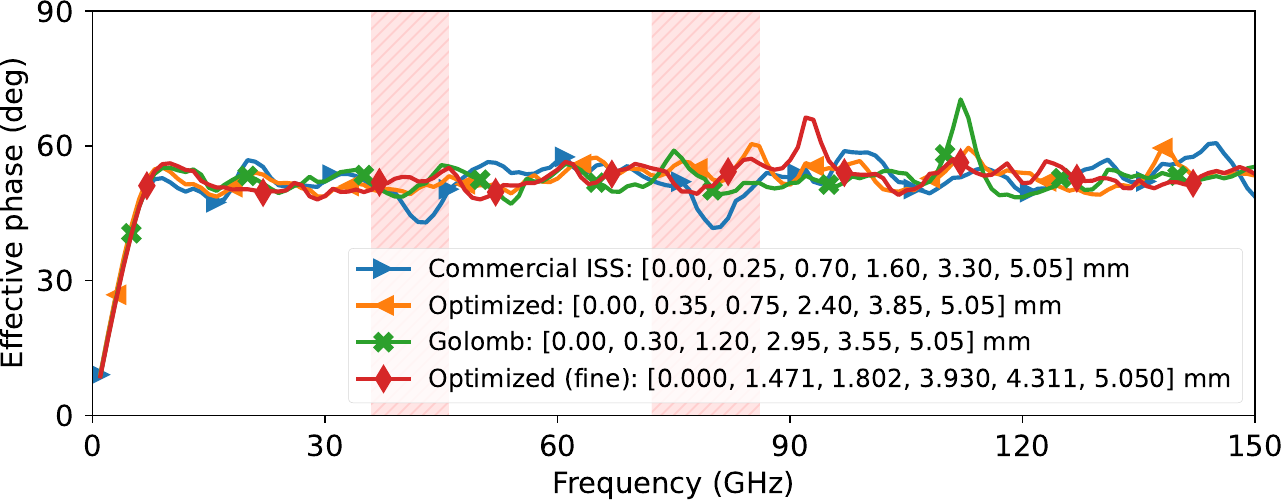}
    \caption{Comparison of the effective phase across frequency for the commercial ISS, optimized ($50\,\mu\mathrm{m}$ steps), Golomb sparse ruler, and fine-resolution optimized ($1\,\mu\mathrm{m}$ steps) line length sets. The effective phase is computed from the nominal line lengths using the CPW model parameters.}
    \label{fig:4.12}
\end{figure}

Before running the MC analysis, we first compare the effective phase of the four line length sets to verify their frequency coverage. Fig.~\ref{fig:4.12} shows the effective phase for the four line length sets, computed using \eqref{eq:2.22}, \eqref{eq:2.23}, and \eqref{eq:2.28}. All four sets provide non-zero effective phase across the calibration band, confirming that each covers the required frequency range. Notably, the fine-resolution optimized set produces an effective phase profile very similar to the standard ($50\,\mu\mathrm{m}$) optimized set, indicating that finer quantization does not yield a noticeable phase improvement in this example. However, the benefit of finer quantization will become apparent later in the form of improved robustness: the uncertainties are spread more evenly across all lines, and the calibration is less sensitive to the removal of individual lines.

\begin{figure}[th!]
    \centering
    \includegraphics[width=1\linewidth]{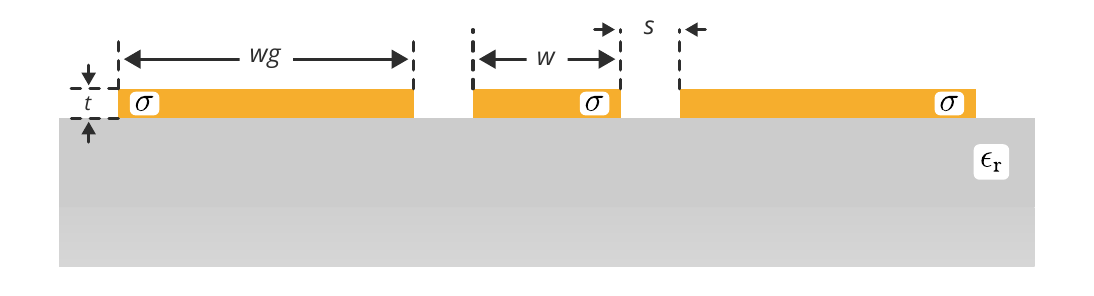}
    \caption{Geometric cross-section parameters of the CPW structure used in the MC analysis. The model uses a lossless Alumina substrate with infinite extent, assuming absorbing boundary conditions. The values used for the MC analysis are listed in Table~\ref{tab:4.2}.}
    \label{fig:4.13}
\end{figure}

To quantify these differences statistically, we perform a MC simulation using a hybrid approach that combines synthetic data generated by an analytical CPW model \cite{Phung2021,Schnieder2003,Heinrich1993} with actual ISS measurements. In the model, all cross-section parameters, including line length, are varied independently across lines. The synthetic line data is embedded into error boxes obtained from multiline TRL calibration of the ISS measurements, and correlated Gaussian noise based on the sample covariance estimated from the VNA measurements is added at each MC iteration \cite{Hatab2023}. The MC analysis was run for 5000 trials to ensure statistical significance. The geometric cross-section parameters of the CPW structure are shown in Fig.~\ref{fig:4.13}. The line length uncertainty used in the MC perturbation is $20\,\mu\mathrm{m}$, consistent with the value assumed in the optimization objective function \eqref{eq:3.6}, but also representative of typical overtravel in probes used for these measurements. The uncertainties for the cross-section parameters are listed in Table~\ref{tab:4.2}, adopted from \cite{Hatab2023}. The measurement noise was modeled using the sample covariance estimated from an ensemble of 500 wave parameter sweeps per standard collected on the VNA at an IFBW of $10\,\mathrm{kHz}$.

\begin{table}[ht!]
    \centering
    \caption{Cross-section dimensions and material properties of the considered CPW structure and their standard uncertainties, adopted from \cite{Hatab2023}.}
    \label{tab:4.2}
    \resizebox{\columnwidth}{!}{%
        \begin{tabular}{cccccc}
            \toprule
            \begin{tabular}[c]{@{}c@{}}Signal\\ width\\ ($\mu$m)\end{tabular} &
            \begin{tabular}[c]{@{}c@{}}Ground\\ width\\ ($\mu$m)\end{tabular} &
            \begin{tabular}[c]{@{}c@{}}Conductor\\ spacing\\ ($\mu$m)\end{tabular} &
            \multicolumn{1}{c}{\begin{tabular}[c]{@{}c@{}}Conductor\\ thickness\\ ($\mu$m)\end{tabular}} &
            \begin{tabular}[c]{@{}c@{}}Relative \\ permittivity\\ (1)\end{tabular} &
            \begin{tabular}[c]{@{}c@{}}Conductor\\ conductivity\\ (S/m)\end{tabular} \\ \midrule
            \begin{tabular}[c]{@{}c@{}}$49.1$\\ $\pm2.55$\end{tabular} &
            \begin{tabular}[c]{@{}c@{}}$273.3$\\ $\pm2.55$\end{tabular} &
            \begin{tabular}[c]{@{}c@{}}$25.5$\\ $\pm2.55$\end{tabular} &
            \begin{tabular}[c]{@{}c@{}}$4.9$\\ $\pm0.49$\end{tabular} &
            \begin{tabular}[c]{@{}c@{}}$9.9$\\ $\pm0.2$\end{tabular} &
            \begin{tabular}[c]{@{}c@{}}$4.11\times10^7$\\ $\pm0.41\times10^7$\end{tabular} \\ \bottomrule
        \end{tabular}
    }
\end{table}

Because line lengths in multiline TRL calibration affect only the eigenvalue formulation, we consider only the normalized error terms obtained via eigendecomposition and analyze the error from the MC analysis. These normalized error terms are derived from the error box model in \eqref{eq:2.1} as follows:
\begin{equation}
    \widetilde{\bs{A}} = \begin{bmatrix} 1 & a_{12} \\ a_{21}/a_{11} & 1\end{bmatrix}; \quad \widetilde{\bs{B}} = \begin{bmatrix} 1 & b_{12}/b_{11} \\ b_{21} & 1\end{bmatrix}
    \label{eq:4.1}
\end{equation}

\begin{figure}[th!]
    \centering
    \includegraphics[width=1\linewidth]{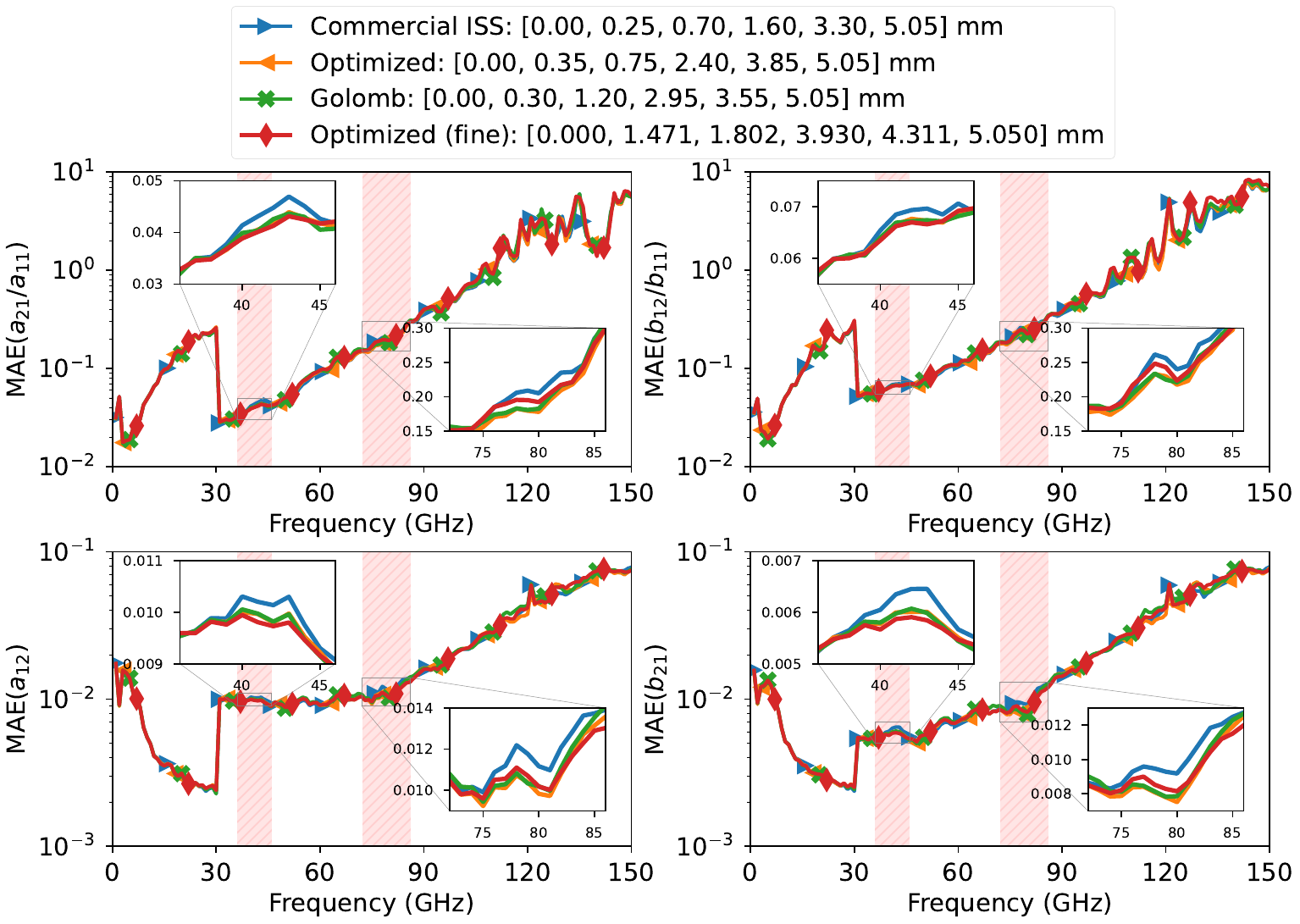}
    \caption{MC analysis results showing the MAE of the normalized error terms across frequency for the commercial ISS, optimized ($50\,\mu\mathrm{m}$ steps), fine-resolution optimized ($1\,\mu\mathrm{m}$ steps), and Golomb sparse ruler line length sets. The highlighted regions ($36$--$46\,\mathrm{GHz}$ and $72$--$86\,\mathrm{GHz}$) indicate where the commercial ISS exhibits elevated MAE.}
    \label{fig:4.14}
\end{figure}

The error boxes used in the simulation were obtained from a reference multiline TRL calibration of the ISS measurements. These serve as the baseline against which the error is quantified. In each MC trial, the synthetic CPW line data is embedded into these measured error boxes and perturbed by correlated noise drawn from the estimated sample covariance, producing realistic raw measurements that capture the frequency-dependent characteristics of the VNA and probe interface. The results, summarized by the mean absolute error (MAE) of the normalized error terms over the MC trials, are shown in Fig.~\ref{fig:4.14}. Two frequency regions of particular interest are highlighted: $36$--$46\,\mathrm{GHz}$ and $72$--$86\,\mathrm{GHz}$, where the commercial ISS exhibits elevated MAE. Although the commercial ISS line lengths yield higher MAE in these two regions, the error is still considered low overall, and all kits perform comparably outside these ranges.

\begin{figure}[th!]
    \centering
    \includegraphics[width=1\linewidth]{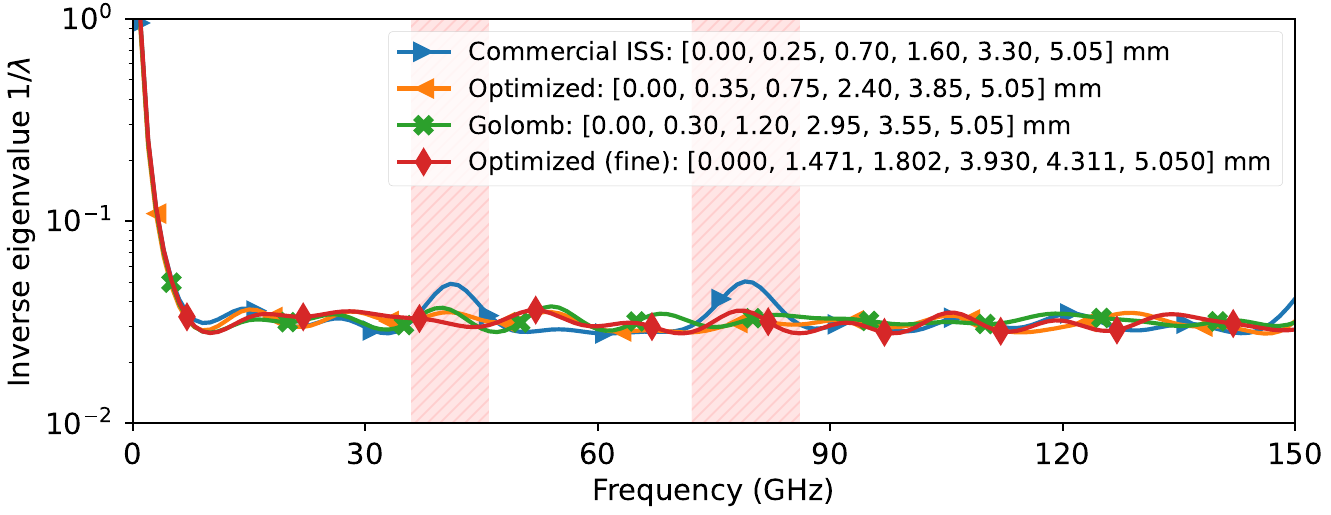}
    \caption{Comparison of the inverse eigenvalue $1/\lambda$ across frequency for the commercial ISS, optimized ($50\,\mu\mathrm{m}$ steps), Golomb sparse ruler, and fine-resolution optimized ($1\,\mu\mathrm{m}$ steps) line length sets.}
    \label{fig:4.15}
\end{figure}

The two highlighted frequency ranges correspond to sensitive regions of the commercial ISS calibration kit, as evident from the effective phase dips in Fig.~\ref{fig:4.12}. This behavior also correlates with the inverse eigenvalue $1/\lambda$ of the multiline TRL calibration, which is proportional to the sensitivity of the eigenvector solution, as discussed in Section~\ref{sec:2c}. Fig.~\ref{fig:4.15} depicts $1/\lambda$ for all considered kits, showing that the frequency ranges with higher MAE in the MC analysis correspond to peaks in $1/\lambda$ for the commercial ISS line lengths. For the optimized and Golomb sets, the effective phase is more uniform across frequency. Their inverse eigenvalue profiles exhibit localized peaks (e.g., around $50\,\mathrm{GHz}$) that are higher than those of the commercial ISS at the same frequencies, but these peaks follow a consistent, low-amplitude ripple pattern rather than the isolated spikes seen in the commercial ISS.

\begin{figure}[th!]
    \centering
    \includegraphics[width=1\linewidth]{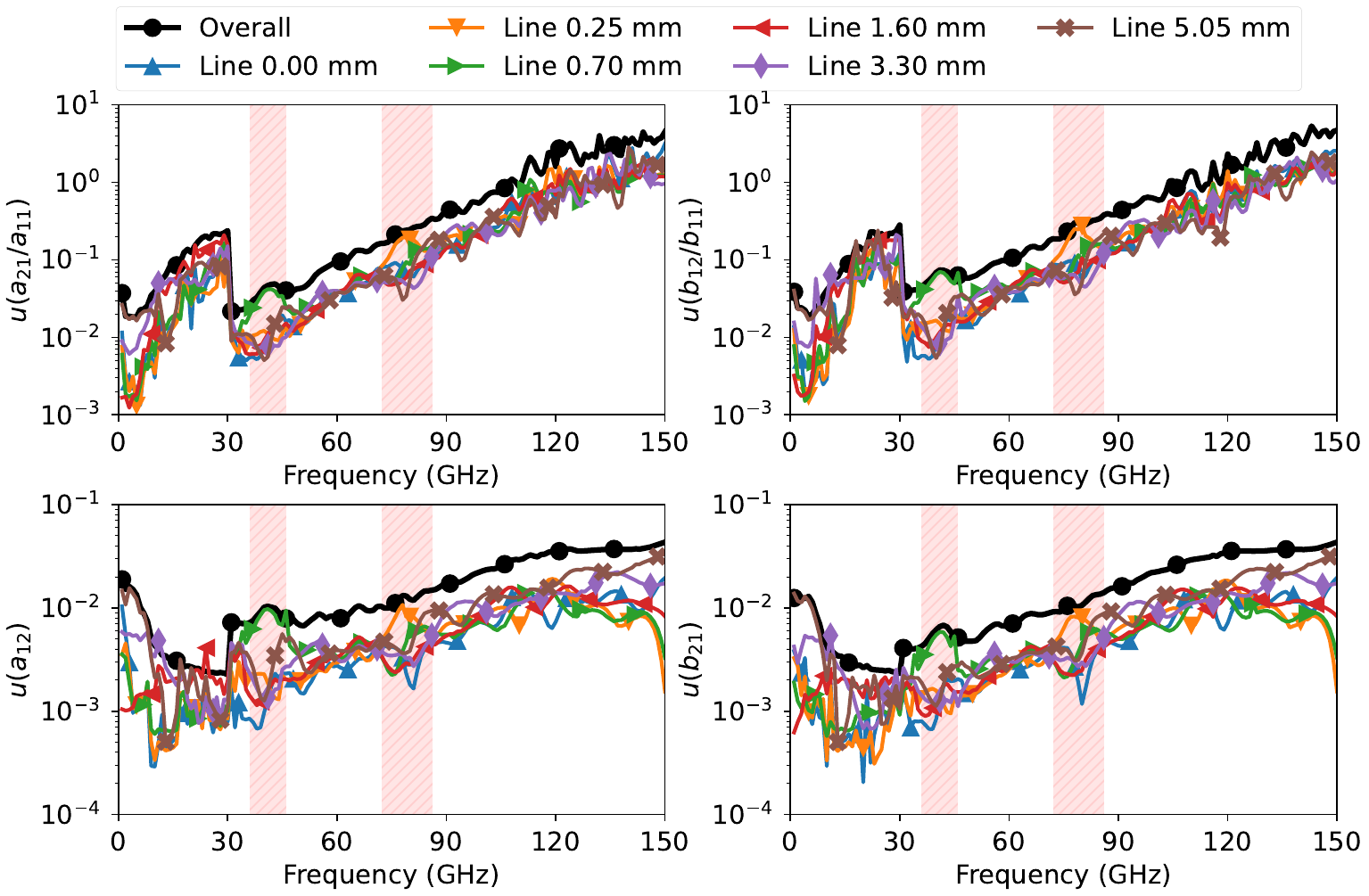}
    \caption{Per-line uncertainty budget of the normalized error terms for the commercial ISS, obtained via linear uncertainty propagation \cite{Hatab2023}. The highlighted regions indicate where the $0.25\,\mathrm{mm}$ and $0.70\,\mathrm{mm}$ lines have the highest individual uncertainty contributions. The corresponding budgets for the optimized, Golomb, and fine-resolution optimized sets are provided in Appendix~\ref{anx:C}.}
    \label{fig:4.16}
\end{figure}

To further analyze the contribution of each individual line to the overall calibration uncertainty, we employ the multiline TRL calibration with linear uncertainty propagation from \cite{Hatab2023}. This framework decomposes the total uncertainty into contributions from each calibration standard via a first-order approximation, allowing us to identify which lines dominate the uncertainty budget as a function of frequency. Fig.~\ref{fig:4.16} shows the per-line uncertainty budget for the commercial ISS. The $0.25\,\mathrm{mm}$ and $0.70\,\mathrm{mm}$ lines exhibit the highest individual uncertainty contributions in the highlighted frequency regions ($36$--$46\,\mathrm{GHz}$ and $72$--$86\,\mathrm{GHz}$), which would explain the elevated MAE observed in the MC analysis at those frequencies. In contrast, the per-line budgets for the optimized, Golomb, and fine-resolution optimized sets (presented in Appendix~\ref{anx:C} for brevity) show that the uncertainty contributions are more evenly distributed across all lines, with no single line dominating the budget at any particular frequency.

\begin{figure}[th!]
    \centering
    \includegraphics[width=1\linewidth]{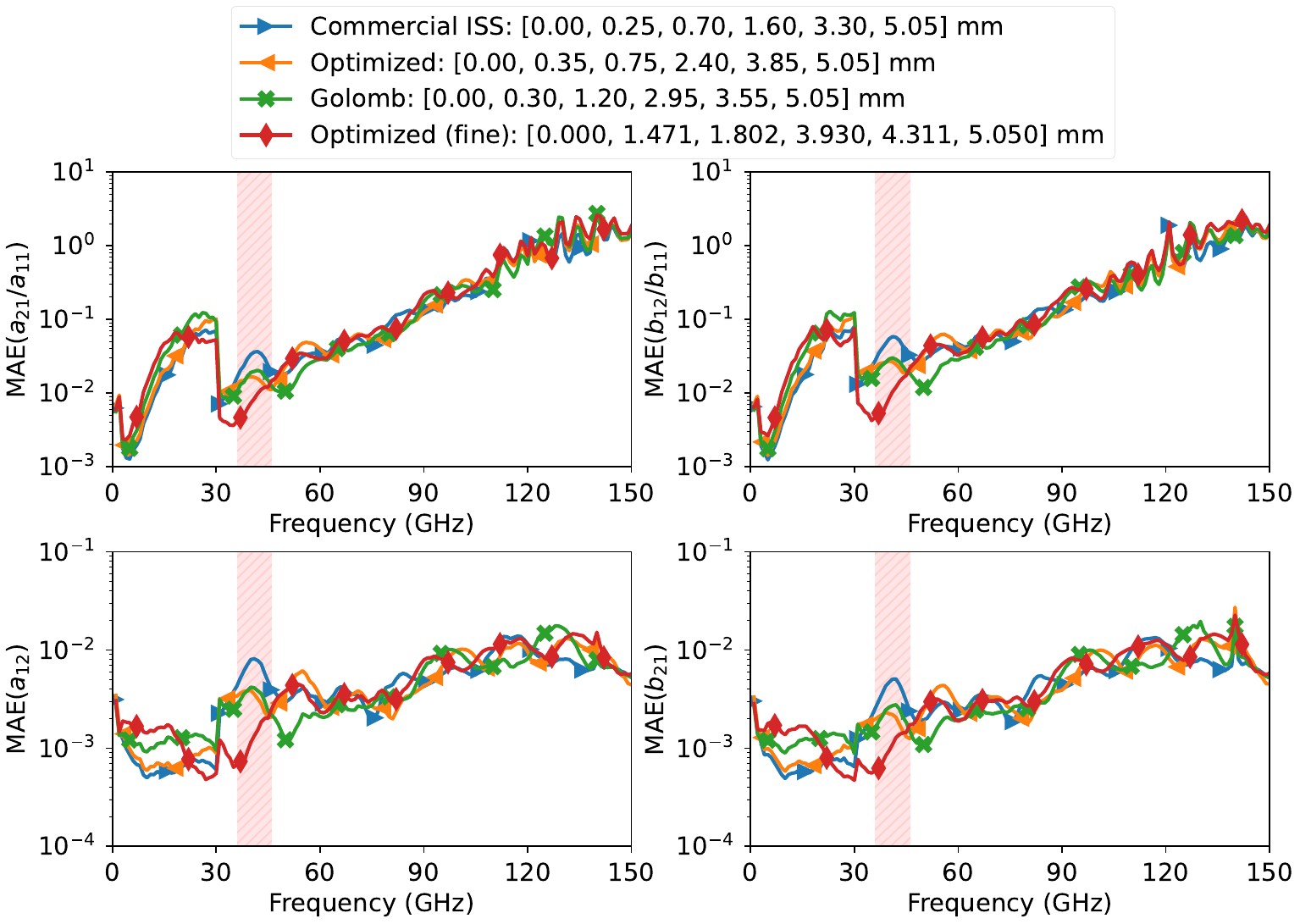}
    \caption{MC analysis with only the 3rd line perturbed while all other lines remain nominal. The 3rd line corresponds to $0.70\,\mathrm{mm}$ (commercial ISS), $0.75\,\mathrm{mm}$ (optimized), $1.20\,\mathrm{mm}$ (Golomb), and $1.802\,\mathrm{mm}$ (fine-resolution optimized). The commercial ISS shows a pronounced MAE peak in the $36$--$46\,\mathrm{GHz}$ range.}
    \label{fig:4.17}
\end{figure}

\begin{figure}[th!]
    \centering
    \includegraphics[width=1\linewidth]{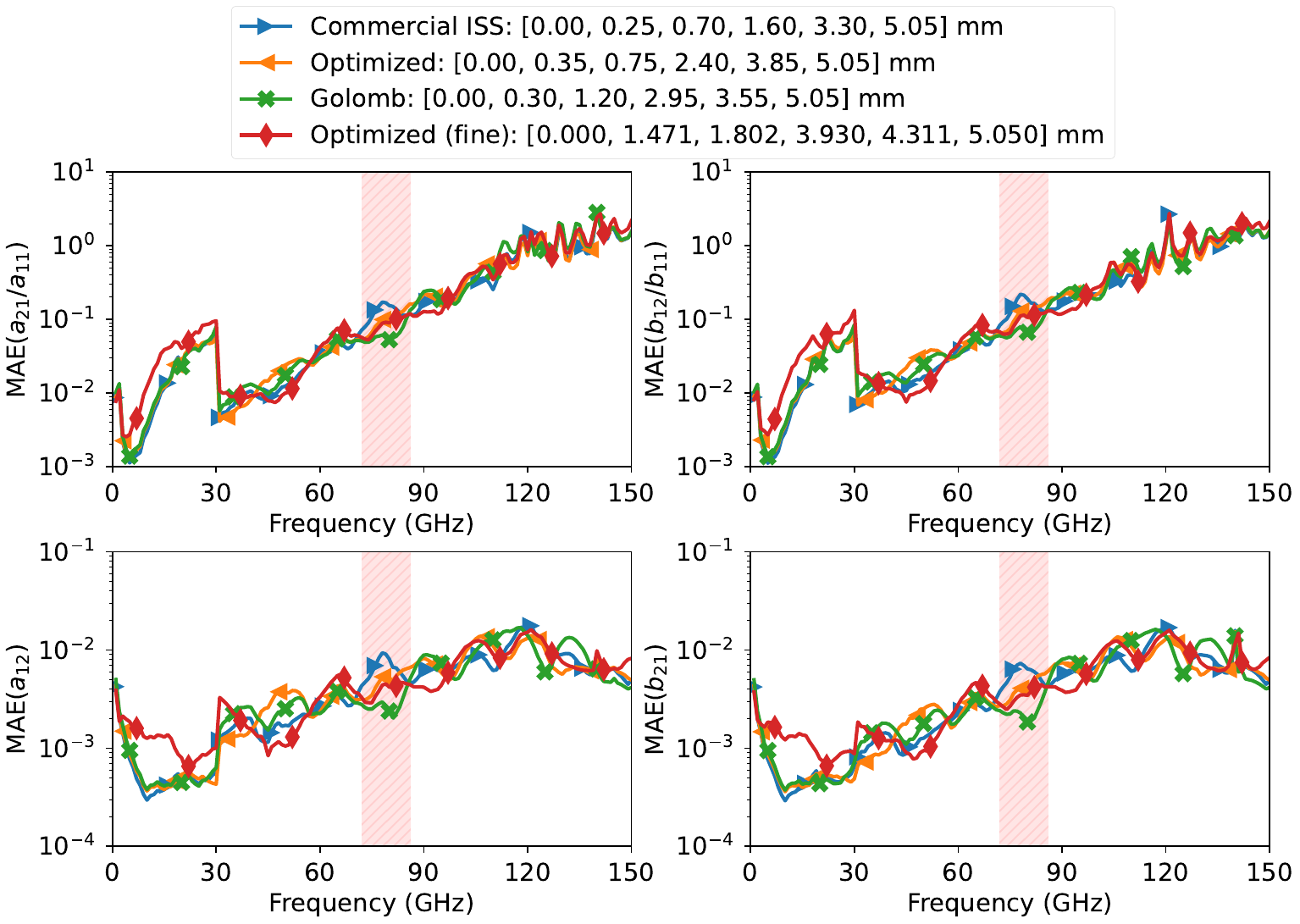}
    \caption{MC analysis with only the 2nd line perturbed while all other lines remain nominal. The 2nd line corresponds to $0.25\,\mathrm{mm}$ (commercial ISS), $0.35\,\mathrm{mm}$ (optimized), $0.30\,\mathrm{mm}$ (Golomb), and $1.471\,\mathrm{mm}$ (fine-resolution optimized). The commercial ISS shows a pronounced MAE rise in the $72$--$86\,\mathrm{GHz}$ range.}
    \label{fig:4.18}
\end{figure}

To confirm the findings from the linear uncertainty analysis, we perform an additional MC analysis in which only the 2nd or 3rd line (in order) of each set is perturbed (cross-section mismatch, length error, and measurement noise) while all other lines remain at their nominal values without perturbation. Fig.~\ref{fig:4.17} shows the MAE when only the 3rd line is perturbed, i.e., the $0.70\,\mathrm{mm}$ line for the commercial ISS, $0.75\,\mathrm{mm}$ for the optimized, $1.20\,\mathrm{mm}$ for the Golomb, and $1.802\,\mathrm{mm}$ for the fine-resolution optimized set. Fig.~\ref{fig:4.18} shows the MAE when only the 2nd line is perturbed, i.e., the $0.25\,\mathrm{mm}$ line for the commercial ISS, $0.35\,\mathrm{mm}$ for the optimized, $0.30\,\mathrm{mm}$ for the Golomb, and $1.471\,\mathrm{mm}$ for the fine-resolution optimized set. The results confirm the per-line uncertainty budget: perturbing only the 3rd line produces a peak in the MAE at the first highlighted region $36$--$46\,\mathrm{GHz}$ for the commercial ISS, while perturbing only the 2nd line leads to an elevated MAE in the second highlighted region $72$--$86\,\mathrm{GHz}$. It should be noted that for any individual line, we do not expect one kit to always deliver lower uncertainty than another; what matters is the overall distribution of uncertainty contributions across lines, which is more balanced for the optimized and Golomb sets than for the commercial ISS.

\begin{figure}[th!]
    \centering
    \includegraphics[width=1\linewidth]{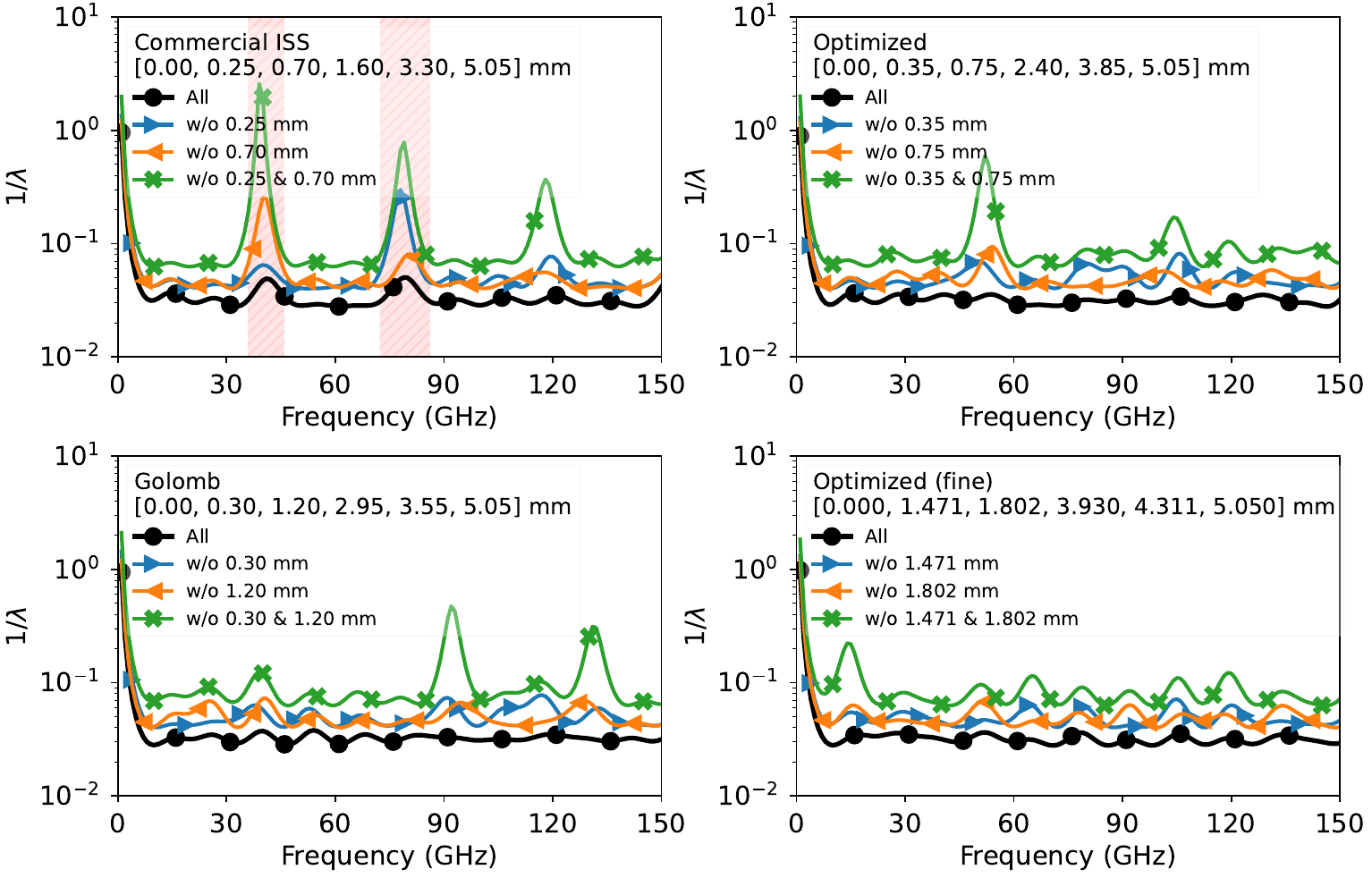}
    \caption{Inverse eigenvalue $1/\lambda$ for all four line length sets under line removal scenarios: nominal (all lines), removing the 2nd line, removing the 3rd line, and removing both the 2nd and 3rd lines.}
    \label{fig:4.19}
\end{figure}

We emphasize that these observations do not imply that the commercial ISS calibration kit is inferior. The key distinction is that the proposed methods, both the optimization and sparse ruler approaches, distribute the uncertainty contributions more evenly across lines. In contrast, the commercial ISS concentrates the uncertainty budget on individual lines: the $0.70\,\mathrm{mm}$ line alone dominates the first highlighted range, while the $0.25\,\mathrm{mm}$ line has a noticeable impact by itself in the second range. This concentration implies that removing either of these lines would lead to a significant increase in calibration sensitivity at those frequencies. To illustrate this, Fig.~\ref{fig:4.19} compares the inverse eigenvalue $1/\lambda$ across all four length sets under scenarios where the 2nd line, the 3rd line, or both are removed. The results show that the commercial ISS is more sensitive to the removal of these specific lines individually: removing either line produces a peak in $1/\lambda$ at the $36$--$46\,\mathrm{GHz}$ and $72$--$86\,\mathrm{GHz}$ ranges, and the situation worsens significantly when both lines are removed, where $1/\lambda$ increases sharply, indicating high sensitivity of the calibration in the absence of these two standards. In comparison, removing either the 2nd or 3rd line from the optimized and Golomb sets still yields a relatively flat $1/\lambda$ response, whereas removing both lines produces a noticeable increase in $1/\lambda$, albeit smaller than that of the commercial ISS. Furthermore, the fine-resolution optimized set exhibits better robustness to line removal than the standard optimized set, particularly when both the 2nd and 3rd lines are removed.

In summary, the commercial ISS is a well-performing calibration kit that produces reliable results across the full frequency band. However, we have demonstrated that the optimized and Golomb sparse ruler line length sets are arguably better choices for frequency coverage up to $150\,\mathrm{GHz}$, as they do not concentrate the uncertainty on individual lines. This more balanced distribution of uncertainty contributions provides greater robustness against line imperfections and makes the calibration less sensitive to the loss or degradation of any single standard.

\section{Design Examples}
\label{sec:5}

\subsection{Exploiting Dispersive Conductors in Transmission Lines}
\label{sec:5a}
Many planar transmission line implementations use conductors with small cross-sectional dimensions, such as thin-film microstrip lines (TFMSL) \cite{Schnieder2001}. Small conductors increase internal inductance, which raises the relative effective permittivity at low frequencies. This effect is further exacerbated by multilayer metallization, for example, electroless nickel immersion gold (ENIG) plating that incorporates ferromagnetic metals such as nickel \cite{Schafsteller2024}. Additionally, frequency dependency in the relative effective permittivity also arises from geometry; for instance, a microstrip is typically more dispersive than a CPW line.

\begin{figure}[th!]
    \centering
    \includegraphics[width=1\linewidth]{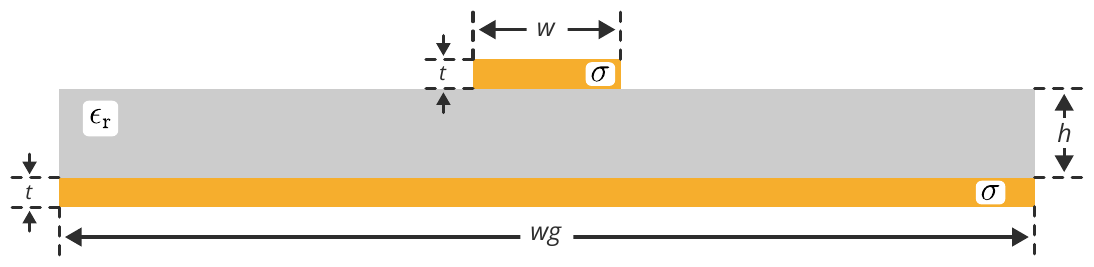}
    \caption{Thin-film microstrip line cross-section parameters from \cite{Schnieder2001}: substrate permittivity $\epsilon_\mathrm{r} = 2.7(1-j0.015)$, conductor conductivity $\sigma = 25\,\mathrm{MS/m}$, metal thickness $t=0.8\,\mu\mathrm{m}$, strip width $w=8\,\mu\mathrm{m}$, substrate height $h=1.7\,\mu\mathrm{m}$, and ground plane width $wg=88\,\mu\mathrm{m}$.}
    \label{fig:5.1}
\end{figure}

In this example, we use the TFMSL structure based on the model from \cite{Schnieder2001} with its cross-section shown in Fig.~\ref{fig:5.1} and parameter values provided in the caption.

\begin{figure}[th!]
    \centering
    \includegraphics[width=1\linewidth]{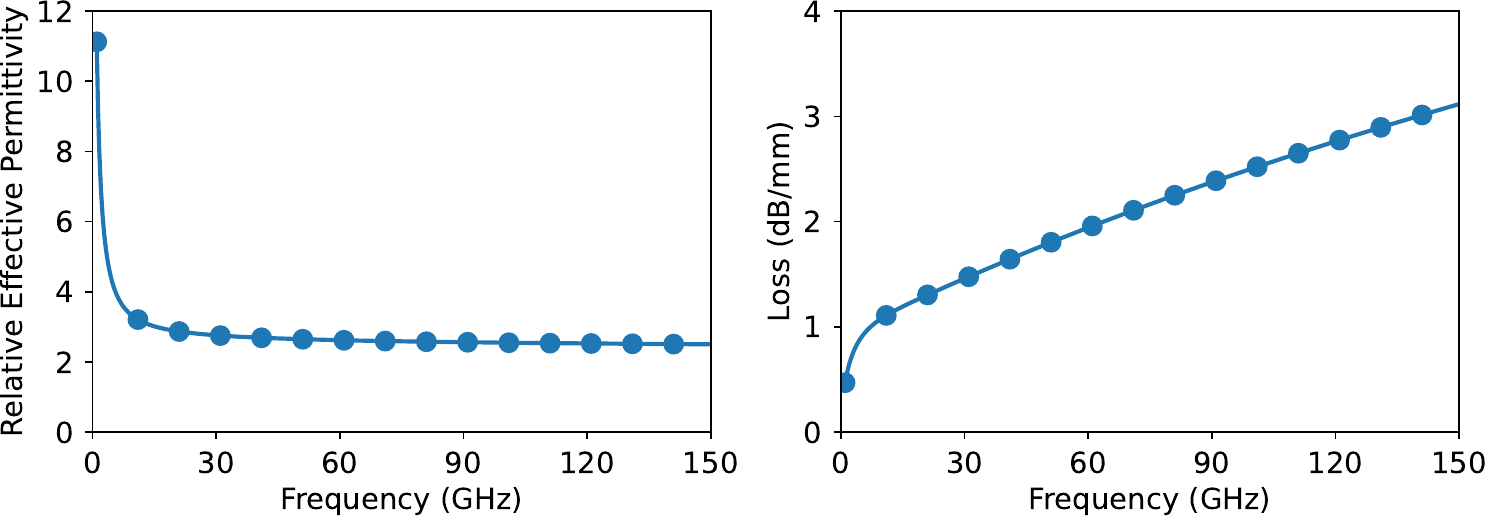}
    \caption{Frequency response of the thin-film microstrip line in Fig.~\ref{fig:5.1}, showing relative effective permittivity and attenuation per unit length.}
    \label{fig:5.2}
\end{figure}

Due to the small dimensions, the relative effective permittivity of this structure is strongly frequency-dependent: it is high at low frequencies and decreases toward the substrate permittivity at higher frequencies. This behavior is shown in Fig.~\ref{fig:5.2}. Because both the substrate and conductors are lossy, the attenuation per unit length also increases noticeably with frequency.

To exploit this dispersion, we compute optimized lengths for two cases. In the first, we assume lossless, frequency-independent relative effective permittivity. In the second, we supply the frequency-dependent relative effective permittivity directly to the optimizer.

As the internal inductance raises the effective permittivity at low frequencies, shorter lines are sufficient under this condition compared to those predicted using only a constant relative effective permittivity. For our constant permittivity case, we used the median value of the frequency-dependent response to avoid the low-frequency outliers caused by internal inductance. With loss included, the optimal length distribution changes accordingly, typically reducing the required number of lines.

\begin{figure}[th!]
    \centering
    \includegraphics[width=1\linewidth]{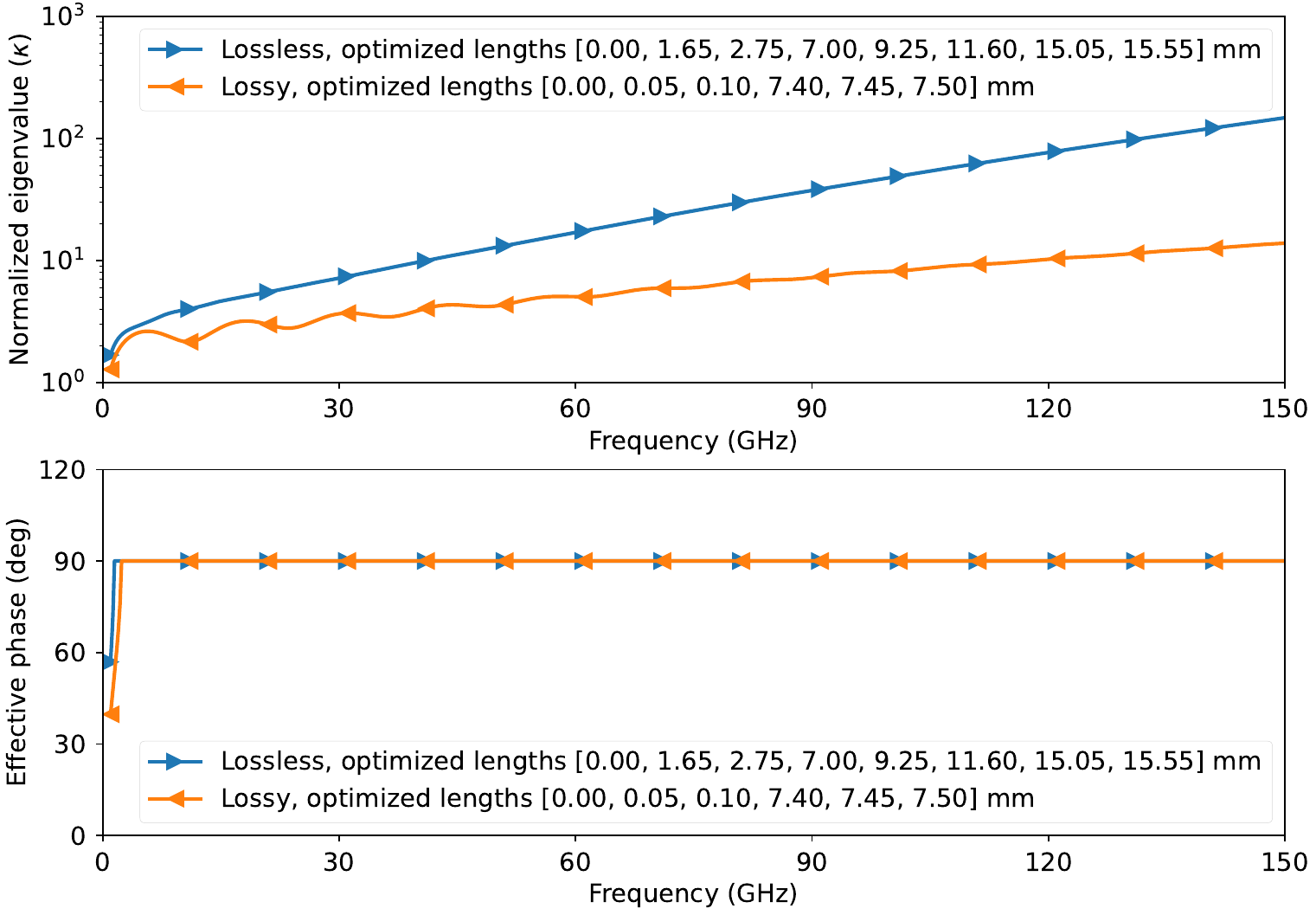}
    \caption{Comparison of effective phase for the thin-film microstrip multiline TRL lines optimized under two scenarios: constant lossless relative effective permittivity versus frequency-dependent dispersive, lossy relative effective permittivity.}
    \label{fig:5.3}
\end{figure}

Fig.~\ref{fig:5.3} presents the effective phase of the computed lines. As expected, using the frequency-dependent model yields substantially shorter lengths and fewer lines while covering the same frequency range. Both runs used the same settings with the loss function in \eqref{eq:3.6} and a $10\,\mu\mathrm{m}$ length standard deviation and quantized lengths in steps of $50\,\mu\mathrm{m}$.

\subsection{On-Wafer Multiline TRL Lengths for THz Frequencies}
\label{sec:5b}
Capabilities for measurements at sub-THz and THz frequencies are getting increasingly important in industry and research \cite{Chen2012, Bauwens2014, Fregonese2020}. Careful design of a multiline TRL calibration kit is therefore essential to span such wide bandwidths. In this example, we compute lengths for a multiline TRL calibration kit implemented as CPW on an Alumina substrate. We focus only on length design; for cross-sectional geometry choices, see, for example, \cite{Spirito2019}. 

In the absence of specific geometry, we assume a typical average relative effective permittivity for CPW on Alumina, $\epsilon_\mathrm{r,eff}\approx 5.2$, and neglect loss. In practice, materials are dispersive, as illustrated in Subsection~\ref{sec:5a}. 

\begin{figure}[th!]
    \centering
    \includegraphics[width=1\linewidth]{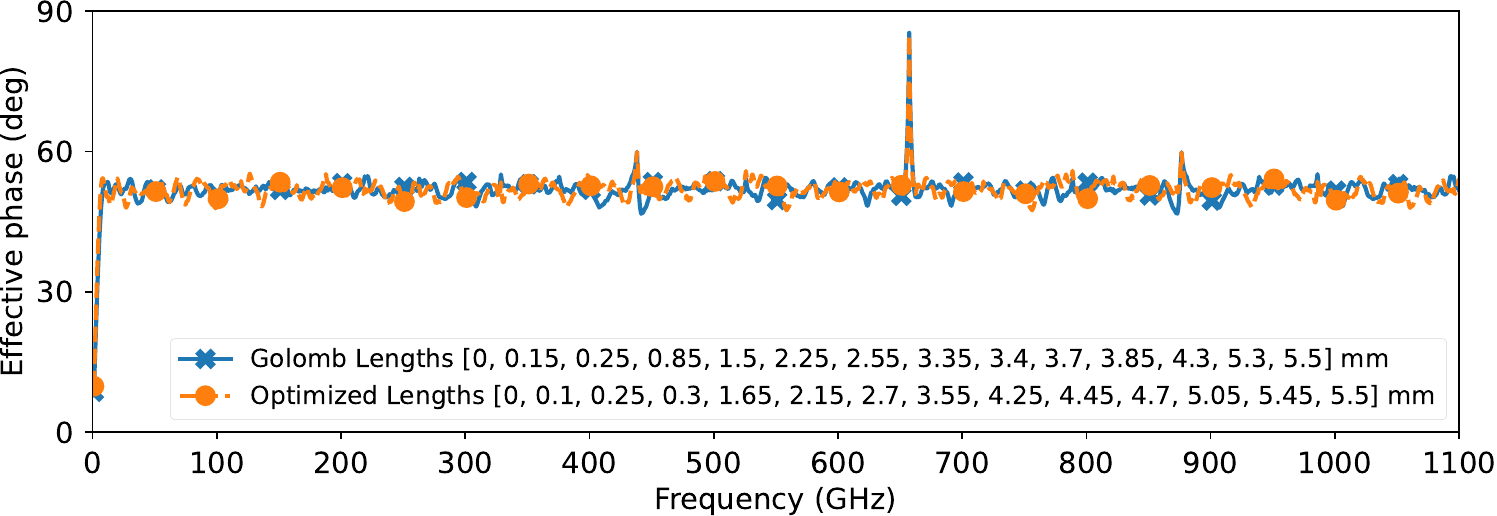}
    \caption{Effective phase up to 1.1\,THz for optimized generic CPW multiline TRL calibration lengths on lossless Alumina substrate.}
    \label{fig:5.4}
\end{figure}

The frequency range is 2\,GHz to 1.1\,THz with a 30$^\circ$ low-frequency phase margin. The computed lines and their effective phase are shown in Fig.~\ref{fig:5.4}. Two methods are used: (i) the optimization method from Subsection~\ref{sec:3a} with the loss function \eqref{eq:3.6} and a $10\,\mu\mathrm{m}$ length standard deviation. The maximum line length is set by the 2\,GHz requirement and the 30$^\circ$ phase margin; the number of lines then follows from that maximum length and the 1.1\,THz upper limit. (ii) a Golomb sparse ruler approach subject to the same constraints.

Fig.~\ref{fig:5.5} shows that 14 lines suffice to cover 2\,GHz to 1.1\,THz with a nearly flat effective phase. Because this analysis assumes a generic lossless CPW on an Alumina substrate, a specific geometry that includes conductor loss and dispersion will require fewer lines, since higher effective permittivity at low frequencies and higher loss at high frequencies both reduce the required maximum length and the number of lines. While it is difficult to predict the exact reduction quantitatively without specifying a particular geometry, the TFMSL example in Subsection~\ref{sec:5a} demonstrates how accounting for loss and dispersion can reduce both the number of lines and the maximum length.

\subsection{Waveguide Multiline Kit Using Longer Lines}
\label{sec:5c}

Defining a TRL calibration kit for a waveguide is straightforward when using the traditional quarter-wavelength line, which spans the waveguide bandwidth, and the quarter-wavelength phase of the line is set in the middle of the band. However, as the frequency increases, the quarter-wavelength line becomes very short and increasingly difficult to manufacture because feature sizes approach typical fabrication tolerances, especially with additive processes such as 3D-printed waveguides \cite{Zhu2023}.

A multiline TRL approach enables the use of longer lines. For example, ``3/4-wave'' or ``5/4-wave'' TRL kits exploit the cyclic nature of TRL: two longer lines are combined to realize a multiline TRL in waveguide \cite{Ridler2009,Ridler2019,Ridler2021}.

Here, we generalize the design of a waveguide multiline TRL calibration kit by imposing a maximum manufacturable line length and then computing the number of lines and their lengths automatically, using the proposed optimization procedure under the imposed constraints.

\begin{figure}[th!]
    \centering
    \includegraphics[width=1\linewidth]{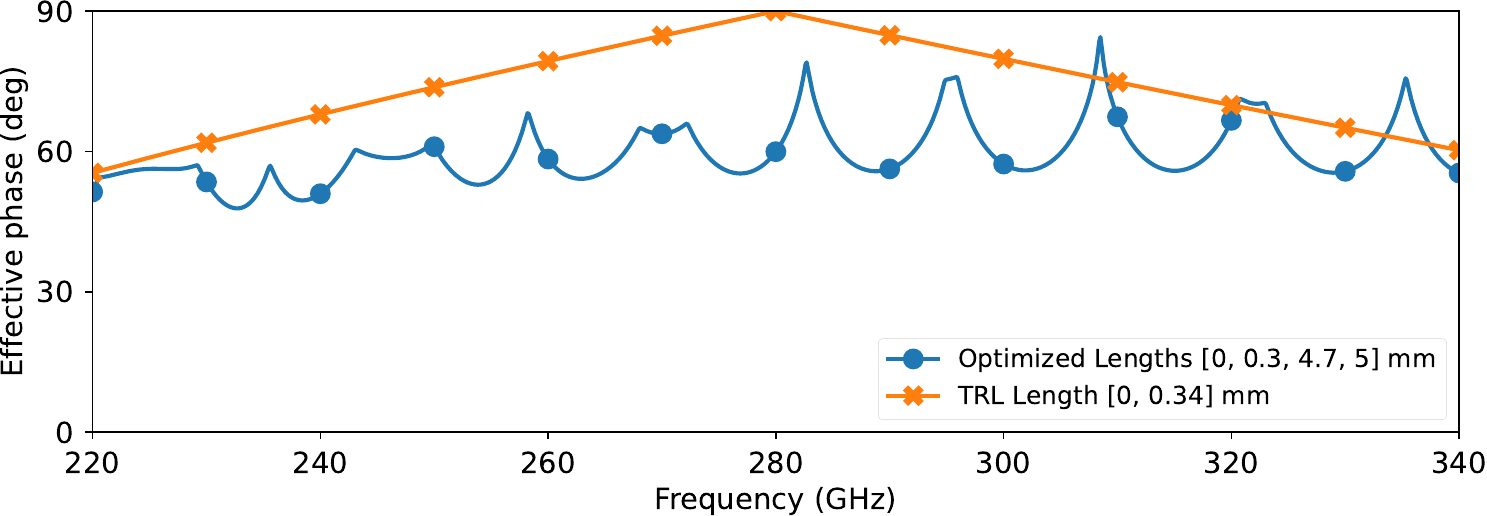}
    \caption{Comparison of effective phase versus frequency for the optimized multiline TRL calibration kit and a conventional quarter-wavelength TRL kit for WM-864 waveguide.}
    \label{fig:5.5}
\end{figure}

We consider the WM-864 waveguide, operating from 220\,GHz to 300\,GHz, with width $864\,\mu\mathrm{m}$ and height $432\,\mu\mathrm{m}$ \cite{IEEE17852013}. The waveguide model used to compute the relative effective permittivity follows \cite{Lomakin2017, Lomakin2018}. To obtain lengths robust to fabrication tolerances, we use the loss function \eqref{eq:3.6} with a $10\,\mu\mathrm{m}$ length standard deviation to model sensitivity to length variation. The maximum line length is limited to $5\,\mathrm{mm}$, which sets the number of lines; changing this limit adjusts the count. The optimized results in Fig.~\ref{fig:5.5} exhibit a flatter effective phase and are compared with those of a conventional quarter-wavelength TRL kit. The optimized multiline TRL kit requires four lines, whereas the quarter-wavelength kit requires two. Although the quarter-wavelength kit already covers the full waveguide bandwidth, the multiline approach offers three practical advantages: first, its longer shims are easier to fabricate, since the quarter-wavelength shim thickness is only $370\,\mu\mathrm{m}$ at the WM-864 band center, which approaches typical machining tolerances. Second, the averaging effect of multiple line pairs reduces sensitivity to individual length errors. Third, the flatter effective phase response indicates that the calibration quality remains more uniform across frequency rather than being concentrated near the band center.

\section{Conclusion}
\label{sec:6}

We present a rigorous analysis and design procedure for selecting optimal lengths for line standards in multiline TRL calibration. First, we develop an effective phase metric that enables fair comparison across length sets, even when they differ in the number of lines. We then introduce two methods for choosing the lengths: a brute-force optimization and a sparse-ruler approach. The optimization method is general and applies to any transmission line type, including dispersive and lossy media, while the sparse-ruler approach is fast and leverages precomputed ruler marks, as Golomb rulers.

We also discussed the recommended number of lines by focusing on bandwidth coverage rather than maximum frequency, which can reduce the required line count when only limited bandwidth is needed. To validate the proposed methods, we presented measurements from four PCB multiline TRL calibration kits on different substrate materials and transmission line structures, all designed using the sparse ruler method and operating up to $150\,\mathrm{GHz}$. These measurements demonstrated that the sparse ruler approach yields stable, flat frequency coverage and that the calibrated device response is robust to the removal of individual lines from the calibration set. We further performed a measurement-based Monte Carlo uncertainty analysis, comparing a commercial ISS calibration kit against optimized and sparse ruler line length sets. The results show that the proposed methods distribute the uncertainty contributions more evenly across individual lines, whereas the commercial ISS concentrates the uncertainty budget on specific lines, making it more sensitive to their removal or degradation. Finally, we provide design examples for computing line lengths, including cases with dispersive materials, THz frequencies, and waveguide-based multiline TRL.

\appendices
\section{Scaling the Weighting Matrix in Multiline TRL Calibration}
\label{anx:A}

The weighting matrix for multiline TRL calibration, as described in \cite{Hatab2022,Hatab2023} and defined in \eqref{eq:2.14}, formulates the eigenvalue problem based on all line measurements. This matrix effectively performs self-weighting of the line pairs. The weighting mechanism can be modified by scaling the matrix entries to achieve specific calibration objectives. Common reasons for such scaling include distributing the weight of repeated lines according to their occurrence, enhancing selectivity by increasing the weighting order beyond the default L1-norm, and implementing customized weighting schemes.

To incorporate scaling, we introduce a nonzero symmetric matrix $\bs{S}$ and apply the Hadamard product \cite{Hogben2014}:
\begin{equation}
    \bs{W}_{\bs{S}} = \bs{S}\odot\bs{W}
    \label{eq:A.1}
\end{equation}
where $\odot$ denotes the Hadamard (element-wise) product. In the special case where $\bs{S}=\bs{1}\bs{1}^T$, with $\bs{1}=\begin{bmatrix}1 & 1 & \cdots\end{bmatrix}^T$, the scaling has no effect and $\bs{W}_{\bs{S}}=\bs{W}$.

The eigenvalue problem of multiline TRL calibration in \eqref{eq:2.14} is updated by simply substituting $\bs{W}_{\bs{S}}$ in place of $\bs{W}$. The eigenvalue $\lambda$ and normalized eigenvalue $\kappa$ are also updated to take the following form:
\begin{equation}
    \lambda_{\bs{S}} = \frac{1}{2}\vc{\bs{W}_{\bs{S}}}^H\vc{\bs{W}}, \quad \kappa_{\bs{S}} =  \frac{2\lambda_{\bs{S}}}{\left\|\vc{\bs{W}_{\bs{S}}}\right\|_1}
    \label{eq:A.2}
\end{equation}

One application for scaling the weighting matrix is to weight repeated lines by their occurrence. For example, consider four lines with the third line repeated, i.e., $[l_1,l_2,l_3,l_3]$. A suitable scaling matrix is defined as follows:
\begin{equation}
    \bs{S} = \bs{q}\bs{q}^T = \begin{bmatrix}
        \xcancel{1} & 1 & 1/2 & 1/2\\
        1 & \xcancel{1} & 1/2 & 1/2\\
        1/2 & 1/2 & \xcancel{1/4} & \xcancel{1/4}\\
        1/2 & 1/2 & \xcancel{1/4} & \xcancel{1/4}
    \end{bmatrix}
    \label{eq:A.3}
\end{equation}
where $\bs{q}=\begin{bmatrix}1 & 1 & 1/2 & 1/2\end{bmatrix}^T$. The crossed entries are ``don't-care'' values and may be set to zero, since the corresponding entries of $\bs{W}$ are by definition zero for line pairs of equal length. The vector $\bs{q}$ contains the weights of the lines based on their occurrence. If all lines are unique, then $\bs{q} = \bs{1}$ and the scaling has no effect.

\begin{figure}[th!]
    \centering
    \includegraphics[width=1\linewidth]{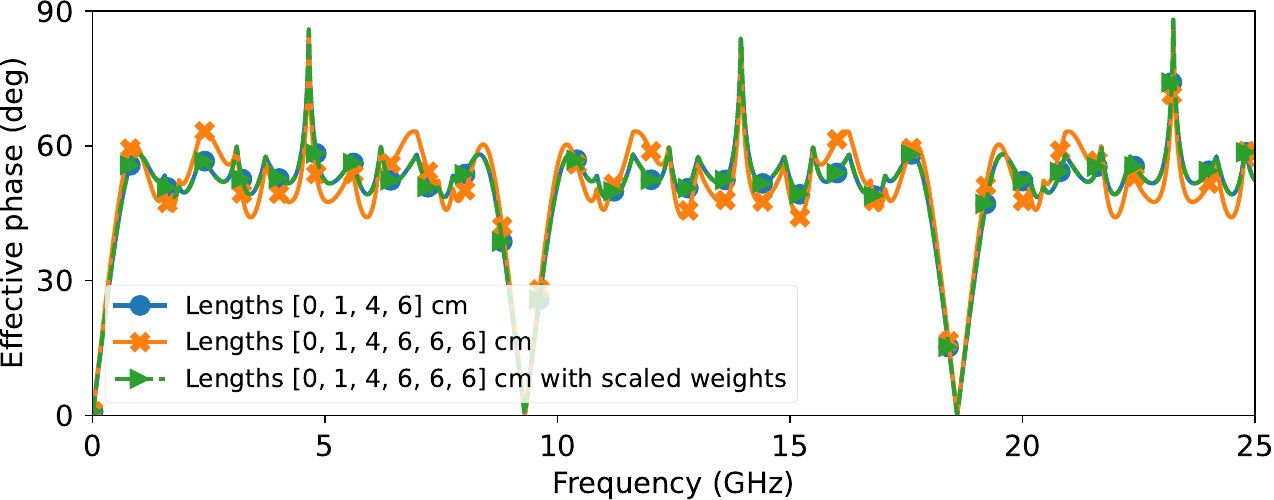}
    \caption{Effective phase comparison in multiline TRL calibration using two different line sets: a four-line set \{0, 1, 4, 6\}\,cm and a six-line set \{0, 1, 4, 6, 6, 6\}\,cm with the 6\,cm line repeated three times. The plot compares standard and scaled weighting approaches for the six-line set. All curves assume a relative effective permittivity of $\epsilon_\mathrm{r,eff}=2.6$.}
    \label{fig:A.1}
\end{figure}

As an example, consider the length sets $[0,1,4,6]\,\mathrm{cm}$ and $[0,1,4,6,6,6]\,\mathrm{cm}$. The second set includes the 6\,cm line three times, hence $\bs{q}=\begin{bmatrix}1 & 1 & 1& 1/3 & 1/3 & 1/3\end{bmatrix}^T$. Fig.~\ref{fig:A.1} shows the effective phase for both sets; the second set is plotted twice, with and without scaling. The scaled version matches the effective phase of the four-line set without redundancy.

\begin{figure}[th!]
    \centering
    \includegraphics[width=1\linewidth]{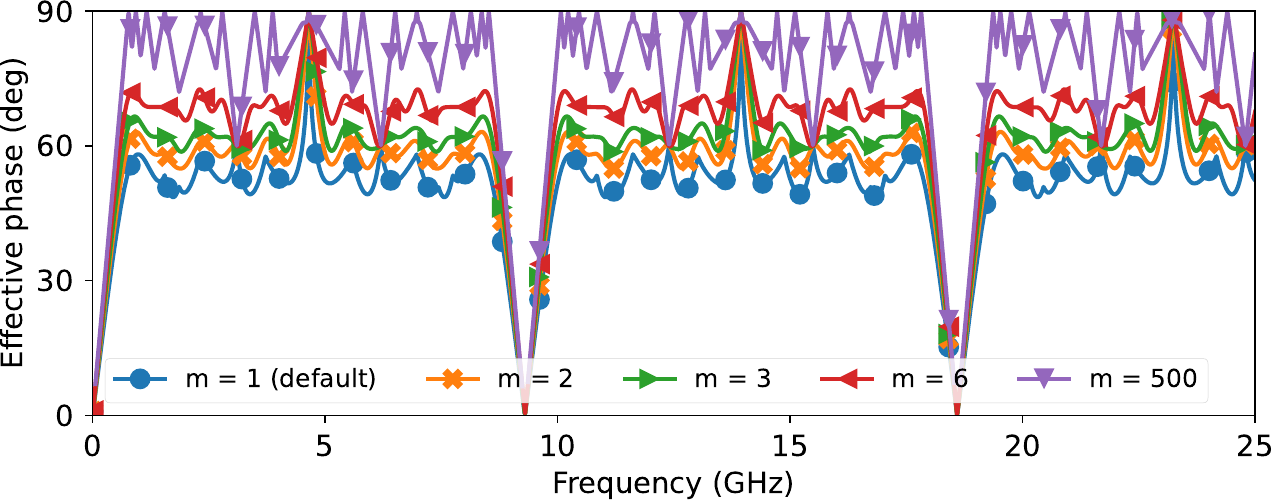}
    \caption{Effective phase comparison for the line set \{0, 1, 4, 6\}\,cm calculated with relative effective permittivity $\epsilon_\mathrm{r,eff}=2.6$, showing the impact of different L$m$-norm weighting orders.}
    \label{fig:A.2}
\end{figure}

Another application of scaling the weighting matrix is to make the weighting more selective. The standard weighting is based on the L$1$-norm, where each line pair is weighted by its eigengap. This can be modified by:
\begin{equation}
    \bs{W}_m = |\bs{W}|^{\odot (m-1)}\odot\bs{W}
    \label{eq:A.4}
\end{equation}
where $|\cdot|$ and $(\cdot)^{\odot (m-1)}$ denote element-wise absolute value and element-wise exponentiation to the power $m-1$, respectively. The default is $m=1$. Higher values of $m$ establish the L$m$-norm for line pair weighting and progressively bias the effective phase toward line pairs with larger phase differences. In the extreme case as $m\rightarrow\infty$, we have L$\infty$-norm weighting, where at each frequency only the line pair with maximum phase difference is selected. Increasing the order too much reduces the averaging effect and biases the result toward line pairs with the largest phase difference. Fig.~\ref{fig:A.2} shows the effective phase for several weighting orders. It should be noted that computing high exponents in software can lead to numerical instability due to finite precision. Therefore, the cases $m=1$ and $m=2$ are considered reasonable and stable numerically.

\section{Computing the Jacobian of the Multiline TRL Eigenvalue with Respect to Line Lengths}
\label{anx:B}

The Jacobian of $\lambda$ with respect to the line lengths is defined as
\begin{equation}
    \bs{J}_\lambda(\bs{l},f) = \begin{bmatrix}
        \frac{\partial\lambda(\bs{l},f)}{\partial l_1} & 
        \frac{\partial\lambda(\bs{l},f)}{\partial l_2} &
        \cdots &
        \frac{\partial\lambda(\bs{l},f)}{\partial l_N}
    \end{bmatrix}.
    \label{eq:B.1}
\end{equation}

To compute this Jacobian, we need to determine each partial derivative with respect to the $i$th length. These derivatives can be obtained directly from \eqref{eq:2.16} in summation form:
\begin{equation}
    \frac{\partial\lambda(\bs{l},f)}{\partial l_i} = \sum\limits_{j\neq i}^{N}\frac{\partial}{\partial l_i}\left|e^{\gamma l_{ij}}-e^{-\gamma l_{ij}}\right|^2,
    \label{eq:B.2}
\end{equation}
where $l_{ij} = l_i - l_j$.

For notational convenience, we define $w_{ij} = e^{\gamma l_{ij}}-e^{-\gamma l_{ij}}$. Using principles from complex-valued calculus \cite{Needham2023}, the partial derivative in \eqref{eq:B.2} can be expressed as
\begin{equation}
    \frac{\partial\lambda(\bs{l},f)}{\partial l_i} = 2\sum\limits_{j\neq i}^{N}\RE{w_{ij}^*\frac{\partial w_{ij}}{\partial l_i}},
    \label{eq:B.3}
\end{equation}
where $\RE{\cdot}$ denotes the real part and $w_{ij}^*$ is the complex conjugate of $w_{ij}$. Computing the derivative of $w_{ij}$ yields
\begin{equation}
    \frac{\partial\lambda(\bs{l},f)}{\partial l_i} = 2\sum\limits_{j\neq i}^{N}\RE{\gamma w_{ij}^*(e^{\gamma l_{ij}}+e^{-\gamma l_{ij}})}.
    \label{eq:B.4}
\end{equation}

\section{Per-Line Uncertainty Budgets for the Optimized, Golomb, and Fine-Resolution Optimized Line Length Sets}
\label{anx:C}

\begin{figure}[th!]
    \centering
    \includegraphics[width=1\linewidth]{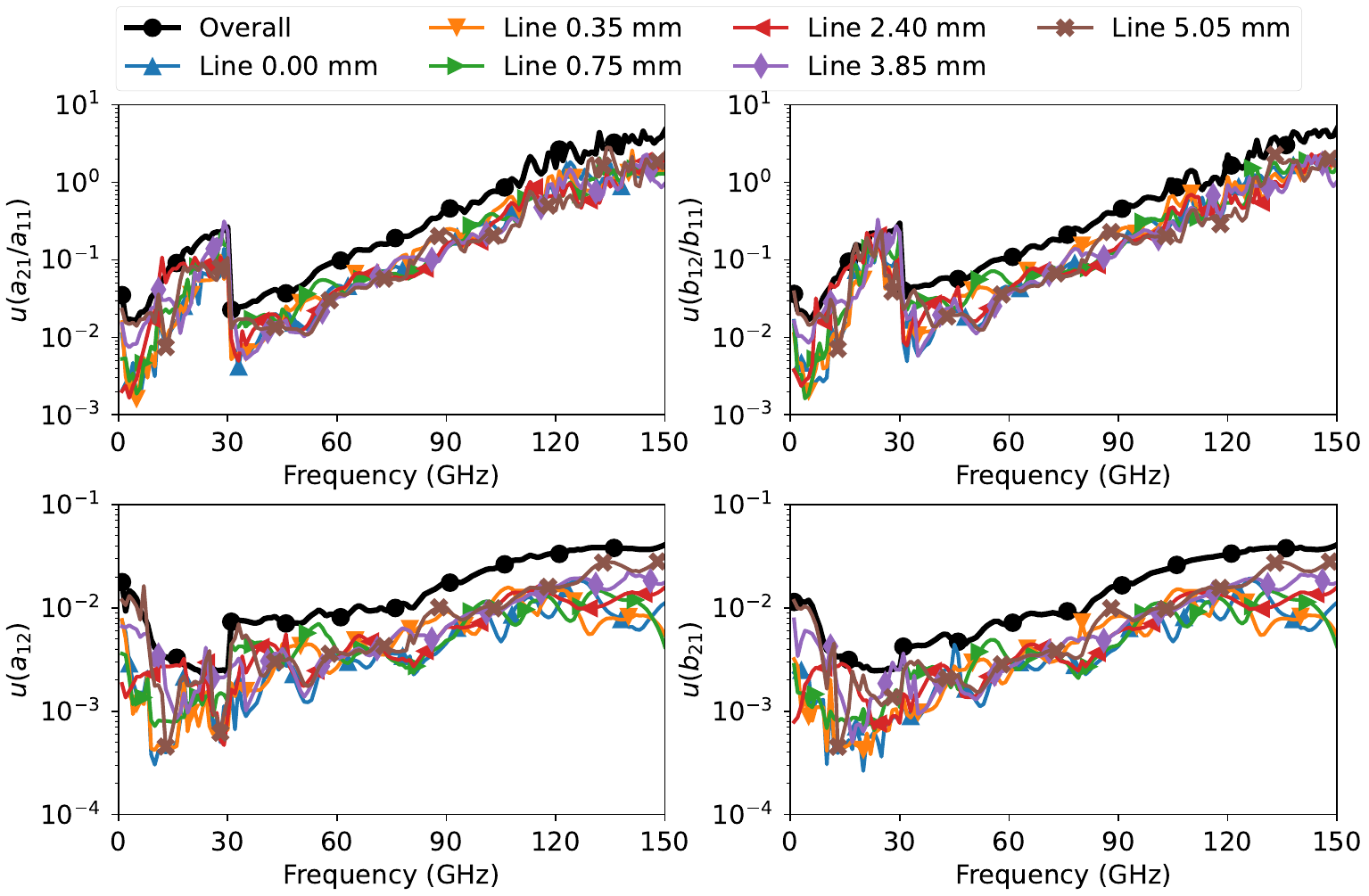}
    \caption{Per-line uncertainty budget of the normalized error terms for the optimized line length set ($50\,\mu\mathrm{m}$ quantization steps), obtained via linear uncertainty propagation \cite{Hatab2023}. Compare with the commercial ISS budget in Fig.~\ref{fig:4.16}.}
    \label{fig:C.1}
\end{figure}

\begin{figure}[th!]
    \centering
    \includegraphics[width=1\linewidth]{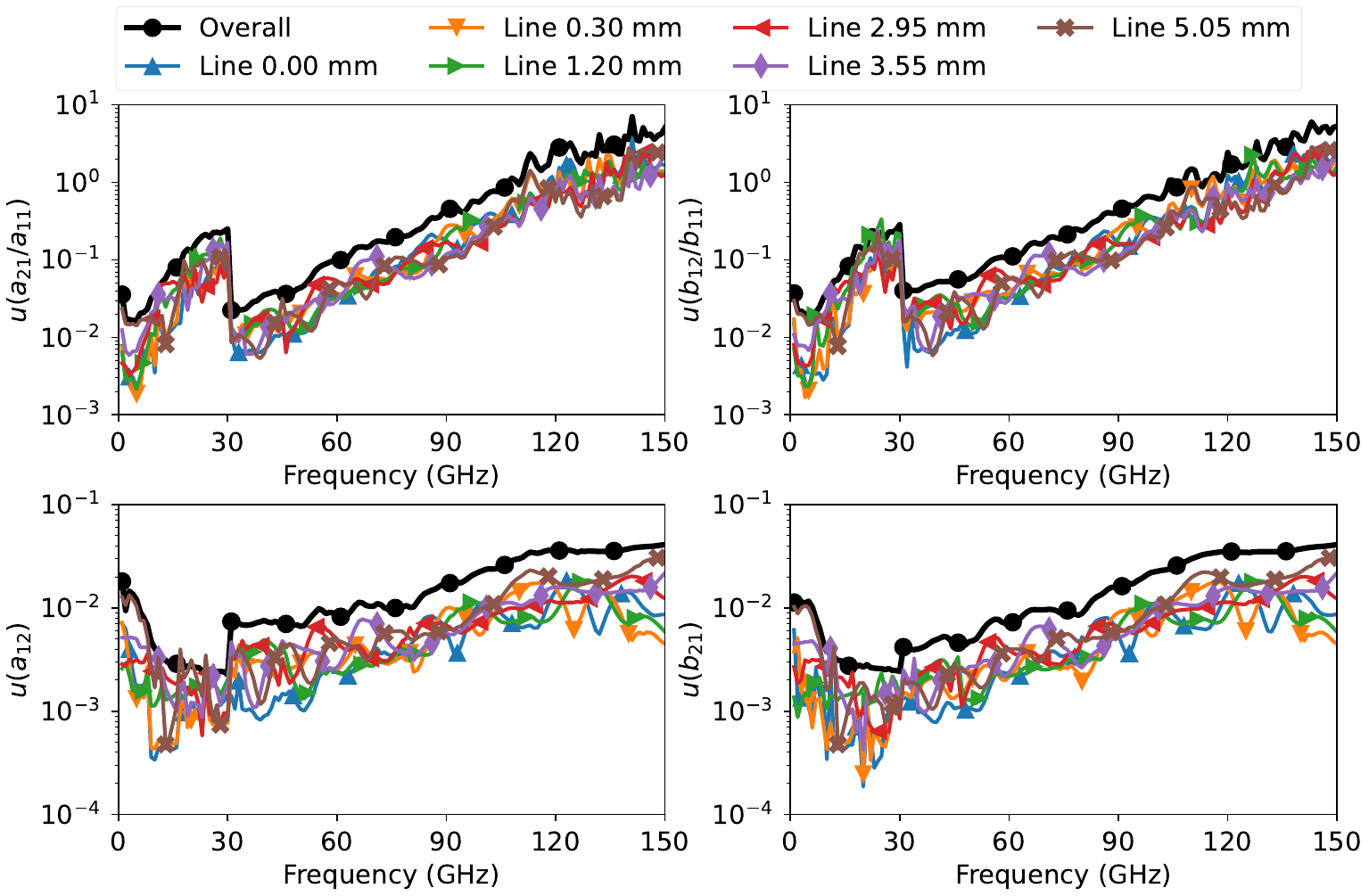}
    \caption{Per-line uncertainty budget of the normalized error terms for the Golomb sparse ruler line length set, obtained via linear uncertainty propagation \cite{Hatab2023}. Compare with the commercial ISS budget in Fig.~\ref{fig:4.16}.}
    \label{fig:C.2}
\end{figure}

\begin{figure}[th!]
    \centering
    \includegraphics[width=1\linewidth]{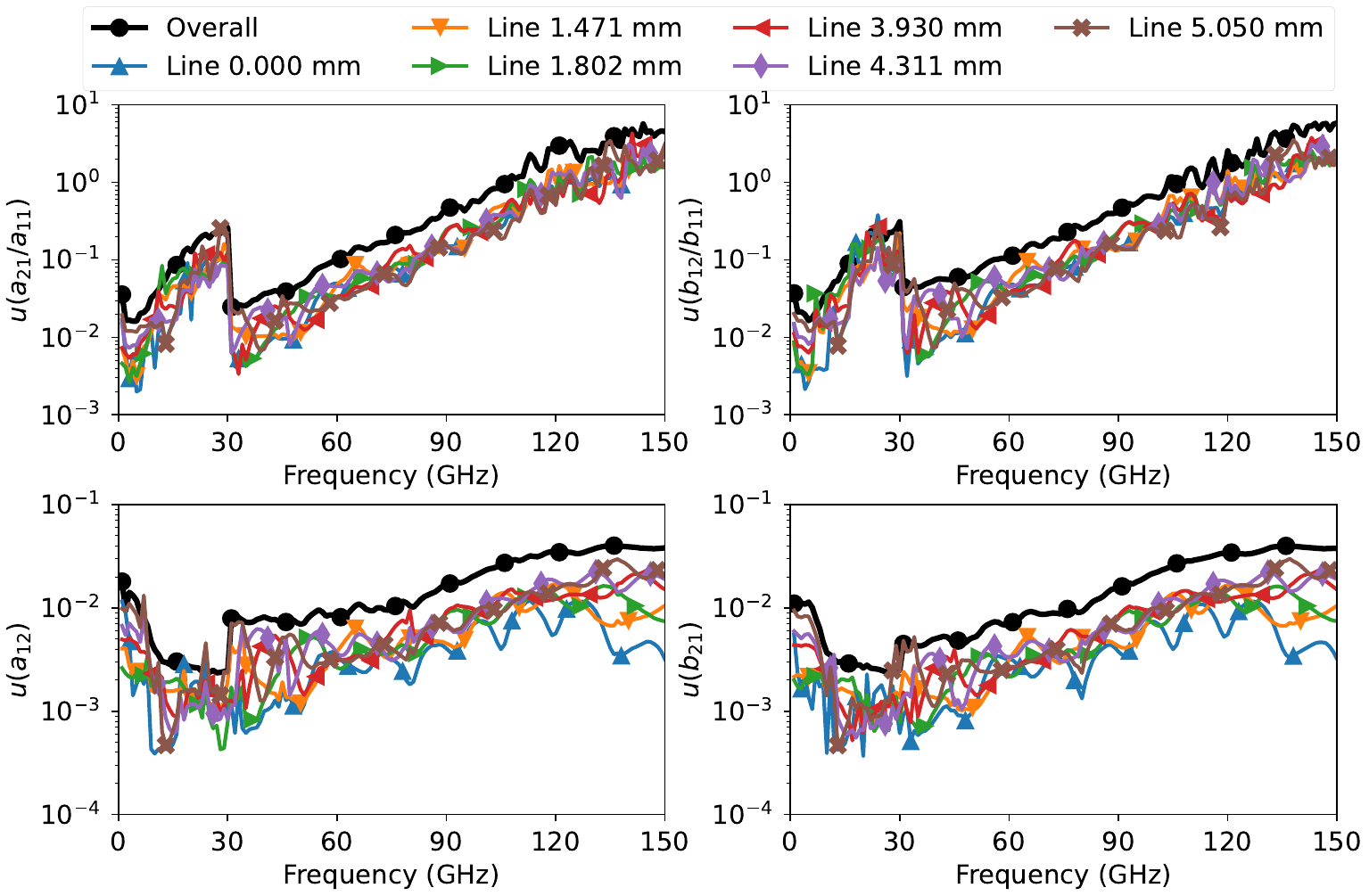}
    \caption{Per-line uncertainty budget of the normalized error terms for the fine-resolution optimized line length set ($1\,\mu\mathrm{m}$ quantization steps), obtained via linear uncertainty propagation \cite{Hatab2023}. Compare with the commercial ISS budget in Fig.~\ref{fig:4.16}.}
    \label{fig:C.3}
\end{figure}

This appendix supplements the per-line uncertainty analysis presented in Section~\ref{sec:4b} by providing the uncertainty budgets for the remaining three line length sets. Figs.~\ref{fig:C.1}--\ref{fig:C.3} show the per-line uncertainty budgets for the optimized ($50\,\mu\mathrm{m}$ steps), Golomb sparse ruler, and fine-resolution optimized ($1\,\mu\mathrm{m}$ steps) line length sets, respectively, computed using the same linear uncertainty propagation framework from \cite{Hatab2023}. In all three cases, the uncertainty contributions are distributed more evenly across the individual lines compared to the commercial ISS (Fig.~\ref{fig:4.16}), with no single line dominating the budget at any particular frequency. This confirms the observation from Section~\ref{sec:4b} that the proposed line length selection methods yield calibration kits with more balanced uncertainty distributions, which contributes to their lower and more uniform MAE in the MC analysis.

\section*{Acknowledgment}
The authors thank AT\&S, Leoben, Austria, for the production of the PCBs used in this paper. 

\bibliographystyle{IEEEtran}
\bibliography{References/references.bib}

\begin{IEEEbiography}[{\includegraphics[width=1in,height=1.25in,clip,keepaspectratio]{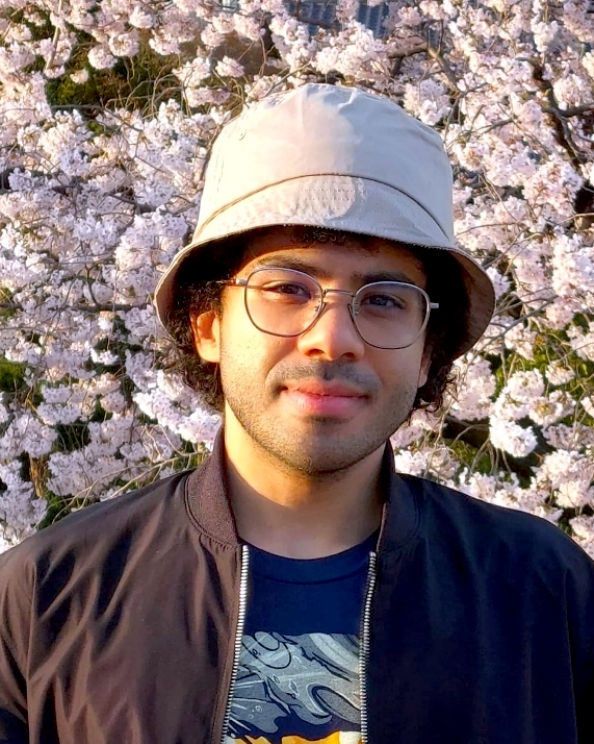}}]{Ziad Hatab} (Member, IEEE) received the Dipl.-Ing. and Dr.techn. degrees in electrical engineering from Graz University of Technology, Graz, Austria, in 2020 and 2024, respectively, at the Institute of Microwave and Photonic Engineering.

In 2020, he joined the Christian Doppler Laboratory for Technology-Guided Electronic Component Design and Characterization in Graz as a researcher. His research focused on vector network analyzer measurement techniques and calibration methods for planar circuit characterization at millimeter-wave and beyond. Since 2024, he has been an R\&D engineer at Keysight Technologies, Santa Rosa, CA, USA, where he works on vector network analyzer system design and advanced measurement techniques at millimeter-wave and sub-terahertz frequencies.
\end{IEEEbiography}

\begin{IEEEbiography}[{\includegraphics[width=1in,height=1.25in,clip,keepaspectratio]{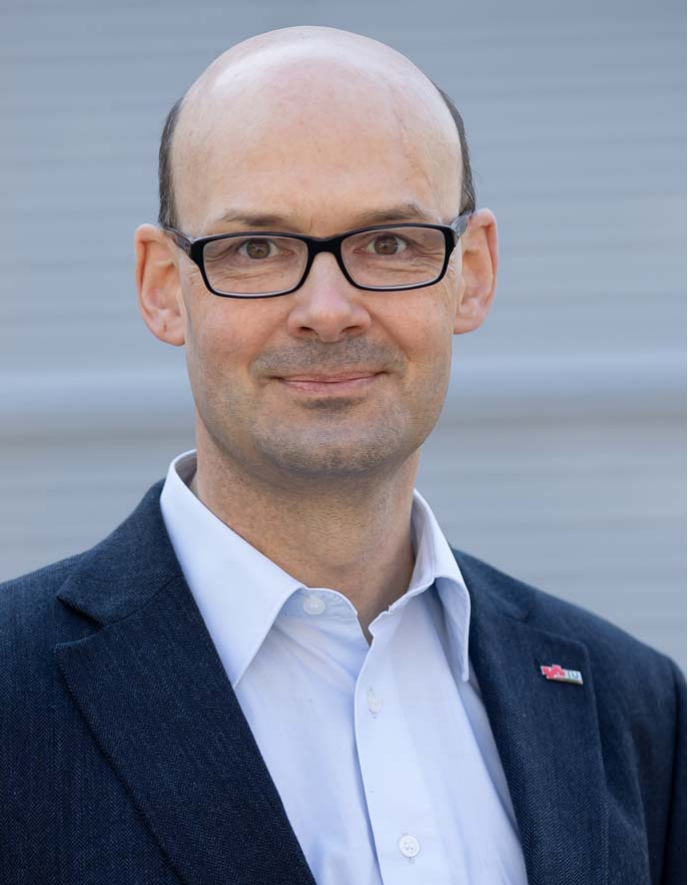}}]{Michael E. Gadringer} (Senior Member, IEEE) is an Associate Professor at the Institute of Microwave and Photonic Engineering at Graz University of Technology, Austria. He received the Dipl.-Ing. and Dr. techn. degrees from Vienna University of Technology, Austria, in 2002 and 2012, respectively. In 2010, he changed to Graz University of Technology. From 2016 to 2021, Michael Gadringer held a tenure-track research and teaching position at this institute. He visited Rohde \& Schwarz GmbH in Munich in 2017 and Infineon Technology AG in 2018 in this context.

During his studies, he was involved in designing analog and digital linearization systems for power amplifiers and behavioral modeling of microwave circuits. His current research focuses on developing broadband microwave and mm-wave systems. Additionally, he is involved in planning and implementing complex measurements, with a focus on calibration techniques and material characterization. Michael Gadringer has authored more than 30 journal articles and 75 conference papers. He holds four worldwide patents and has co-edited the book “RF Power Amplifier Behavioral Modeling,” published by Cambridge University Press.

Michael Gadringer was a member of the IEEE 1765 standard working group on the recommended practice for estimating the Error Vector Magnitude of digitally modulated signals. In addition, he is contributing to the IEEE P2822 working group on the recommended practice for Microwave, Millimeter-wave, and THz on-wafer calibrations, deembedding, and measurements.
The IEEE Instrumentation and Measurement Society selected him as a 2020 IEEE TIM Outstanding Reviewer. Since August 2022, he has served as an Associate Editor of the IEEE Transactions on Instrumentation and Measurements. For this service, the IEEE Instrumentation and Measurement Society recognized him as one of the outstanding Associate Editors of the IEEE Transactions on Instrumentation and Measurement of 2025.

\end{IEEEbiography}

\begin{IEEEbiography}[{\includegraphics[width=1in,height=1.25in,clip,keepaspectratio]{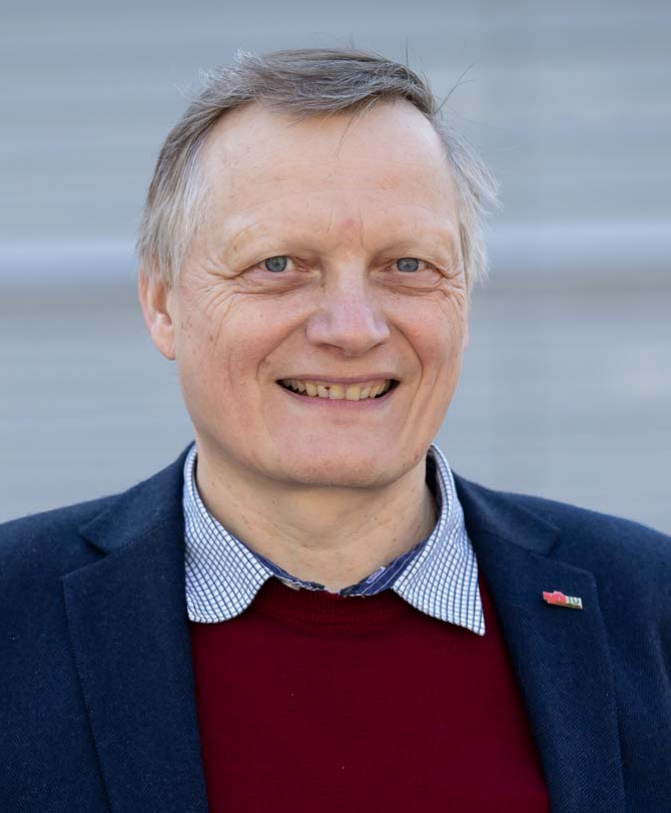}}]{Wolfgang Böosch} (Fellow, IEEE) received his Dipl.Ing. degree from the Technical University of Vienna, Austria, in 1985, his Ph.D. degree from the Graz University of Technology, Austria, in 1988, and his M.B.A. degree from the School of Management, University of Bradford, U.K., in 2004. In 2010, he joined Graz University of Technology to establish a new Institute for Microwave and Photonic Engineering. For nine years, he was the Dean of the Faculty of Electrical and Information Engineering, which currently incorporates 13 Institutes and 20 Full Professors covering the areas of Energy generation and distribution, Electronics, and Information Engineering. He is responsible for the strategic development, budget, and personnel of the Faculty.

Prior to this, he was the Chief Technology Officer (CTO) of the Advanced Digital Institute, Shipley, U.K. He was also the Director of Business and Technology Integration for RFMD, U.K. For almost ten years, he was with Filtronic plc, U.K., as the CTO of Filtronic Integrated Products and the Director of the Global Technology Group. Before joining Filtronic, he held positions with the European Space Agency (ESA), The Netherlands, working on amplifier linearization techniques, with MPR-Teltech, Burnaby, BC, Canada, working on MMIC technology projects, and with the Corporate Research and Development Group of M/A-COM, Boston, MA, USA, where he worked on advanced topologies for high-efficiency power amplifiers. For four years, he was with DaimlerChrysler Aerospace (now Hensoldt), Ulm, Germany, working on T/R modules for airborne radar.

He is a Fellow of the IEEE and the IET. He has published more than 180 papers and holds 4 patents.
\end{IEEEbiography}
\vfill

\end{document}